\documentclass{aa}  

\usepackage{threeparttable}
\usepackage{graphicx}
\usepackage{adjustbox}
\usepackage{txfonts}
\usepackage[colorlinks=true, linkcolor={blue!70!black}, citecolor={blue!70!black}, urlcolor={blue!70!black}]{hyperref}

\usepackage{amsmath}

\newcommand{\asym}[3]{\ensuremath{#1^{+#2}_{-#3}}}

\begin{document}

   \title{Exploring the gaseous halos of $z\,>\,3$ radio galaxies with UVES and JWST/NIRSpec}

   \subtitle{}

   \author{Jelena Ritter\inst{1,2}\thanks{E-mail: jritter@mpa-garching.mpg.de} \and Wuji Wang\inst{3}
          \and Dominika Wylezalek\inst{2} \and Carlos De Breuck\inst{4} \and Joël Vernet\inst{4} \and Fabrizio Arrigoni Battaia\inst{1}
          }

   \institute{Max-Planck-Institut f\"ur Astrophysik, Karl-Schwarzschild-Str. 1, D-85748 Garching bei M\"unchen, Germany
   \and
   Astronomisches Rechen-Institut, Zentrum f\"ur Astronomie der Universit\"at Heidelberg, M\"onchhofstr. 12-14, D-69120 Heidelberg, Germany
   \and
    IPAC, California Institute of Technology, 1200 E California Blvd., Pasadena, CA 91125, USA
   \and
   European Southern Observatory, Karl-Schwarzschild-Straße 2, D-85748 Garching-bei-M\"unchen, Germany
             }

   \date{Received xxxx; Accepted xxxx}

  \abstract
  {High-redshift radio galaxies (HzRGs) are among the most massive galaxies in the Universe and sites of extreme active galactic nuclei (AGN) feedback processes, powering energetic radio jets. They are typically embedded in giant Ly$\alpha$ halos that are known to extend over $100\,\text{kpc}$ into the circumgalactic medium (CGM). In this paper, we target the Ly$\alpha$ halos around four high-redshift radio galaxies in a redshift range of 3.1 < z < 4.5 using high-resolution spectroscopy from the Ultraviolet Echelle Spectograph (UVES) at the VLT, focusing on absorption features in the Ly$\alpha$ emission that trace neutral hydrogen (H\,\textsc{i}) systems. We compare the UVES data to Multi Unit Spectroscopic Explorer (MUSE) observations of the same targets and find that the higher spectral resolution of UVES ($\Delta v \approx 12\,\text{km} \text{s}^{-1}$) allows for a more complete identification of absorbers and reveals the splitting of deep absorbers into multiple components. We identify between 6 and 14 absorbers for each  target in our sample with column densities of $N_{\text{H\,\textsc{i}}} = 10^{12}-10^{17}\,\text{cm}^{-2}$. About 70\,\% of the absorbers can be spatially resolved along the radio jet axis, showing minimal variation in column densities over extents of more than 30 kpc. Our results indicate that a fraction of the absorbers may be physically associated with the host systems. Complementary JWST/NIRSpec observations of two of the targets, 4C+03.24 and TN\,J0205, reveal potential outflows in the ionized interstellar medium (ISM). We discuss a kinematic link between the [\ion{O}{iii}]-emitting gas and the cool halo gas as traced by Ly$\alpha$, suggesting a common outflow origin. }

   \keywords{galaxies: active–galaxies: evolution–galaxies: high-redshift–galaxies: halos
               }

   \maketitle
%

\section{Introduction} \label{sec:intro}
High-redshift radio galaxies (HzRGs) are massive galaxies (with stellar masses between $10^{11}\,\text{M}_{\odot}$ and $10^{12}\,\text{M}_{\odot}$, \citealt{Rocca-Volmerange2004, Seymour2007}), residing in overdense protocluster environments \citep[e.g.][]{Venemans2002, Wylezalek2013, Noirot2016, Noirot2018}. They host powerful active galactic nuclei (AGN), which significantly impact the evolution of the host galaxy through feedback processes \citep{HeckmannBest2014, HarrisonRamosAlmeida2024}. HzRGs exhibit both radiative mode feedback, where galactic outflows result from AGN-driven winds or radiation pressure pushing on dust and gas, and kinetic mode feedback, where collimated jets and outflows transfer energy to the surrounding medium \citep[e.g.][]{Fabian2012, KingPounds2015, Harrison2017, Nesvadba2017a, Nesvadba2017b}. This leads to the expulsion and redistribution of gas within the host galaxy and its surrounding environment. Additionally, strong photoionization by the central AGN impacts gas cooling, further influencing the buildup of the host galaxy's stellar mass.
This circumstance makes HzRGs unique and ideal targets for studying the evolution of massive galaxies and the role of AGN feedback processes. 

An important characteristic of HzRGs is
the giant gaseous halos in which they are embedded and which reach out into the galaxies' circumgalactic medium (CGM) (see \citet{Tumlinson2017} for a review) and often extend beyond the radio source \citep{vanOjik1997}. This extended halo structure provides a unique window into AGN-driven feedback and gas dynamics on circumgalactic scales. The gaseous halos can be observed in various emission lines, the most prominent and brightest among which is Ly$\alpha\,\lambda 1216$ which can reach luminosities of $\gtrapprox 10^{44}\,\text{erg}\,\text{s}^{-1}$ \citep[e.g][]{Villar-Martin2007, MileyDeBreuck2008, Shukla2022, Wang2023}. Additionally, the halos show \ion{He}{ii}\,$\lambda 1640$ emission and are metal-enriched, detected in \ion{C}{iv}\,$\lambda\lambda1548,1551$, N\,\textsc{v}\,$\lambda\lambda1238,1242$ as well as in [O\textsc{iii}]\,$\lambda\lambda4959,5007$ and [O\textsc{ii}]\,$\lambda\lambda3726,3729$ \citep[e.g.][]{McCarthy1993_review, DeBreuck2000, Villar-Martin2007_SF, Humphrey2008, Kolwa2019}. Apart from atomic gas emission, molecular gas, in particular carbon monoxide (CO), which is used as a tracer of molecular hydrogen, has been observed in HzRGs, in some cases even on halo scales \citep{Klamer2005, Ivison2012, Emonts2014,Emonts2015, Emonts2016, Gullberg2016, Li2021, Falkendal2021, DeBreuck2022, Emonts2023}.

The Ly$\alpha$ nebulae of HzRGs are ubiquitously observed to show deep extended absorption features \citep{vanOjik1997, Jarvis2003, Wilman2004, Humphrey2008b,Silva2018b, Kolwa2019, Wang2021}, first discovered by \citet{Rottgering1995}. This absorption of the Ly$\alpha$ emission in HzRGs has been interpreted as arising from neutral hydrogen gas (H\,\textsc{i} absorbers) in the CGM of the hosts, surrounding the emission region. The absorbers allow us to indirectly probe the neutral gas, without having to study H\,\textsc{i} in emission. While H\,\textsc{i} 21\,cm absorption against the radio continuum is commonly used to detect  neutral gas in low-redshift radio galaxies \citep[e.g.][]{Morganti2018}, such observations are challenging at high redshift and detections remain rare \citep[e.g.][]{Rottgering1999, Cody2003, Chandra2004, Curran2013}.

 The absorbers in the Ly$\alpha$ profiles are found to be spatially extended over almost the entire Ly$\alpha$ emission region, with projected spatial extents of up to 50\,kpc and covering factors of approximately unity \citep{vanOjik1997,Humphrey2008b, Swinbank2015, Silva2018a, Silva2018b}. The estimated gas masses of H\,\textsc{i} can reach  log($M_{\text{H}\textsc{i}}/M_{\odot}) \sim 9$ \citep{Silva2018a}. 
 
By modeling the absorption features, the neutral hydrogen gas can be characterized in terms of column densities and kinematics. Most absorbers are found to be blueshifted with respect to the systemic redshift of the HzRGs \citep{vanOjik1997, Jarvis2003, Humphrey2008b}, likely indicating outflowing gas. In a few cases, redshifted absorbers have been detected, which could be a signature of infalling gas, feeding the central AGN \citep{Kolwa2019}. 

A common theory explaining the nature of the absorbing gas has been developed over the years. \citet{Binette2000} first proposed that the absorption arises from giant expanding shells of gas, based on the blueshifted velocities, large spatial extents and the fact that the absorbing material appears spatially distinct from the Ly$\alpha$ emission region. The expansion could be driven by energetic AGN or stellar feedback processes. The idea was further developed and confirmed using a larger sample of targets \citep{Jarvis2003, Wilman2004, Silva2018a, Kolwa2019, Wang2021}. The absorbing gas is found to extend both along as well as perpendicular to the radio jet axis, thus providing evidence for the large-scale shell theory \citep{Humphrey2008b, Silva2018a, Silva2018b}. \citet{Wilman2004} found that the absorbers are either weak, with column densities of $N_{\text{H}\textsc{i}}\approx 10^{13}-10^{15}\,\text{cm}^{-2}$ or strong with column densities $\gtrsim 10^{18}\,\text{cm}^{-2}$. Hydrodynamical simulation works suggest that the strong column density absorbers could form behind the radio jet's bow shock and get fragmented by the propagating jet \citep{Krause2002,Krause2005}. 

When modeling the Ly$\alpha$ profiles of HzRGs, it is generally assumed that the Ly$\alpha$ emission region is spatially separate from the absorbers and that the Ly$\alpha$ photons are primarily absorbed or scattered out of our line of sight \citep{Binette2000, Jarvis2003, Wilman2004, Swinbank2015, Silva2018b, Kolwa2019, Wang2021, Wang2023}. While Ly$\alpha$ radiative transfer modeling is successful in reproducing some of the line shapes observed in high-redshift galaxies \citep[e.g.][]{Verhamme2006, Dijkstra2006, Gronke2015, Park2022}, profiles as seen in HzRGs with several absorption features and deep absorbers split into several components \citep[e.g.][]{Jarvis2003}, are not fully accounted for by radiative transfer simulations. In studies of Ly$\alpha$ emission in HzRGs, it is therefore commonly assumed that absorption superimposed on the intrinsic Ly$\alpha$ emission, which is likely influenced by radiative transfer effects, is the primary factor shaping the profiles.

Scenarios for the evolution of the absorbing gas shells up to the present have been proposed and discussed. In a study of 18 HzRGs, \citet{vanOjik1997} found that smaller radio sources exhibit Ly$\alpha$ absorption more frequently than larger radio sources. Following these findings, \citet{RöttgeringPentericci1999} proposed an explanation based on an evolutionary scenario. While propagating through the halo, the radio jets interact with the gas, leading to the formation of cocoons and shocks that eventually ionize the H\,\textsc{i} gas, thereby reducing the observed absorption. \citet{Krause2002, Krause2005, Swinbank2015} argue that the outflowing shells might become thermally and gravitationally unstable and fragment over time. 

In this paper, we present the results of observations of four HzRGs in a redshift range of 3.1 < z < 4.5. In particular, we focus on the absorption features observed in their Ly$\alpha$ halos. We present VLT Ultraviolet echelle spectrograph (UVES) observations, which allow us to analyze the absorbers at high spectral resolution of $\Delta v \approx 12\,\text{km} \text{s}^{-1}$. In Section \ref{sec:MUSE_compar} we compare this resolution to complementary VLT Multi Unit Spectroscopic Explorer (MUSE) observations of the four targets to assess how the different spectral capabilities affect the characterization of the absorbers. Additional JWST/NIRSpec integral field unit (IFU) observations of two of the four targets in our sample allow us to compare the observations of the Ly$\alpha$ halos to the ionized phase in the ISM of the HzRGs, as traced by [\ion{O}{iii}]. 

The paper is structured as follows. In Section \ref{sec:obs}, we present the observations and data reductions. The line fitting procedure of the UVES spectra is outlined in Section \ref{sec:linefitting}. The results are presented in Section \ref{sec:results}, along with the spatially resolved fitting of the Ly$\alpha$ profiles and the comparison of the Ly$\alpha$ profiles to the NIRSpec [\ion{O}{iii}] detections. We discuss and interpret our findings in Section \ref{sec:discussion} and finally summarize the results in Section \ref{sec:conclusions}. 

Throughout this work, we assume a flat $\Lambda\text{CDM}$ cosmology with $\text{H}_0 = 70\,\text{km}\,\text{s}^{-1}\,\text{Mpc}^{-1}$ and $\Omega_{\text{m}} = 0.3$.

\section{Observations and data reduction} \label{sec:obs}
\subsection{UVES observations}
\begin{table*}[!ht]
    \caption{Details of the targets in the sample.}
    \label{tab:UVES_data}
    \centering
    \begin{tabular}{c c c c c}
    \hline
    Target Name & Redshift & RA\tablefootmark{*} (J2000) & DEC\tablefootmark{*} (J2000) & Position angle\tablefootmark{**} \\
    & z & hh:mm:ss & dd:mm:ss & degree \\
    \hline
    4C+03.24 & 3.5657\tablefootmark{\textdagger} & 12:45:38.37 & +03:23:21.0 & 156 \\
    4C+04.11 & 4.5077\tablefootmark{\textdaggerdbl} & 03:11:47.97 & +05:08:03.74 & 162 \\
    MRC 0316-257 & 3.1238\tablefootmark{\textdaggerdbl} & 03:18:12.07 & -25:35:10.22 & 55 \\
    TN J0205+2242 & 3.5060\tablefootmark{\textdaggerdbl} & 02:05:10.69 & +22:42:50.4 & 153 \\
    \hline
    \end{tabular}
    \tablefoot{\tablefoottext{*}{Coordinates of the central AGN.}
    \tablefoottext{**}{Radio axis position angle (from North to East) as specified for the UVES observations.}
    \tablefoottext{\textdagger}{Redshift determined from the [\ion{O}{iii}] emission line \citep{Nesvadba2017a}.}
    \tablefoottext{\textdaggerdbl}{Redshift determined from the \ion{He}{ii} emission line (\citet{Kolwa2023} and \citet{Wang2023} for MRC\,0316).}
    }

\end{table*}

The four HzRGs of this work's sample were observed with the UVES instrument at the VLT under the program 108.21WL (PI: Wuji Wang) between November 2021 and March 2022. These four targets are part of a larger sample of eight HzRGs which have all been previously observed using VLT/MUSE. The selection criteria for the eight HzRGs included redshifts z > 2.9 to ensure the Ly$\alpha$ emission line falls within the MUSE spectral range, Dec < 25° for observability from the southern hemisphere and the presence of known extended bright Ly$\alpha$ emission nebulae (> 10 ", \citealt{Wang2023}). The four HzRGs targeted in this study were then specifically followed up with UVES observations, as they were among the sources in the eight-object sample that lacked existing high spectral resolution data. Three additional objects in the larger sample have been observed at high spectral resolution and studied in previous works \citep{Jarvis2003, Wilman2004, Kolwa2019}. In this paper, we focus exclusively on the newly obtained UVES data. 

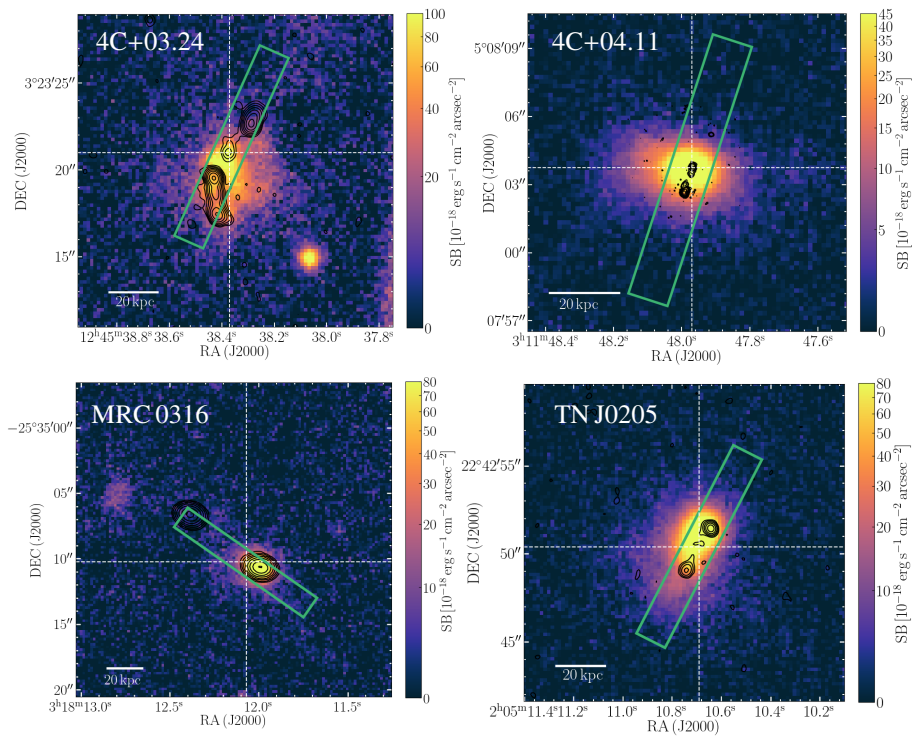
\begin{figure*}
\sidecaption
  \includegraphics[width=12cm]{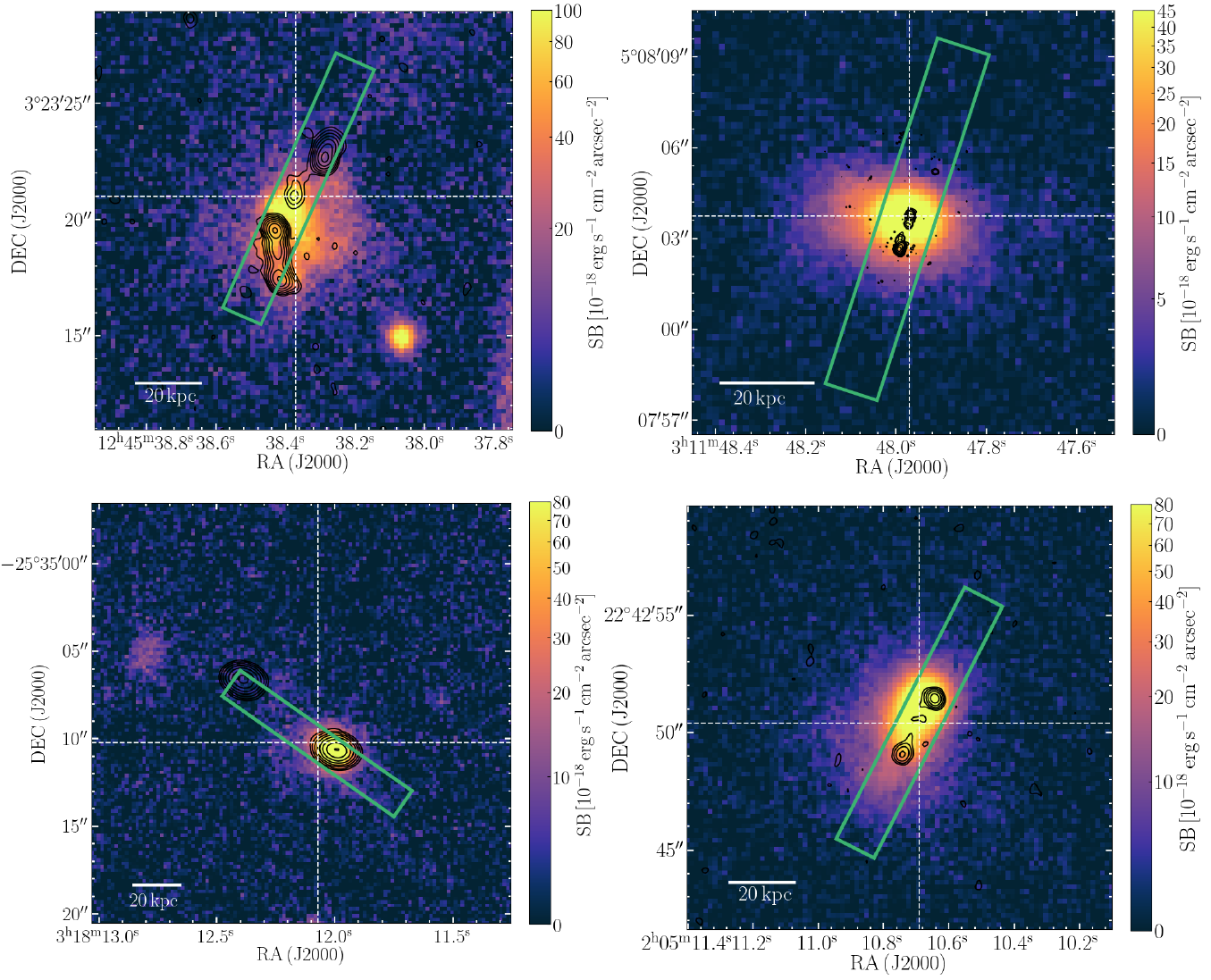}
      \begin{tikzpicture}[overlay, remember picture]
        \node[text=white] at (-10.2,9.25) {4C+03.24};
        \node[text=white] at (-4.2,9.25) {4C+04.11};
        \node[text=white] at (-10.2,4.3) {MRC\,0316};
        \node[text=white] at (-4.2,4.3) {TN\,J0205};
    \end{tikzpicture}
     \caption{MUSE surface brightness maps of the narrowband Ly$\alpha$ emission. The wavelength ranges in the observed frame for the narrowbands are 5550-5560$\,\text{\AA}$ for 4C+03.24, 6705-6710$\,\text{\AA}$ for 4C+04.11, 5018-5037$\,\text{\AA}$ for MRC\,0316 and 5475-5481$\,\text{\AA}$ for TN\,J0205. The radio contours (starting at $0.2$\,mJy/beam and increasing by a factor of 2), showing the extent of the radio emission, are overplotted in black lines. The green rectangle shows the position of the UVES slit on the sky. The white crosshairs indicate the position of the AGN. We note that the bright emission southwest of 4C+03.24 is likely a foreground star as its continuum resembles a blackbody \citep{Nesvadba2017a}.}
     \label{fig:MUSE_SB}
\end{figure*}

The UVES slit was centered on the central AGN position and oriented along the radio jet axis. The AGN coordinates and position angles of the radio jet axes are given in Table \ref{tab:UVES_data}. In Figure \ref{fig:MUSE_SB}, we show the position and orientation of the slits on the MUSE Ly$\alpha$ surface brightness maps. The black contours show the extent of the radio lobes. The radio data were recorded with the Very Large Array (VLA) in the case of 4C+03.24, MRC\,0316 and TN\,J0205 \citep[e.g.][]{Carilli1997, vanOjik1996, Nesvadba2007} and with the Multi-Element Radio Linked Interferometer Network (MERLIN) in the case of 4C+04.11 \citep{Parijskij2014}. 

A slit width of $1.8\,"$ was chosen, which corresponds to a resolving power of about $\lambda/\Delta\lambda = 26000$ and a spectral resolution of $\Delta\lambda \approx  0.19 - 0.27\,\text{\AA}$ over the wavelength range of the red arm or $\Delta v \approx 12\,\text{km}\,\text{s}^{-1}$ in terms of velocity (see UVES User manual\footnote{\label{UVES_manual}UVES User Manual, latest version: \url{https://www.eso.org/sci/facilities/paranal/instruments/uves/doc/ESO_514367_User_Manual_117.pdf}}).
The length of the slit on the sky is $12\,"$. The CCD read-out mode for the observations is a $2 \times 3$ (spatial $\times$ spectral) binning mode. The spatial pixel scale after the binning is $0.36\,"$. Additional binning is performed in the dispersion direction to improve the signal-to-noise ratio (SNR), resulting in pixel scales of 0.09\,$\text{\AA}$ for 4C+04.11, MRC\,0316 and TN\,J0205 and 0.14\,$\text{\AA}$ for 4C+03.24. 

The observations were divided into five Observation Blocks (OBs) of about 50 minutes exposure time each. For the target MRC\,0316 we include an additional OB, which was observed since one of the OBs did not fulfill the seeing conditions requested in the proposal. For 4C+03.24, only 4 OBs are included, as one OB shows a much lower SNR than the other four. The final spectra thus have total exposure times on target of about 4 hours and 10 minutes for 4C+04.11 and TN\,J0205, 5 hours for MRC\,0316 and 3 hours and 20 minutes for 4C+03.24. The exposure times, the spectral resolution around the Ly$\alpha$ emission region, average seeing disk diameters over all included OBs and other details of the UVES observations are summarized in Table \ref{tab:UVES_obs}. The spectra are not flux calibrated as no standard star was observed during the observing run.

\begin{table*}[!ht]
\caption{Details of the UVES observations.}
\label{tab:UVES_obs}
\begin{center}
\begin{tabular}{ c c c c c   }
 \hline
 Target Name&  UT date & exposure time & spectral resolution\tablefootmark{*}  &seeing\tablefootmark{**}  \\
    & dd/mm/yyyy &  hours & [\text{\AA}] & arcsec \\
 \hline
 4C+03.24  & 08/02/2022 - 01/03/2022  & 3.34 & 0.21 & $0.63\pm 0.06$ \\
 4C+04.11 &  02/12/2021 - 08/12/2021 &4.17 & 0.25 &$0.52\pm 0.07$\\
 MRC 0316-257 & 01/11/2021 - 30/12/2021  & 5.01& 0.19& $1.05\pm 0.24$\\
 TN J0205+2242   & 07/12/2021 - 27/12/2021 &4.17& 0.21 & $0.93\pm 0.20$\\
 \hline
\end{tabular}

\tablefoot{\tablefoottext{*}{Spectral resolution around the observed Ly$\alpha$ line.}
    \tablefoottext{**}{Seeing averaged over all included OBs with uncertainties given by the standard error of the mean.}
    }	   
\end{center}
\end{table*}

\subsection{UVES data reduction} \label{sec:UVES_data_red}
We develop an optimized data reduction method for the following spectral analysis. First, cosmic rays are removed in the raw science frames of each exposure using the \textsc{lacosmic}\footnote{\url{https://lacosmic.readthedocs.io/en/stable/}} package \citep{LA-cosmic, astro-scrappy} with the Laplacian SNR threshold set to a value of 3. The UVES data reduction pipeline is executed on the cleaned science frame via \textsc{esorex} (ESO Recipe Execution Tool, \citealt{Esorex2015}), a command-line driven utility to run the pipeline recipes, with the processed calibration files provided as input and the extraction method set to 2d. The pipeline performs the bias subtraction and flat-fielding and uses order-definition frames and ThAr arc frames to locate and define the orders and perform the wavelength calibration. By setting the extraction method to 2d, the 2-dimensional long-slit spectra, containing both spatial and spectral information, are extracted. To obtain the optimal signal extraction region for the final spectra, cumulative spatial flux profiles are evaluated around the Ly$\alpha$ wavelength range and the signal region is chosen as the spatial region in the 2d spectra, where the cumulative flux reaches between 10\,\% and 90\,\% of its maximum value. The median over the sky background, corresponding to regions above and below the signal region, is subtracted from the signal region. The sky-subtracted signal region is summed to obtain the bias-subtracted, flat-fielded, wavelength-calibrated, sky-subtracted 1d spectra. Different exposures are median-combined for each target. 
We call the spectra extracted in the described way the UVES master spectra.

\subsection{Ancillary MUSE data}
The four observed high-redshift radio galaxies listed in Table \ref{tab:UVES_obs} have previously been observed (PI: Joël Vernet and MUSE WFMAO commissioning observations for 4C+03.24) and analyzed, with focus on the tomography of the Ly$\alpha$ emission nebulae, using the MUSE instrument at VLT \citep{Vernet2017,Wang2021, Wang2023}. Using an integral field spectrograph allows to map the CGM and the spatial extent of the gaseous  Ly$\alpha$ emission nebulae. The observations were taken in Wide-Field-Mode (WFM) with a fov of $60\times60\,\text{arcsec}^2$ and Adaptive Optics (AO) was only employed for 4C+03.24. 4C+04.11, MRC\,0316 and TN\,J0205 have total exposure times of about 4.24\,h and 4C+03.24 of 1.25\,h. They cover an extended wavelength range of $4650 - 9300\,\text{\AA}$. 
The spatial sampling is $0.2\,"\,\text{pix}^{-1}$, the spectral sampling is 0.125\,nm\,$\text{pix}^{-1}$ and the resolving power is $\lambda/\Delta\lambda = 1750 - 3750$ for the covered wavelength range, corresponding to a resolution of  $\Delta v \approx 171-90\,\text{km}\,\text{s}^{-1}$ \citep{Wang2023}. The UVES spectra therefore provide a spectral resolution that is about 8 to 14 times higher compared to the spectral resolution of MUSE. The MUSE data reduction is detailed in \citet{Wang2023}. 

\subsection{Ancillary JWST/NIRSpec data}
4C+03.24 and TN\,J0205 have been observed using the NIRSpec IFU instrument onboard JWST as part of the JWST Cycle 1 General Observers program with proposal ID 1970 (PI: Wuji Wang). The observations zoom into the central $\sim$ $25 \times 25$\,$\text{kpc}^2$ surrounding the AGNs. 4C+03.24 and TN\,J0205 both have total exposure times of about 4 hours. For the observations a filter/disperser combination of F170LP/G235H was chosen with a resolving power of $\sim\,2700$ and a wavelength range of $1.66 - 3.17\,\mu \text{m}$ \citep{Böker2022}, covering [\ion{O}{ii}]$\lambda\lambda3726,3729$, H${\beta}$, [\ion{O}{iii}]$\lambda\lambda4959,5007$ and H${\alpha}$ at the redshifts of the targets. The spectral resolution is about $110\,\text{km}\,\text{s}^{-1}$. 

The data reduction performed on the JWST/NIRSPec data cubes is similar to the procedure described in \citet{Vayner2023}, \citet{Wang2024} and \citet{Wang2025}. To process the data, the JWST Science Calibration pipeline \footnote{\url{https://github.com/spacetelescope/jwst}} \citep{Bushouse2023} is used. 

The first and second stage are implemented following the standard pipeline, but the \textit{cube build} method that creates a 3d cube from the 2d spectra was applied with the \textit{emsm} approach instead of the default \textit{drizzle}, as using the \textit{emsm} method reduces low-frequency ripples due to undersampling \citep{Vayner2023, Wang2024}. 
To combine the cubes at each dithered position, the Python package \textsc{reproject} \footnote{\url{https://pypi.org/project/reproject/}} was implemented, following the script from \citet{Vayner2023}. The pixel scale after resampling is 0.05\,".


\section{Line Fitting Procedure} \label{sec:linefitting}
In order to characterize the absorption features in the halo emission of the four HzRGs, we implement a line fitting procedure that has been used successfully in previous works \citep[e.g.][]{Jarvis2003,Swinbank2015, Silva2018b, Kolwa2019, Wang2021}. We fit the Ly$\alpha$ profiles with an underlying Gaussian emission (consisting of one or two components) and superimposed Voigt profiles \citep[e.g.][]{Tepper-Garcia2006} for the absorption features. The use of a Gaussian profile for the underlying intrinsic emission is supported by previous studies \citep[e.g.][]{ArrigoniBattaia2019, Chang2023}.
The Gaussian model can be written as
\begin{equation}\label{Gaussian}
    F_{\lambda}=\frac{F}{\sigma_{\lambda}\sqrt{2\pi}}\text{exp} \left [ - \frac{1}{2} \left(\frac{\lambda-\lambda_0}{\sigma_{\lambda}}\right)^2\right].
\end{equation}
Here, $F_{\lambda}$ is the flux density at wavelength $\lambda$, $F$ is the integrated flux of the emission line, $\sigma_{\lambda}$ the line width, $\lambda_0$ is the center of the line and $\lambda$ the wavelength. 
The absorption can be quantified in terms of the optical depth $\tau_\lambda$ as proportional to $\text{e}^{-\tau_\lambda}$. More specifically the Voigt-Hjerting function, also called the Voigt profile, is denoted as
\begin{equation}
    \tau_\lambda= \frac{N\sqrt{\pi}e^2f{\lambda_0}^2}{\Delta \lambda_D m_e c^2}H(a,x),
\end{equation}
with the column density $N$ of the absorbing gas, the electron charge $e$, the electron mass $m_e$, the oscillator strength $f$ (values taken from \citealt{Kramida2023}) and the speed of light $c$. $\Delta \lambda_\text{D}$ depends on the Doppler parameter $b$, characterizing the width of the line, as $\Delta \lambda_\text{D}=\frac{b}{c}\lambda_0$. $H(a,x)$ is the Hjerting function, as approximated in \citet{Tepper-Garcia2006} for column densities up to $10^{22}\,\text{cm}^{-2}$ with $x\equiv\frac{\lambda-\lambda_0}{\Delta \lambda_D}$ and $a\equiv\frac{\lambda_0^2\Gamma}{4\pi c\Delta\lambda_D}$. $\Gamma$ is the Lorentzian width. The Voigt profiles ($\text{exp}({-\sum_{i=1}^{n} \tau_{\lambda,n}})$ with $n$ the number of absorbers) are then multiplied with the Gaussian profiles. In order to match the resolution of the UVES instrument, we convolve this with the line-spread function (LSF) of our observations using the Fast Fourier Transform (FFT) within the \textsc{scipy} library \citep{Virtanen2020}. The UVES LSF is determined to have an average Gaussian width of $\langle \sigma_\lambda \rangle \approx 0.10\,\text{\AA}$ or a full width at half maximum of $\langle \text{FWHM}\rangle \approx 0.24\,\text{\AA}$ from fitting a Gaussian profile to some of the strong skylines. The final fitting function is thus given by: 
\begin{equation}\label{Fit_function_final}
F=\left( \sum\limits_{j=1}^m F_ {\lambda,m} \times \text{exp}({-\sum_{i=1}^{n} \tau_{\lambda,n}}) \right) \circledast \mathrm{LSF}
\end{equation}
When modeling the absorbing gas, we assume that the gas clouds cover the extended emission line region and have a covering factor close to unity (C $\approx$ 1.0). This assumption is supported by observations of similar absorbers with MUSE and has therefore been adopted in previous works \citep[e.g.][]{Swinbank2015, Kolwa2019, Wang2021}. Moreover, the majority of absorbers in our UVES data span the full spatial extent of the Ly$\alpha$ emission region (see Figure \ref{fig:spatial_column_densities}), and several show saturated absorption troughs that reach zero flux (e.g. absorbers 2 and 3 in MRC\,0316, absorber~3 in TN\,J0205, and multiple systems in 4C+04.11, see Figure \ref{fig:UVES_fit}). All of this suggests a high covering fraction of the absorbing gas.

We first employ non-linear least-squares fitting, using the Python package \textsc{lmfit} \footnote{\url{https://lmfit.github.io/lmfit-py/}} \citep{Newville_lmfit} and minimizing the residual. The least-squares optimization is used to determine suitable initial parameter values and to identify the preferred model complexity. The final parameter inference is obtained from Markov Chain Monte Carlo (MCMC) sampling. We describe the details of the procedure in the following.

During the initial least-squares fitting, we fit the underlying Gaussian and the convolved Voigt profile simultaneously to the data, following the model derived above (see Equation \ref{Fit_function_final}). We start the fitting using only one underlying Gaussian emission component and then add another component, if using two components provides better fit results according to the Bayesian Information Criterion (BIC; \citealt{Liddle2007})\footnote{The results are also consistent using the Akaike Information Criterion (AIC) \citep{Liddle2007}}. For 4C+04.11, the blue wing of the Ly$\alpha$ line is heavily affected by absorbers, which makes it difficult to determine the underlying shape of the Gaussian. We therefore start by fitting a Gaussian model using only the red wing as input, following \citet{Wang2021}. For 4C+04.11, MRC\,0316 and TN\,J0205, we identify two Gaussian components and for 4C+03.24 one. We refer to the component closest to the systemic redshift as the systemic component, and to the others as blueshifted or redshifted depending on their velocity offset with respect to the systemic redshift. The continuum is fitted with a zero-order polynomial (the intercept) and left free to vary.

The position of the absorption troughs and the number of absorbers are identified visually. During the initial least-squares fitting, we start out with a low number of absorbers and gradually add more absorbers and narrow absorption features to the fitting. Different models with different number of absorbers are then compared using BIC to determine the preferred model complexity. The parameters of the absorbers are constrained based on previous studies of Ly$\alpha$ absorption systems in HzRGs. Observations suggest that these absorbers either have lower column densities between $10^{13}\text{cm}^{-2}$ and $10^{15}\,\text{cm}^{-2}$ or high column densities larger than $10^{18}\,\text{cm}^{2}$ \citep{Wilman2004}. The column density is therefore constrained to lie between $10^{12}\text{cm}^{-2}$ and $10^{20}\text{cm}^{-2}$ following \citet{Kolwa2019}. The lower limit on the Doppler parameter is set by the resolution of the UVES instrument as $b_\text{min} \propto \sigma_\lambda= 12\,\text{km}\, {\text{s}}^{-1} / \left(2\,\sqrt{2\,\text{ln}(2)}\right ) \approx 5\,\text{km}\,{\text{s}}^{-1}$. The upper limit is set to $b_\text{max}=400\,\text{km}\,{\text{s}}^{-1}$ following the argument in \citet{Kolwa2019}, that the absorbing gas is relatively quiescent compared to other regions exhibiting line widths with FWHM exceeding $1000\,\text{km} {\text{s}}^{-1}$. The redshifts of the absorbers are determined from visual inspection of the spectra and are free to vary by $\pm\,0.001$. These limits are applied as parameter bounds in the least-squares fitting and as priors in the subsequent MCMC sampling.

Likely due to the high number of fit parameters, \textsc{lmfit} did not return a covariance matrix from the least-squares fitting, and uncertainties on the fitted parameters could not be obtained this way. We therefore perform MCMC sampling to obtain parameter estimates and credible intervals using the Python package \textsc{emcee} \footnote{\url{https://emcee.readthedocs.io/en/stable/}}\citep{Foreman-Mackey2013} that implements the affine-invariant Ensemble Sampler proposed by \citet{Goodman-Weare2010}. For the MCMC analysis we adopt the preferred model complexity determined from the least-squares fitting and sample the full model parameter space simultaneously. The MCMC chains are initialized around the maximum-likelihood solution from the least-squares fit and explore the parameter space within the priors motivated above (see also Table \ref{tab:MCMC_constraints}). The predictive models shown in the figures (e.g. Figure \ref{fig:UVES_fit}) are derived from the MCMC posterior distributions. The shaded regions indicate the 16th–84th percentile range of the model realizations, representing the uncertainties.

\section{Results}\label{sec:results}
In this section, we first compare the resolution of the UVES spectra to complementary MUSE data. We then present the results of the Ly$\alpha$ line fitting, starting with the analysis of the UVES master spectra. We examine the spatial profiles along the radio jet axis and conclude by comparing these findings with JWST observations of 4C+03.24 and TN\,J0205.

\subsection{Comparison with MUSE} \label{sec:MUSE_compar}
\begin{figure*}[!ht]
    \centering
    \begin{tikzpicture}
        \node[anchor=north west,inner sep=0] at (0,0) {\includegraphics[width=0.48\textwidth]{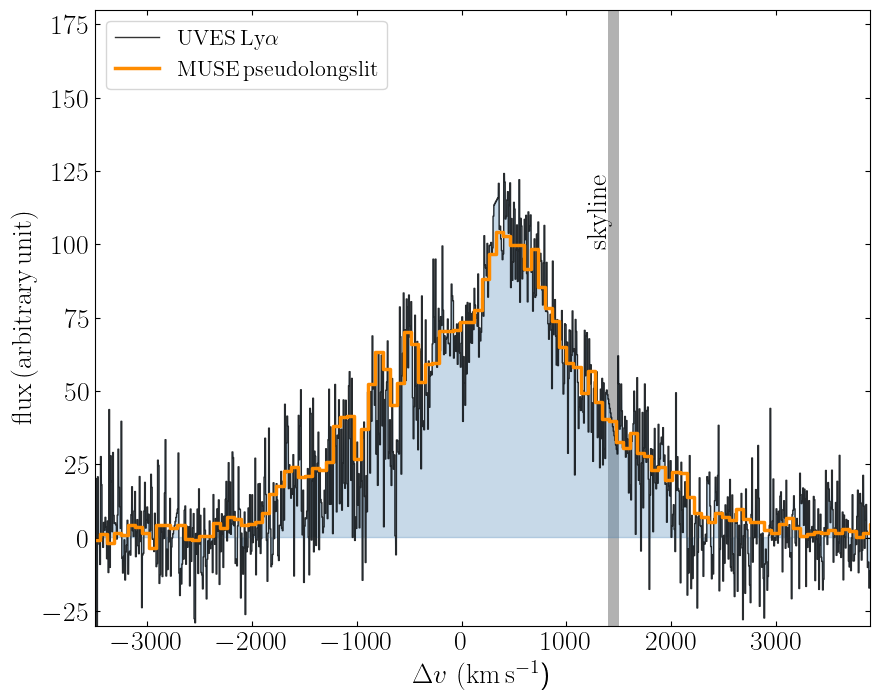}};
        \node[align=left, text=black, text width=0.1\textwidth, inner sep=1pt,font=\small\sffamily] at (8,-0.5) {4C+03.24};
    \end{tikzpicture}
    \quad
    \begin{tikzpicture}
        \node[anchor=north west,inner sep=0] at (0,0) {\includegraphics[width=0.48\textwidth]{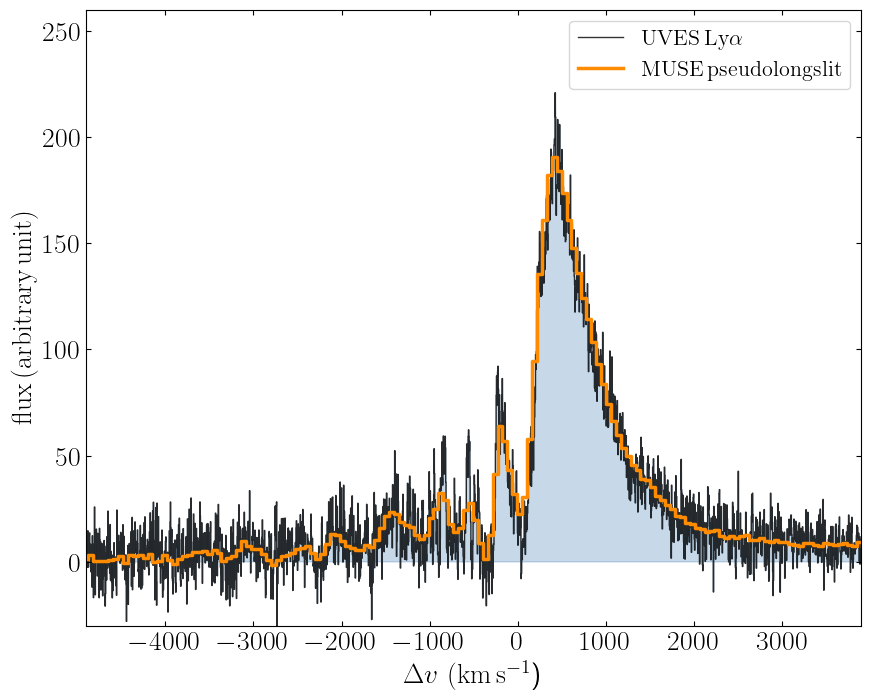}};
        \node[align=left, text=black, text width=0.1\textwidth, inner sep=1pt,font=\small\sffamily] at (2.0,-0.5) {\text{4C+04.11}};
    \end{tikzpicture}
    \begin{tikzpicture}
        \node[anchor=north west,inner sep=0] at (0,0) {\includegraphics[width=0.48\textwidth]{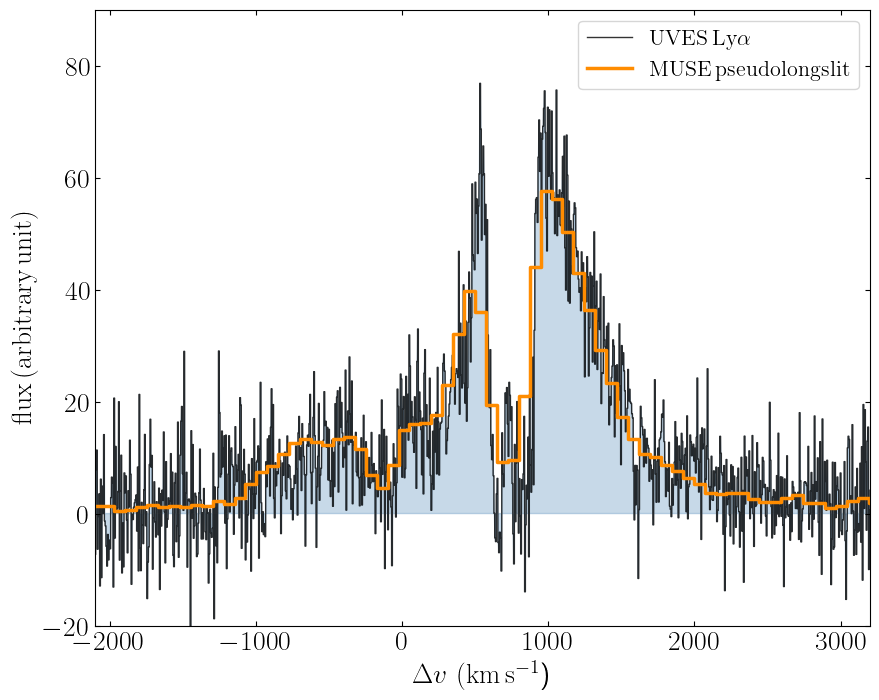}};
        \node[align=left, text=black, text width=0.1\textwidth, inner sep=1pt,font=\small\sffamily] at (2,-0.5) {\text{MRC\,0316}};
    \end{tikzpicture}
    \quad
    \begin{tikzpicture}
        \node[anchor=north west,inner sep=0] at (0,0) {\includegraphics[width=0.48\textwidth]{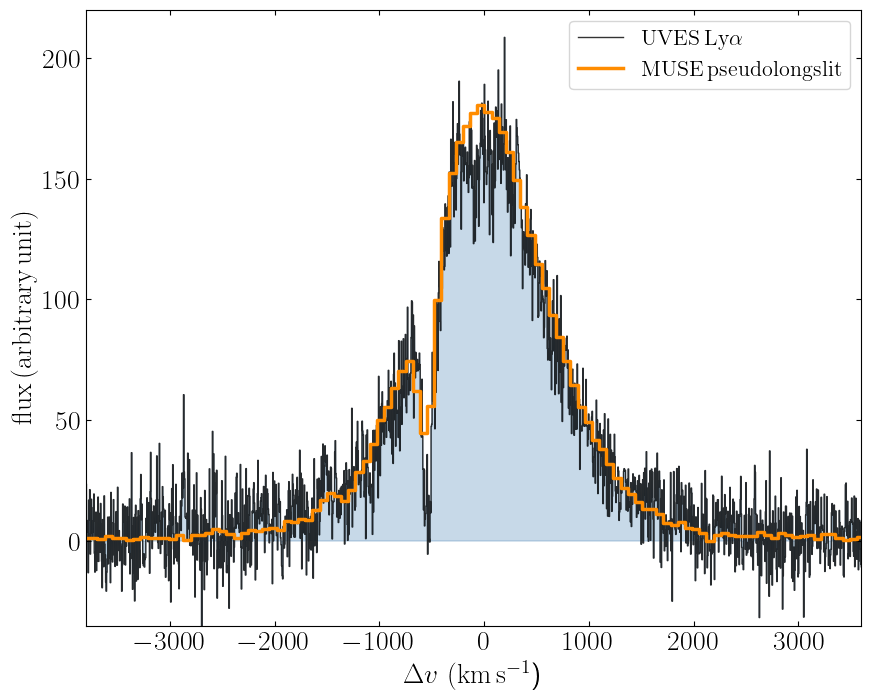}};
        \node[align=left, text=black, text width=0.1\textwidth, inner sep=1pt,font=\small\sffamily] at (2.0,-0.5) {\text{TN\,J0205}};
    \end{tikzpicture}

    \caption{The UVES spectra of the four targets around the Ly$\alpha$ emission with the MUSE pseudo-longslit spectra overplotted in orange. The grey shaded area in the spectrum of 4C+03.24 marks the position of the strong skyline around 5577$\,\text{\AA}$.} 
    \label{fig:MUSE_pseudo}
\end{figure*}
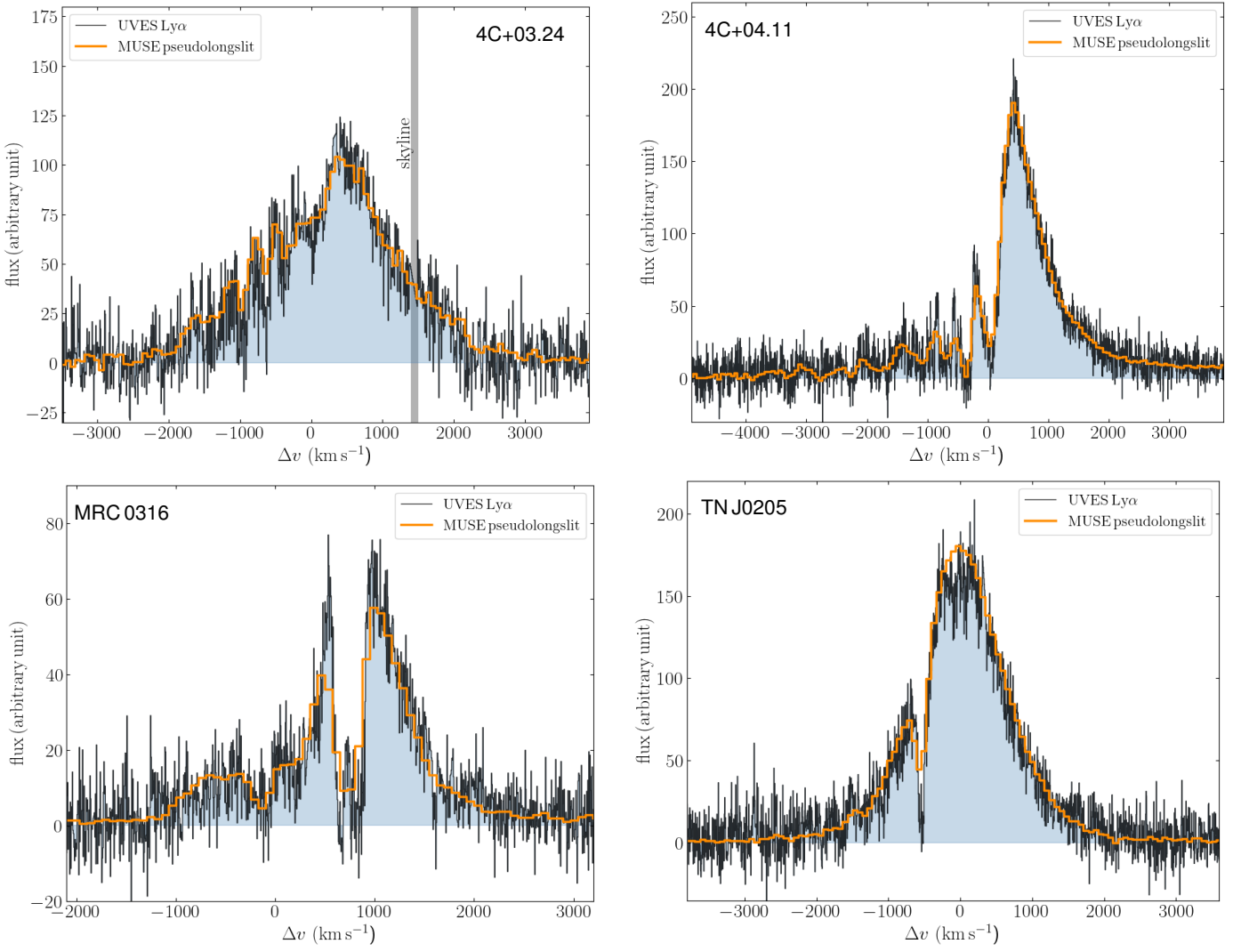

The MUSE data of the four targets in this work's sample have previously been analyzed, in \citet{Vernet2017, Wang2021, Kolwa2023, Wang2023}. To compare the UVES data to the MUSE data, we extract pseudo-longslit spectra from the MUSE data, simulating a slit width and length as used in the instrument setting for the UVES observations (slit width of $1.8\,\text{"}$ and a height of $12\,\text{"}$). To extract the 2-dimensional position-velocity pseudo-longslit spectra, the Python-package \textsc{pvextractor}\footnote{\url{https://pvextractor.readthedocs.io/en/latest/api.html\#module-pvextractor}} is used. The center and orientation of the slit in the extraction is chosen according to the values specified for the UVES observations as the center of the target and the position angle of the radio jet, which are listed in Table \ref{tab:UVES_data}. For better visual comparison, the MUSE pseudo-longslit spectra are scaled to the UVES spectra by interpolating the UVES data onto the wavelength grid of the MUSE data in a region alongside the main Ly$\alpha$ peak and calculating a scaling factor accordingly. The MUSE pseudo-longslit spectra and UVES spectra are overplotted in Figure \ref{fig:MUSE_pseudo}. UVES allows to observe much narrower absorption features, which get broadened due to the moderate resolution in the case of the MUSE instrument. For MRC\,0316, the deep absorber observed with MUSE at around $800\,\text{km}\,\text{s}^{-1}$ splits into two separate, almost fully saturated absorbers when observed with UVES. A similar result was previously found in \citet{Jarvis2003} for MRC\,0200+015, where a $N_{\text{H I}} \sim$ $10^{19}\,\text{cm}^{-2}$ absorber (observed with the ESO Multi Mode Instrument (EMMI): \citealt{vanOjik1997}) split into two $N_{\text{H I}} \sim$ $10^{14}\,\text{cm}^{-2}$ absorbers when observed with UVES. Some narrow absorbers, e.g. the very narrow feature on the red wing of the Ly$\alpha$ emission line of MRC\,0316 at around $1600\,\text{km}\,\text{s}^{-1}$ could not be identified using the MUSE observations. Additionally, for some of the deep features, visible in the spectra (e.g. the feature at around 0\,$\text{km}\,\text{s}^{-1}$ for 4C+04.11) for which previously only one absorber was identified (compare \citealt{Wang2021}), BIC and AIC now favor additional absorbers. For TN\,J0205, the dips observed on the emission line in the UVES spectra, but not in the MUSE data, are likely due to differences in observational setup and flux scaling.

\subsection{Line fitting results}
\begin{figure*}[!ht]
    \centering
    \begin{tikzpicture}
        \node[anchor=north west,inner sep=0] at (0,0) {\includegraphics[width=0.48\textwidth]{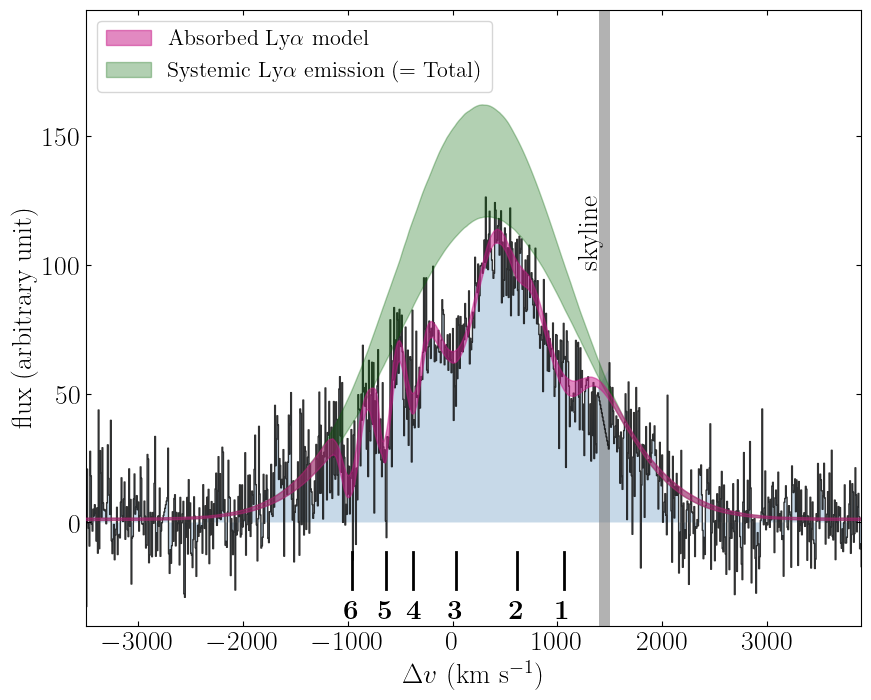}};
        \node[align=left, text=black, text width=0.1\textwidth, inner sep=1pt,font=\small\sffamily] at (8,-0.5) {4C+03.24};
    \end{tikzpicture}
    \quad
    \begin{tikzpicture}
        \node[anchor=north west,inner sep=0] at (0,0) {\includegraphics[width=0.48\textwidth]{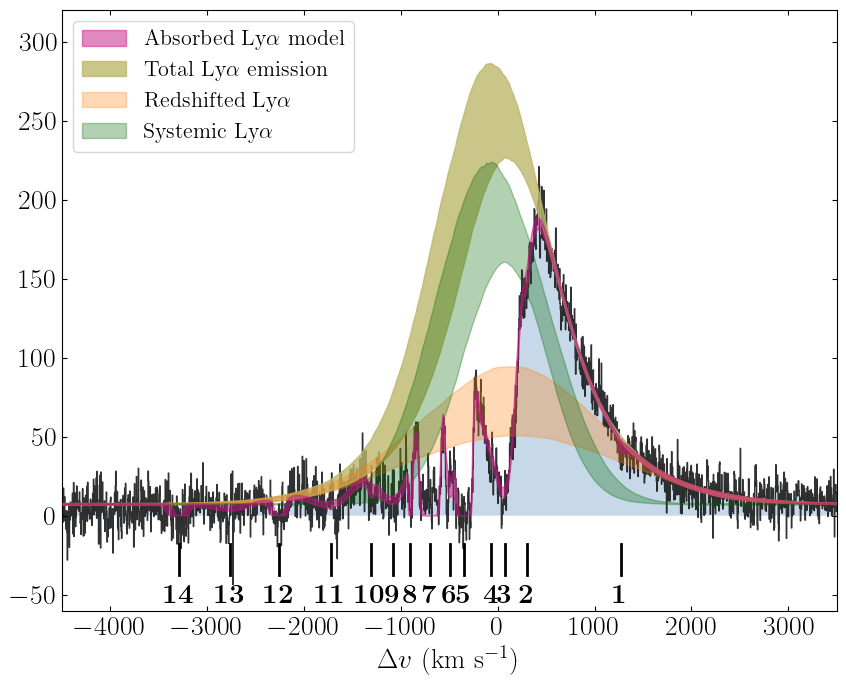}};
        \node[align=left, text=black, text width=0.1\textwidth, inner sep=1pt,font=\small\sffamily] at (8,-0.5) {\text{4C+04.11}};
    \end{tikzpicture}
    \begin{tikzpicture}
        \node[anchor=north west,inner sep=0] at (0,0) {\includegraphics[width=0.48\textwidth]{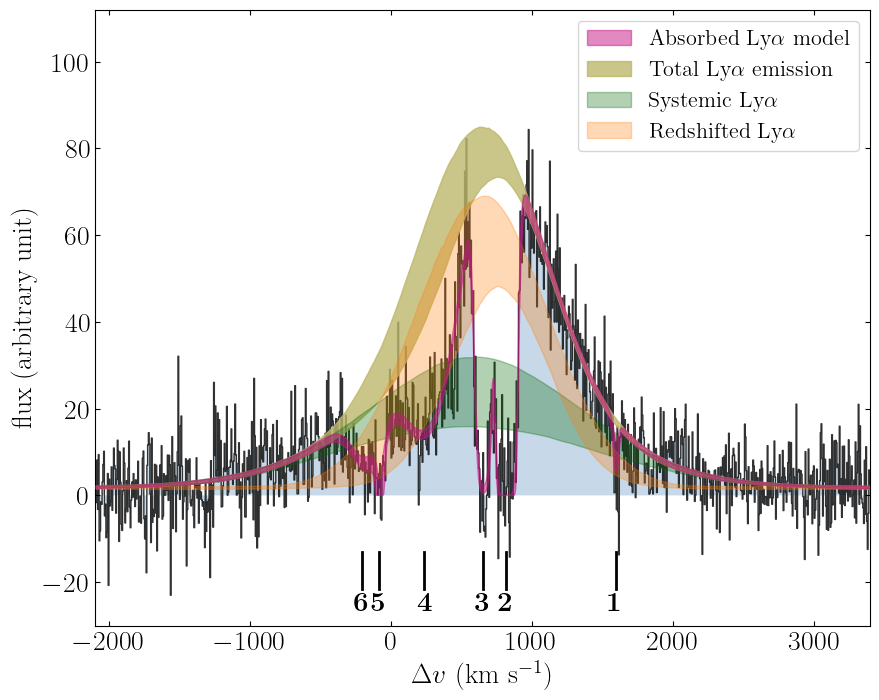}};
        \node[align=left, text=black, text width=0.1\textwidth, inner sep=1pt,font=\small\sffamily] at (2.3,-0.5) {\text{MRC\,0316}};
    \end{tikzpicture}
    \quad
    \begin{tikzpicture}
        \node[anchor=north west,inner sep=0] at (0,0) {\includegraphics[width=0.48\textwidth]{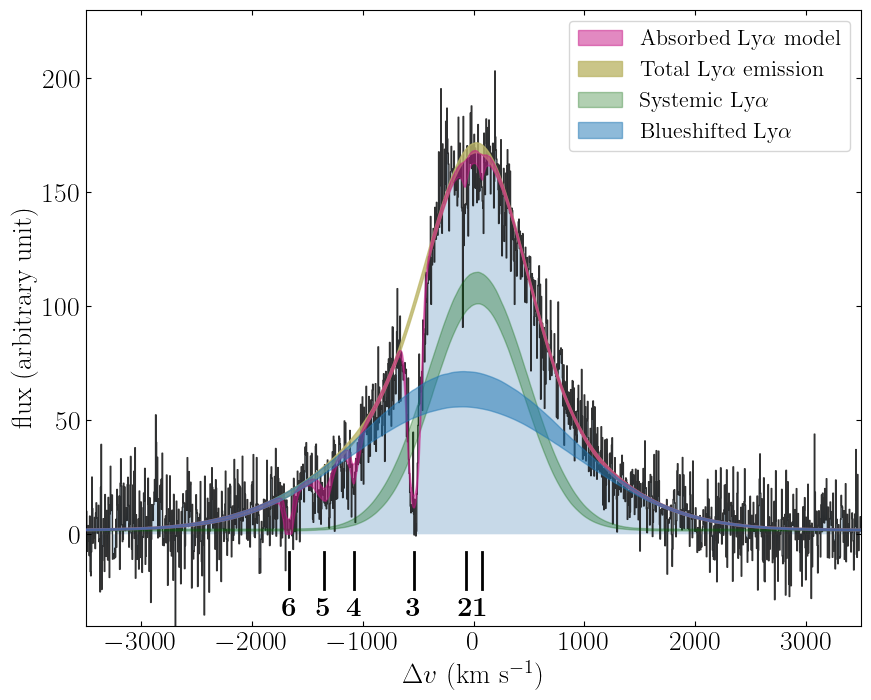}};
        \node[align=left, text=black, text width=0.1\textwidth, inner sep=1pt,font=\small\sffamily] at (2.3,-0.5) {\text{TN\,J0205}};
    \end{tikzpicture}

    \caption{UVES Ly$\alpha$ master spectra of the four targets. The shaded regions indicate the posterior predictive 68\% credible intervals of the Ly$\alpha$ model and its Gaussian emission components. Colours correspond to the full absorbed model (magenta), intrinsic Gausian emission (olive), systemic component (green), blueshifted component (blue), and redshifted component (orange). The systemic component is the one closest to the $0\,\text{km}\,{\text{s}}^{-1}$ as determined from the systemic redshift of the target. The position of the absorbers are marked with black vertical lines. The grey shaded area in the spectrum of 4C+03.24 marks the position of the strong skyline around 5577$\,\text{\AA}$, which was interpolated for the fitting.}
    \label{fig:UVES_fit}
\end{figure*}

We identify 6 absorbers for 4C+03.24, MRC\,0316 and TN\,J0205. 4C+04.11 shows
absorption features out to large velocity shifts from the Ly$\alpha$ line center and we
identify altogether 14 absorbers for this target. The fitted spectra can be seen
in Figure \ref{fig:UVES_fit}. The posterior parameter estimates for the Gaussian emission components and Voigt absorbers are listed in Tables \ref{tab: 4C0324_results}, \ref{tab:4C0411_results}, \ref{tab:MRC0316_results}, and \ref{tab:TNJ0205_results}.

The line centers for the underlying Gaussian emission of 4C+04.11 and TN\,J0205 are found to be close to the systemic velocity shift of the host galaxy, determined from the \ion{He}{ii} emission line at the position of the central AGN. This is to be expected if the Ly$\alpha$ emission gas is located in the potential well of the radio galaxy. For 4C+03.24 and MRC\,0316, the Ly$\alpha$ line is however redshifted, by about 300\,$\text{km}\,\text{s}^{-1}$ for 4C+03.24 and about 580 and 710 \,$\text{km}\,\text{s}^{-1}$ for the two velocity components of MRC\,0316. Such velocity offsets between Ly$\alpha$ emission and the systemic redshift are commonly observed and can arise from radiative transfer effects and bulk gas motions, in addition to uncertainties in the systemic redshift estimate. For 4C+03.24, \ion{He}{ii} emission could not be identified and the systemic redshift is therefore based on the [\ion{O}{iii}] emission from \citet{Nesvadba2017a}, which also agrees with the [\ion{O}{iii}] redshift at the host galaxy position in our NIRSpec data. 

We find that most of the absorbers are blueshifted with respect to the systemic redshift of the HzRG and the Ly$\alpha$ line center, which is in agreement with previous studies of HzRGs \citep[e.g.][]{Jarvis2003, Humphrey2008b, Kolwa2019}. This suggests that the absorbing gas is outflowing relative to the Ly$\alpha$ emission line region.
For MRC\,0316 and 4C+04.11, we identify absorbers (absorbers 1 for both targets) that are clearly redshifted with respect to both the Ly$\alpha$ line center and the systemic redshift of the AGN ($\Delta\,v_{\text{sys}}\sim1600\,\text{km}\text{s}^{-1}$ and $\Delta\,v_{\text{sys}}\sim1270\,\text{km}\text{s}^{-1}$ respectively). Both of these absorbers are found to be very narrow with low column densities log($N_{\text{H}\textsc{i}}/\text{cm}^{-2}$) of approximately 13.4 and 12.9 and could potentially trace inflowing gas. However, we cannot exclude the possibility that they are intervening metal absorbers at unrelated redshifts. This will be further discussed in Section \ref{sec:Origin_absorbers}. 

For our sample of four HzRGs, about 90\,\% of the absorbers are optically thin with column densities of $N_{\text{H\,\textsc{i}}} = 10^{12}-10^{15}\,\text{cm}^{-2}$. Only overall four absorbers have column densities $10^{15}\,\text{cm}^{-2}< N_{\text{H\textsc{i}}} < 10^{17}\,\text{cm}^{-2}$ and we do not find any high column densities $> 10^{17}\,\text{cm}^{-2}$. About  60\,\% of the identified absorbers have Doppler parameters $b < 80\,\text{km}\,\text{s}^{-1}$. The high spectral resolution of UVES allows to detect these narrow absorbers more accurately compared to instruments with lower spectral resolution, such as MUSE.

There is a known degeneracy between column density and Doppler parameter, such that different combinations of these two parameters may lead to similar Voigt profiles, as discussed previously in \citet{Silva2018b}. Inspection of the MCMC corner plots (Figures \ref{fig:corner_4C0324}, \ref{fig:corner_MRC0316}, and \ref{fig:corner_TNJ0205} in Appendix \ref{chap:appendix_MCMC}) shows that several parameters are only weakly constrained and exhibit extended posterior tails and degeneracies. This behavior is expected given the large number of free parameters in the multi-component fits, the limited SNR in the data, and the intrinsic degeneracy between column density and Doppler parameter. In particular, we note that parameters associated with absorbers in the outer wings of the Ly$\alpha$ line are not as well constrained due to the decreased SNR in these regions.

We further note that some of the broad Ly$\alpha$ absorption features may in fact represent blends of multiple narrower components that cannot be fully resolved at the given spectral resolution. This effect is expected to become more pronounced toward higher redshift, where the number density of intergalactic absorbers increases. The inferred Doppler parameters should therefore be interpreted with caution.

\begin{table*}[htbp]
\caption[Best Ly$\alpha$ fit results for 4C+03.24]{Median posterior parameter estimates with 68\% credible intervals (16th–84th percentiles) from MCMC sampling of the Ly$\alpha$ emission and Voigt absorption model of 4C+03.24.}
\label{tab: 4C0324_results}
\centering
\small
\setlength{\tabcolsep}{5pt}
\renewcommand{\arraystretch}{1.15}
\begin{tabular}{l c c c c}
\hline
\multicolumn{5}{c}{Emission fitting results} \\
\hline
Line & Line center (rest) & Line center (obs.) & Line flux & Line width \\
 & $\lambda_0$ [\AA] & $\lambda$ [\AA] & $F$ [arb. unit] & FWHM [$\mathrm{km\,s^{-1}}$] \\
\hline
Ly$\alpha$ & 1215.67 & \asym{5555.98}{0.78}{0.63} & \asym{5174}{742}{512} & \asym{1973}{77}{84} \\
\hline
\multicolumn{5}{c}{Absorption fitting results} \\
\hline
 & Absorber & Absorber redshift & Column density & Doppler parameter \\
 &  & $z$ & $\log(N_{\mathrm{HI}}/\mathrm{cm}^{-2})$ & $b$ [$\mathrm{km\,s^{-1}}$] \\
\hline
 & 1 & \asym{3.5819}{0.0004}{0.0004} & \asym{14.22}{0.11}{0.11} & \asym{249}{31}{26} \\
 & 2 & \asym{3.5751}{0.0003}{0.0003} & \asym{13.72}{0.42}{0.50} & \asym{191}{89}{72} \\
 & 3 & \asym{3.5660}{0.0002}{0.0002} & \asym{14.40}{0.16}{0.17} & \asym{280}{52}{45} \\
 & 4 & \asym{3.5599}{0.0002}{0.0002} & \asym{13.94}{0.11}{0.11} & \asym{101}{19}{15} \\
 & 5 & \asym{3.5556}{0.0003}{0.0002} & \asym{14.03}{0.18}{0.11} & \asym{87}{38}{36} \\
 & 6 & \asym{3.5507}{0.0002}{0.0002} & \asym{14.15}{0.14}{0.17} & \asym{90}{16}{23} \\
\hline
\end{tabular}
\end{table*}

\begin{table*}[htbp]
\caption[Best Ly$\alpha$ fit results for 4C+04.11]{Median posterior parameter estimates with 68\% credible intervals (16th–84th percentiles) from MCMC sampling of the Ly$\alpha$ emission and Voigt absorption model of 4C+04.11.}
\label{tab:4C0411_results}
\centering
\small
\setlength{\tabcolsep}{5pt}
\renewcommand{\arraystretch}{1.15}
\begin{tabular}{l c c c c}
\hline
\multicolumn{5}{c}{Emission fitting results} \\
\hline
Line & Line center (rest) & Line center (obs.) & Line flux & Line width \\
 & $\lambda_0$ [\AA] & $\lambda$ [\AA] & $F$ [arb. unit] & FWHM [$\mathrm{km\,s^{-1}}$] \\
\hline
Ly$\alpha$ &
1215.67 &
\asym{6695.78}{2.09}{2.25} &
\asym{5130}{1460}{1180} &
\asym{1179}{150}{142} \\

Ly$\alpha$ (r.) &
1215.67 &
\asym{6699.17}{1.27}{2.03} &
\asym{3471}{903}{764} &
\asym{2343}{334}{236} \\
\hline
\multicolumn{5}{c}{Absorption fitting results} \\
\hline
 & Absorber & Absorber redshift & Column density & Doppler parameter \\
 &  & $z$ & $\log(N_{\mathrm{HI}}/\mathrm{cm}^{-2})$ & $b$ [$\mathrm{km\,s^{-1}}$] \\
\hline
 & 1  & \asym{4.5313}{0.0005}{0.0006} & \asym{12.88}{0.44}{0.55} & \asym{63}{108}{44} \\
 & 2  & \asym{4.5132}{0.0003}{0.0004} & \asym{12.89}{0.42}{0.53} & \asym{53}{45}{28} \\
 & 3  & \asym{4.5090}{0.0002}{0.0002} & \asym{14.25}{0.16}{0.12} & \asym{93}{21}{17} \\
 & 4  & \asym{4.5068}{0.0003}{0.0007} & \asym{14.76}{0.07}{0.11} & \asym{254}{26}{26} \\
 & 5  & \asym{4.5013}{0.0002}{0.0001} & \asym{15.47}{0.36}{0.39} & \asym{44}{9}{6} \\
 & 6  & \asym{4.4988}{0.0004}{0.0002} & \asym{14.34}{0.17}{0.15} & \asym{73}{34}{24} \\
 & 7  & \asym{4.4949}{0.0001}{0.0001} & \asym{15.58}{0.30}{0.47} & \asym{57}{14}{5} \\
 & 8  & \asym{4.4911}{0.0001}{0.0001} & \asym{14.05}{1.08}{0.39} & \asym{14}{11}{6} \\
 & 9  & \asym{4.4880}{0.0004}{0.0004} & \asym{14.66}{0.16}{0.17} & \asym{206}{52}{54} \\
 & 10 & \asym{4.4837}{0.0005}{0.0005} & \asym{13.32}{0.62}{0.84} & \asym{37}{124}{27} \\
 & 11 & \asym{4.4761}{0.0006}{0.0006} & \asym{14.48}{0.16}{0.20} & \asym{222}{65}{43} \\
 & 12 & \asym{4.4662}{0.0003}{0.0004} & \asym{15.07}{0.62}{0.56} & \asym{47}{23}{14} \\
 & 13 & \asym{4.4570}{0.0007}{0.0007} & \asym{14.31}{0.25}{0.47} & \asym{248}{88}{60} \\
 & 14 & \asym{4.4471}{0.0005}{0.0005} & \asym{16.81}{1.47}{1.89} & \asym{32}{28}{20} \\
\hline
\end{tabular}
\tablefoot{Ly$\alpha$ is the systemic emission component (the one closest to $0\,\text{km}\,{\text{s}}^{-1}$), and Ly$\alpha$ (r.) is the emission component redshifted with respect to the systemic component.}
\end{table*}

\begin{table*}[htbp]
\caption[Best Ly$\alpha$ fit results for MRC\,0316]{Median posterior parameter estimates with 68\% credible intervals (16th–84th percentiles) from MCMC sampling of the Ly$\alpha$ emission and Voigt absorption model of MRC\,0316.}
\label{tab:MRC0316_results}
\centering
\small
\setlength{\tabcolsep}{5pt}
\renewcommand{\arraystretch}{1.15}
\begin{tabular}{l c c c c}
\hline
\multicolumn{5}{c}{Emission fitting results} \\
\hline
Line & Line center (rest) & Line center (obs.) & Line flux & Line width \\
 & $\lambda_0$ [\AA] & $\lambda$ [\AA] & $F$ [arb. unit] & FWHM [$\mathrm{km\,s^{-1}}$] \\
\hline
Ly$\alpha$     & 1215.67 & \asym{5022.87}{0.61}{0.91} & \asym{761}{188}{215} & \asym{1942}{316}{204} \\
Ly$\alpha$ (r.)& 1215.67 & \asym{5025.12}{1.19}{1.23} & \asym{915}{367}{227} & \asym{930}{149}{110} \\
\hline
\multicolumn{5}{c}{Absorption fitting results} \\
\hline
 & Absorber & Absorber redshift & Column density & Doppler parameter \\
 &  & $z$ & $\log(N_{\mathrm{HI}}/\mathrm{cm}^{-2})$ & $b$ [$\mathrm{km\,s^{-1}}$] \\
\hline
 & 1 & \asym{3.1458}{0.0001}{0.0001} & \asym{13.37}{0.18}{0.25} & \asym{22}{11}{5} \\
 & 2 & \asym{3.1350}{0.00004}{0.00004} & \asym{15.02}{0.34}{0.19} & \asym{50}{7}{7} \\
 & 3 & \asym{3.1328}{0.0001}{0.0001} & \asym{14.45}{0.06}{0.05} & \asym{51}{5}{4} \\
 & 4 & \asym{3.1272}{0.0002}{0.0002} & \asym{14.57}{0.09}{0.13} & \asym{223}{25}{26} \\
 & 5 & \asym{3.1227}{0.0001}{0.0001} & \asym{14.29}{1.04}{0.41} & \asym{22}{15}{10} \\
 & 6 & \asym{3.1215}{0.0003}{0.0003} & \asym{14.29}{0.13}{0.16} & \asym{135}{32}{28} \\
\hline
\end{tabular}
\tablefoot{Ly$\alpha$ is the systemic emission component (the one closest to $0\,\mathrm{km\,s^{-1}}$), and Ly$\alpha$ (r.) is the emission component redshifted with respect to the systemic component.}
\end{table*}

\begin{table*}[htbp]
\caption[Best Ly$\alpha$ fit results for TN\,J0205]{Median posterior parameter estimates with 68\% credible intervals (16th–84th percentiles) from MCMC sampling of the Ly$\alpha$ emission and Voigt absorption model of TN\,J0205.}
\label{tab:TNJ0205_results}
\centering
\small
\setlength{\tabcolsep}{5pt}
\renewcommand{\arraystretch}{1.15}
\begin{tabular}{l c c c c}
 \hline
 \multicolumn{5}{c}{Emission fitting results} \\
 \hline
 Line & Line center (rest) & Line center (obs.) & Line flux & Line width\\
     & $\lambda_0$ [\text{\AA}] & $\lambda$ [\text{\AA}] & $F$ [arb. unit] & FWHM [$\text{km}\,\text{s}^{-1}$]\\
 \hline
Ly$\alpha$     & 1215.67 & \asym{5478.48}{0.19}{0.19} & \asym{2082}{257}{245} & \asym{1006}{54}{56} \\
Ly$\alpha$ (bl.)& 1215.67 & \asym{5476.15}{0.44}{0.50} & \asym{2716}{234}{243} & \asym{2260}{128}{109} \\
 \multicolumn{5}{c}{} \\
 \hline
\multicolumn{5}{c}{Absorption fitting results} \\
\hline
 & Absorber & Absorber redshift & Column density & Doppler parameter \\
 &  & $z$ & $\log(N_{\mathrm{HI}}/\mathrm{cm}^{-2})$ & $b$ [$\mathrm{km\,s^{-1}}$] \\
\hline
 & 1 & \asym{3.5071}{0.0004}{0.0005} & \asym{12.44}{0.30}{0.29} & \asym{55}{132}{30} \\
 & 2 & \asym{3.5049}{0.0005}{0.0004} & \asym{12.56}{0.30}{0.34} & \asym{66}{76}{38} \\
 & 3 & \asym{3.49792}{0.00004}{0.00004} & \asym{14.19}{0.03}{0.03} & \asym{60}{4}{4} \\
 & 4 & \asym{3.4898}{0.0002}{0.0002} & \asym{13.50}{0.14}{0.18} & \asym{53}{19}{23} \\
 & 5 & \asym{3.4858}{0.0003}{0.0003} & \asym{13.72}{0.13}{0.18} & \asym{75}{21}{21} \\
 & 6 & \asym{3.4809}{0.0002}{0.0002} & \asym{14.14}{0.91}{0.26} & \asym{40}{20}{19} \\
 \hline
\end{tabular}
\tablefoot{Ly$\alpha$ is the systemic emission component (the one closest to $0\,\text{km}\,{\text{s}}^{-1}$), and Ly$\alpha$ (bl.) is the emission component blueshifted with respect to the systemic component.}
\end{table*}

\subsection{Spatially resolved Ly$\alpha$ fitting} \label{sec:spatial}

The UVES data are long-slit spectra obtained using a slit-length on the sky of $12\,"$. The final 2d UVES spectra span a spatial extent of approximately $10.9\,"$ with a spatial pixel scale of about $0.36\,"$ due to i.a. projection effects or rebinning of the spectra during the data reduction. 
By extracting 1-dimensional spectra from different regions along the length of the slit, we can analyze the spatial distribution of the Ly$\alpha$ emission and the absorption features. To this end, the 2d spectra of the different OBs obtained during the data reduction are median combined to obtain one combined 2d spectrum for each target. We extract 1d spectra from different spatial regions with similar SNR along the slit in the 2d spectra following the procedure described in Section \ref{sec:UVES_data_red}.

As the UVES slit was centered on the position of the AGN and oriented along the radio jet axis, we extract spatial regions approximately corresponding to the regions where the radio jet is either predominantly receding from us or approaching us. The jet kinematics are determined from the [\ion{O}{iii}] velocity shifts of SINFONI observations for 4C+03.24, MRC\,0316 and TN\,J0205 \citep{Nesvadba2007,Nesvadba2008,Nesvadba2017a}. 
For 4C+04.11 the jet kinematics was determined in \citet{Parijskij2014} by using high-resolution radio polarization measurements. For 4C+03.24, TN\,J0205 and MRC\,0316, we extract spectra from three regions in the 2d long-slit spectra. One region where the jet is predominantly approaching, one region close to the central AGN position where the jet is predominantly receding and one region further away from the AGN where the jet is receding. As 4C+04.11 shows a more compact Ly$\alpha$ emission region, spectra from two regions are extracted, where the jet is predominantly approaching and receding. The flux of the spatial spectra is normalized to the spatial extent along the slit that they cover.
We fit the spatial spectra of 4C+04.11, MRC\,0316 and TN\,J0205 with two Gaussian components, consistent with the modeling of the master spectra, while we adopt a single Gaussian component for 4C+03.24.

We first perform a least-squares fit, using the best-fit parameters of the master spectra as initial values. If some of the absorbers identified in the master spectra cannot be identified in the spatially resolved spectra, we set their column density to zero in the fitting. MCMC sampling is then applied, initialized around the least-squares solution and adopting priors similar to before. The absorbers appear to show similar velocity shifts across the extent of the Ly$\alpha$ emission region, suggesting that they do not exhibit strong velocity gradients.

\begin{figure}[htbp]
    \centering
    \includegraphics[width= 0.99\hsize]{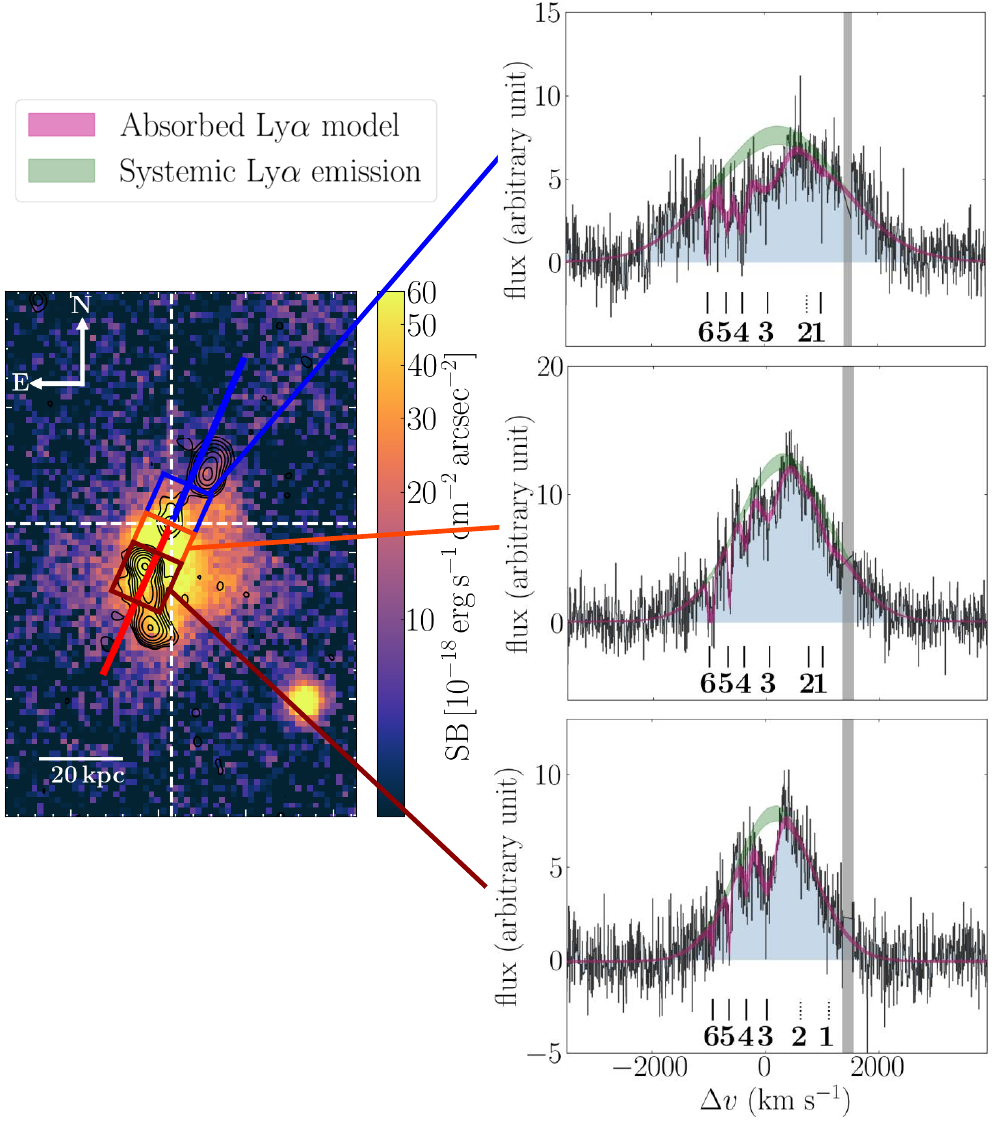}
    \caption{Spatial resolved analysis of the UVES spectra of 4C+03.24. Left: MUSE surface brightness map of the Ly$\alpha$ emission showing the receding and approaching radio jet side with red and blue lines. The dark red, orange and blue squares approximately correspond to the three regions on the sky from which the UVES spectra are extracted. The white crosshairs indicate the position of the AGN. Right: the UVES spectra extracted from the three regions with the magenta shaded region indicating the 68\% credible interval of the posterior model. The absorbers not identified in the respective spectrum are marked with dotted black lines. We call the region (and the corresponding UVES spectra) marked in blue the "approaching", the one marked in orange the "central receding" and the one marked in dark red the "outer receding" part. The regions cover a spatial extent along the radio jet of about 10.6 kpc, 7.9 kpc and 13.2 kpc respectively.}
    \label{fig:4C0324_spatial}
\end{figure}

\begin{figure}[htbp]
    \centering
    \includegraphics[width= 0.99\hsize]{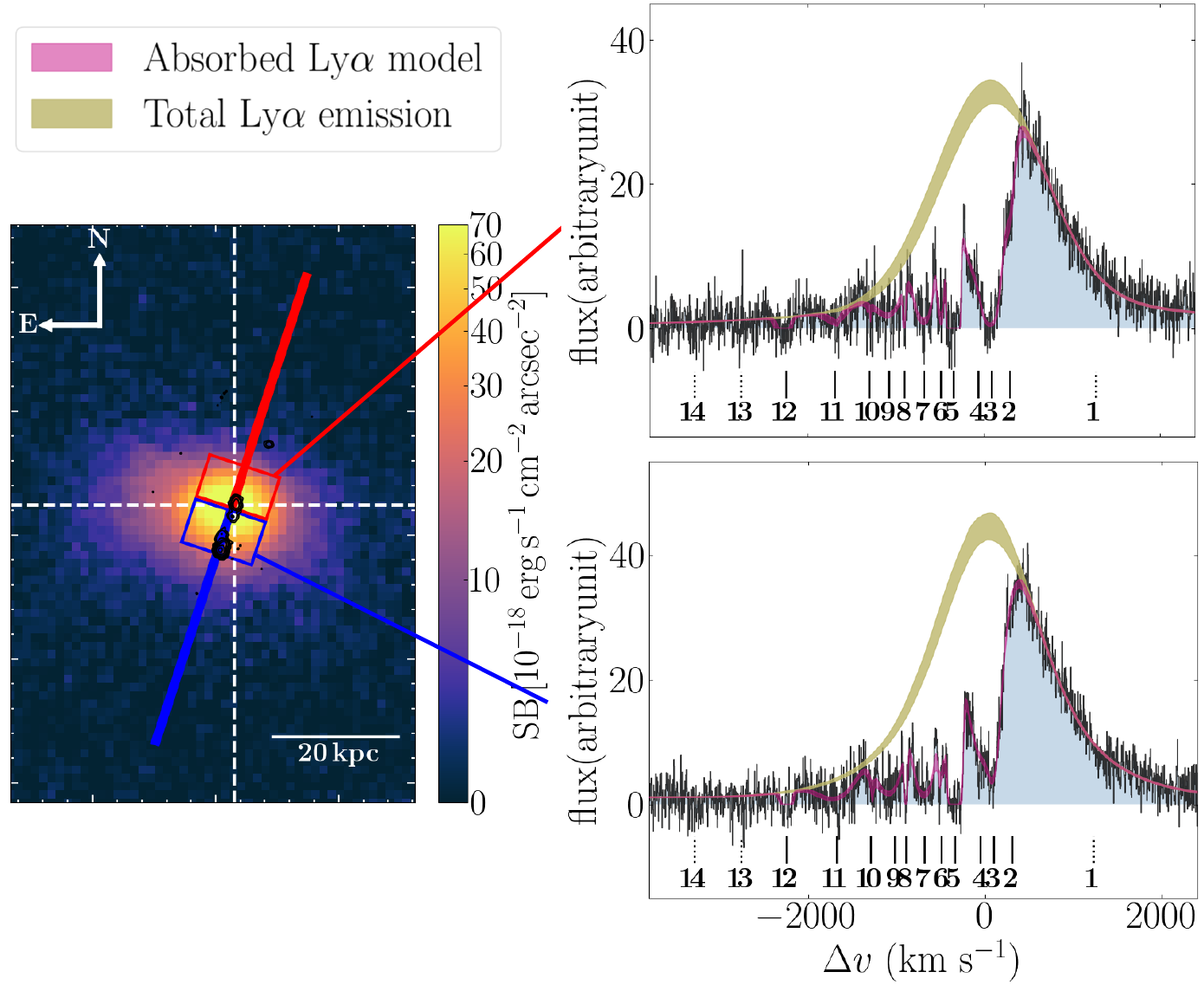}
    \caption{Similar to Fig. \ref{fig:4C0324_spatial}, but for 4C+04.11. We only distinguish two regions for this target, namely the "approaching" and "receding" jet side. Both regions cover a spatial extent along the radio jet of about 7.2 kpc.}
    \label{fig:4C0411_spatial}
\end{figure}

\begin{figure}[htbp]
    \centering
    \includegraphics[width= 1.015\hsize]{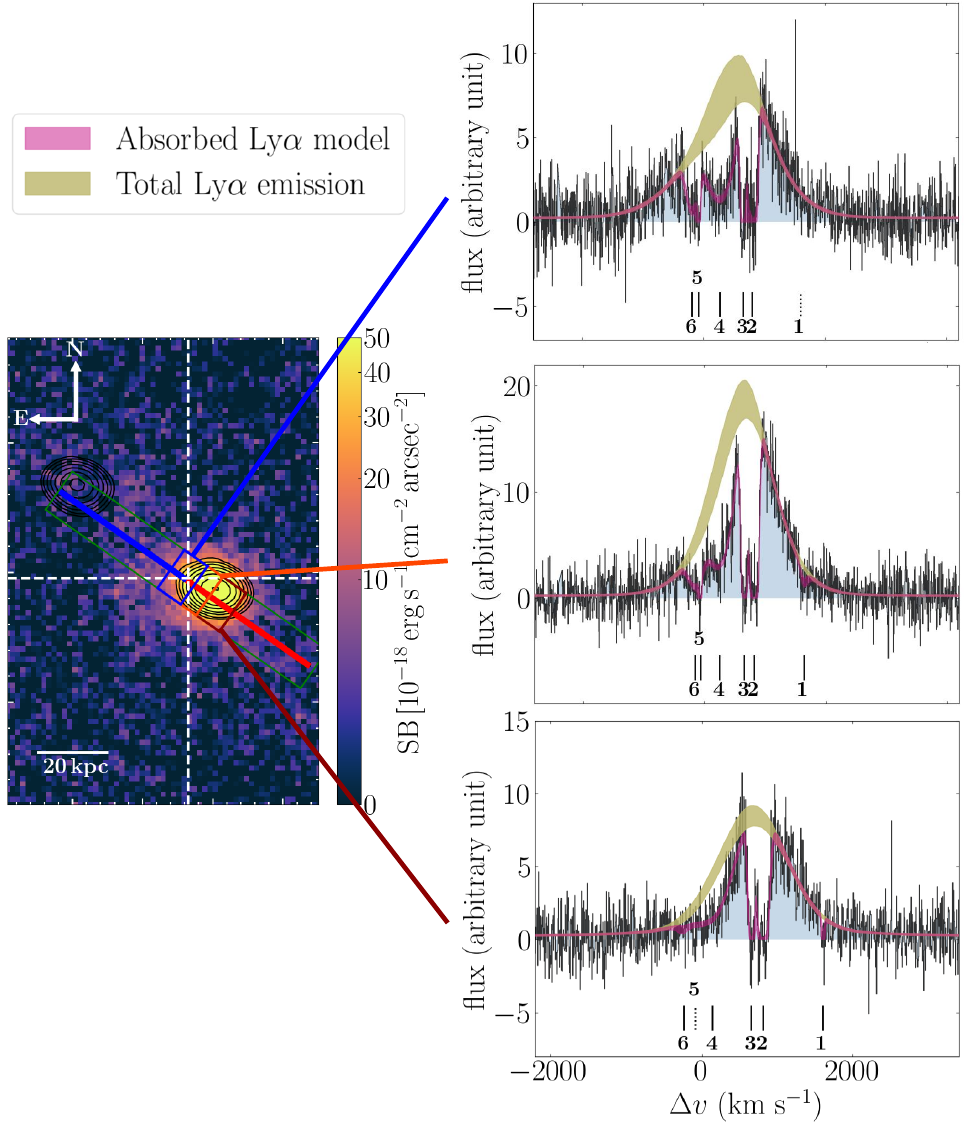}
    \caption{Similar to Fig. \ref{fig:4C0324_spatial}, but for MRC\,0316. The regions cover a spatial extent along the radio jet of about 8.3 kpc, 5.5 kpc and 8.3 kpc respectively.}
    \label{fig:MRC0316_spatial}
\end{figure}

\begin{figure}[htbp]
    \centering
    \includegraphics[width=\hsize]{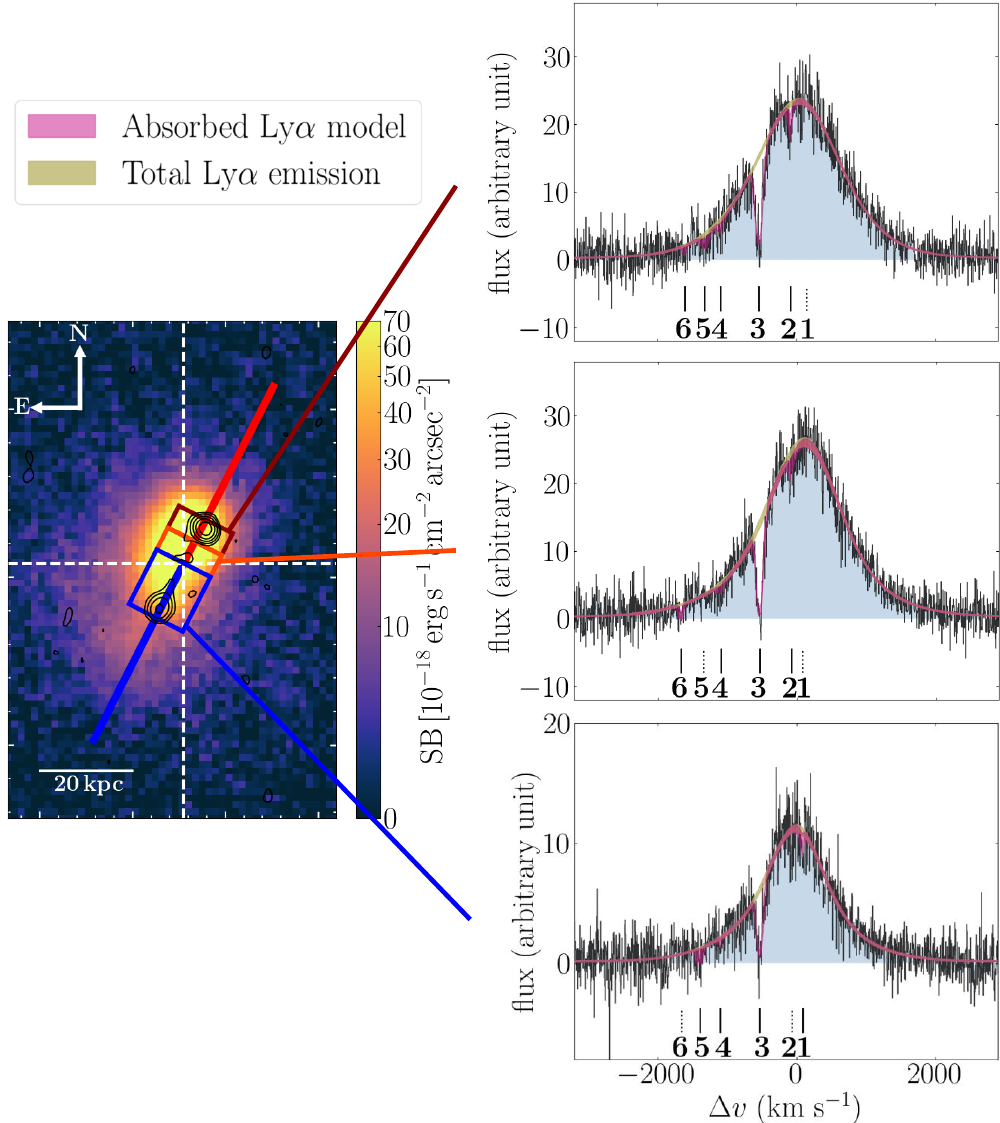}
    \caption{Similar to Fig. \ref{fig:4C0324_spatial}, but for TN\,J0205. The regions cover a spatial extent along the radio jet of about 5.3 kpc, 5.3 kpc and 13.3 kpc respectively.}
    \label{fig:TNJ0205_spatial}
\end{figure}

\begin{figure}[htbp]
    \centering
    \resizebox{\columnwidth}{!}{%
        \begin{tikzpicture}
            \node[anchor=north west,inner sep=0] at (0,0) {\includegraphics[width=0.3\textwidth]{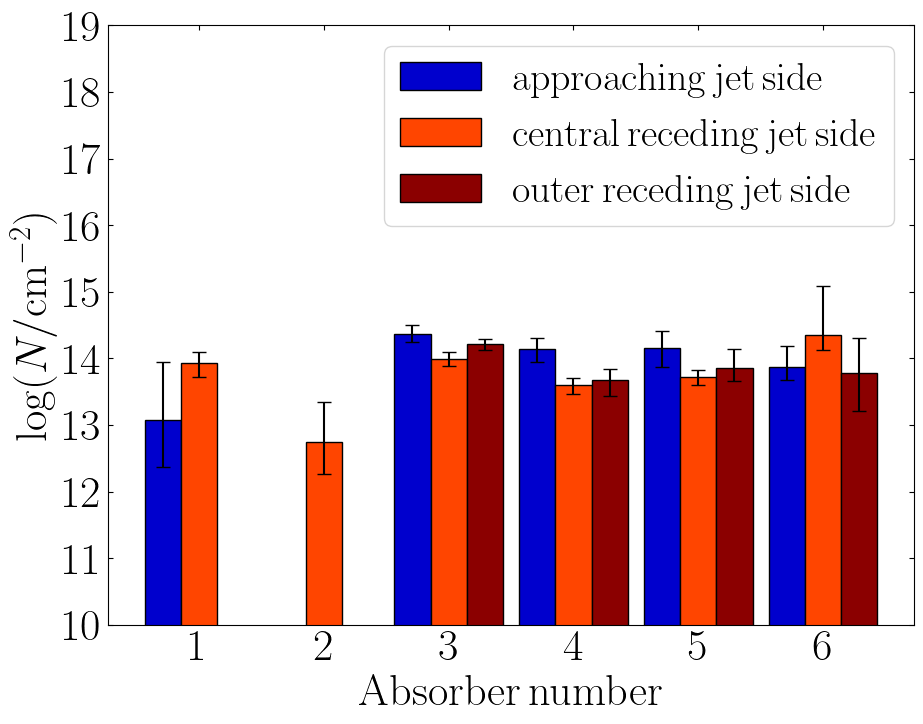}};
            \node[align=left, text=black, text width=0.1\textwidth, inner sep=1pt,font=\small\sffamily] at (1.7,-0.4) {4C+03.24};
            
            \node[anchor=north west,inner sep=0] at (5.5,0) {\includegraphics[width=0.3\textwidth]{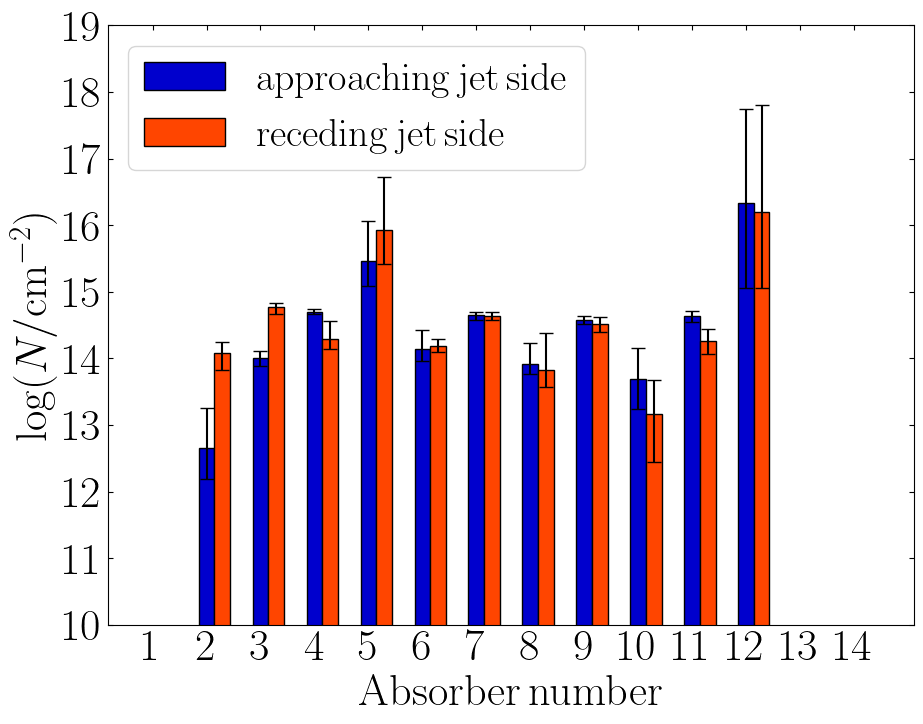}};
            \node[align=left, text=black, text width=0.1\textwidth, inner sep=1pt,font=\small\sffamily] at (10.3,-0.4) {\text{4C+04.11}};
            
            \node[anchor=north west,inner sep=0] at (0,-4.5) {\includegraphics[width=0.3\textwidth]{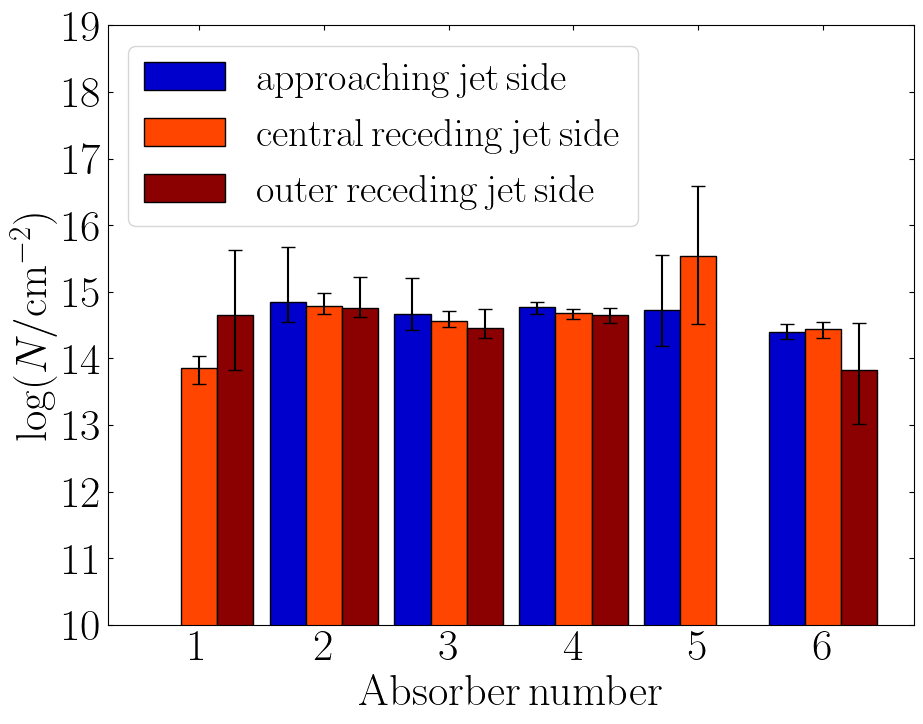}};
            \node[align=left, text=black, text width=0.1\textwidth, inner sep=1pt,font=\small\sffamily] at (4.9,-4.9) {\text{MRC\,0316}};
            
            \node[anchor=north west,inner sep=0] at (5.5,-4.5) {\includegraphics[width=0.3\textwidth]{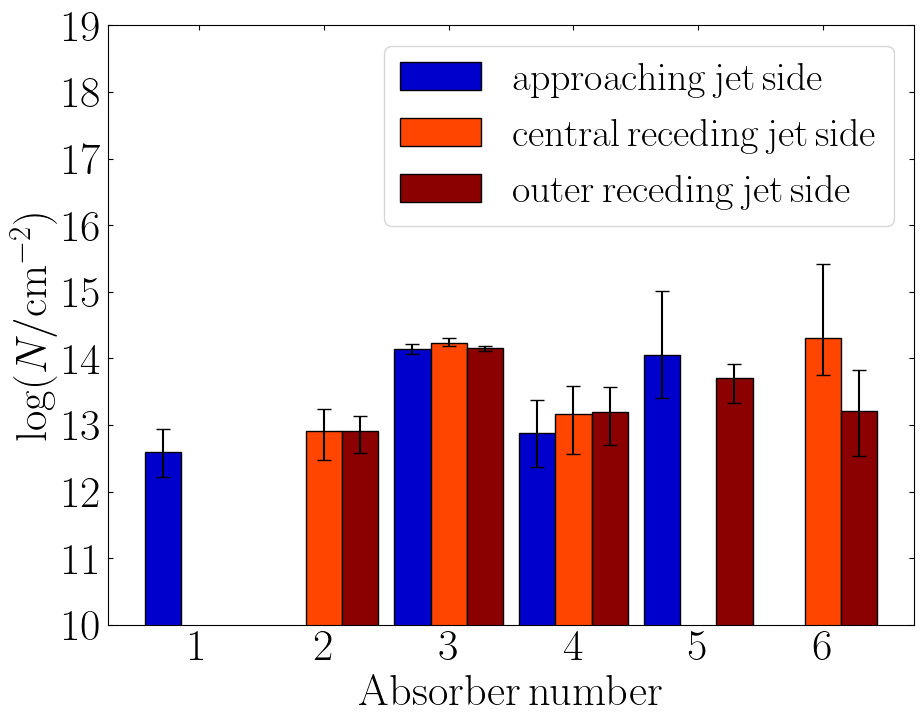}};
            \node[align=left, text=black, text width=0.1\textwidth, inner sep=1pt,font=\small\sffamily] at (7.3,-4.9) {\text{TN\,J0205}};
        \end{tikzpicture}   }
    \caption[Spatial variation in absorber column densities] 
    {Histogram of the logarithmic column density plotted against the absorber number for the different spatial spectra. The coloured bars show the median posterior values of $\log(N_{\mathrm{HI}}/\mathrm{cm}^{-2})$ for each absorber, while the error bars indicate the 16th–84th percentile uncertainties derived from the MCMC sampling. For 4C+03.24, MRC\,0316 and TN\,J0205, we call the different spatial regions the "outer receding jet side", the "central receding jet side" and the "approaching jet side" (see also Figures \ref{fig:4C0324_spatial}, \ref{fig:MRC0316_spatial} and \ref{fig:TNJ0205_spatial}). For 4C+04.11, we analyze two spatial regions, namely the "approaching" and the "receding jet side" (see Figure \ref{fig:4C0411_spatial}). 
    \label{fig:spatial_column_densities}}
\end{figure}

The approaching jet and receding jet sides and the spectra with the model results are shown in Figures \ref{fig:4C0324_spatial}, \ref{fig:4C0411_spatial}, \ref{fig:MRC0316_spatial} and \ref{fig:TNJ0205_spatial}.
66\,\% (21 out of 32) of the absorbers identified in our UVES master spectra are clearly detected over the entire extent of the spatial regions from which the spectra are extracted.
Figure \ref{fig:spatial_column_densities} shows a comparison between the column densities in the receding and approaching jet side. In an analysis of the MUSE data for the target 4C+04.11, \citet{Wang2021} identified a column density gradient from the southwest to the northeast region in a direction perpendicular to the radio jet axis with an increase of about 1 dex over the range of 24 kpc for one of the absorbers, as well as a small gradient along the jet axis. In our analysis, this absorber corresponds to absorber 3, for which  we identify a similar gradient along the jet axis, as log($N/\text{cm}^{-2}$) increases towards the receding jet side by about 0.7 dex. We can therefore tentatively confirm the column density gradient. Under the assumption of the expanding shell scenario this gradient could
be explained by the geometry of the gas shell as well as possible jet-gas interactions
\citep[see][]{Wang2021}. Our analysis of the other absorbers and targets suggests that this is not a general trend in HzRGs. Most absorbers exhibit little variation in column densities across the Ly$\alpha$ emission region. We also do not see any trend in column densities towards the receding or approaching jet sides. \vspace{-0.1 in}

\subsection{Comparison with [\ion{O}{iii}] emission} \label{sec:jwst}
The JWST/NIRSpec observations zoom into the central $\sim 25 \times 25 \,\text{kpc}^2$ of the HzRGs 4C+03.24 and TN\,J0205 and thus allow us to map the interstellar gas close to the central AGN with $\lesssim 1$\,kpc resolution. The NIRSpec data cover several important emission lines, in particular [\ion{O}{ii}]\,$\lambda\lambda3726,3729$, H${\beta}$,  [\ion{O}{iii}]\,$\lambda\lambda4959,5007$ or H${\alpha}$. Hereafter, the forbidden singly and doubly ionized oxygen lines will be referred to as [\ion{O}{ii}] and [\ion{O}{iii}].  As [\ion{O}{iii}] has the highest SNR in comparison and as it is often used in studies of AGN narrow-line regions \citep[e.g.][]{Harrison2014, Wylezalek2017}, we use this doublet line as indicator of the ionized gas morphology.
For both of the targets [\ion{O}{iii}] shows a very complicated morphology and velocity structure. Multiple narrowband images of the [\ion{O}{iii}]\,$\lambda5007$ emission, corresponding to different velocity ranges relative to the systemic, can be found in the appendix, Figures \ref{fig:4C0324_narrowband_OIII} and \ref{fig:TNJ0205_narrowband_OIII}. 

We examine the [\ion{O}{iii}] emission of the two targets in our sample to compare the morphology and kinematics of the ionized gas with those of the Ly$\alpha$ emission and absorption systems. 
As JWST is calibrated using vacuum wavelengths and UVES using air wavelengths, we convert the JWST data into air wavelengths for better comparison, following \cite{Wang2021}, using the equation from \citet{Morton2000}\footnote{$\lambda_{\text{air}}=\lambda_{\text{vac}}/n$, with $n=1+8.34254\cdot10^{-5} + 2.406147 \cdot 10^{-2}/(130-s^2) + 1.5998 \cdot 10^{-4}/(38.9-s^2), s=10^{4}/\lambda_{\text{vac}}$ with $\lambda_{\text{vac}}$ in $\text{\AA}$.}. To account for the differences in spatial resolution between UVES and NIRSpec, we convolve the NIRSpec datacubes to match the seeing-limited resolution of UVES. For comparison, we then extract rectangular apertures from the convolved NIRSpec datacubes, which are chosen to match the regions on the sky corresponding to several 1d UVES spectra extracted along the slit. 
The UVES spectra are extracted following the same procedure as described in Section \ref{sec:UVES_data_red}. 

\begin{figure*}[!ht]
    \centering
    \begin{adjustbox}{width=0.85\linewidth,center}
        \includegraphics{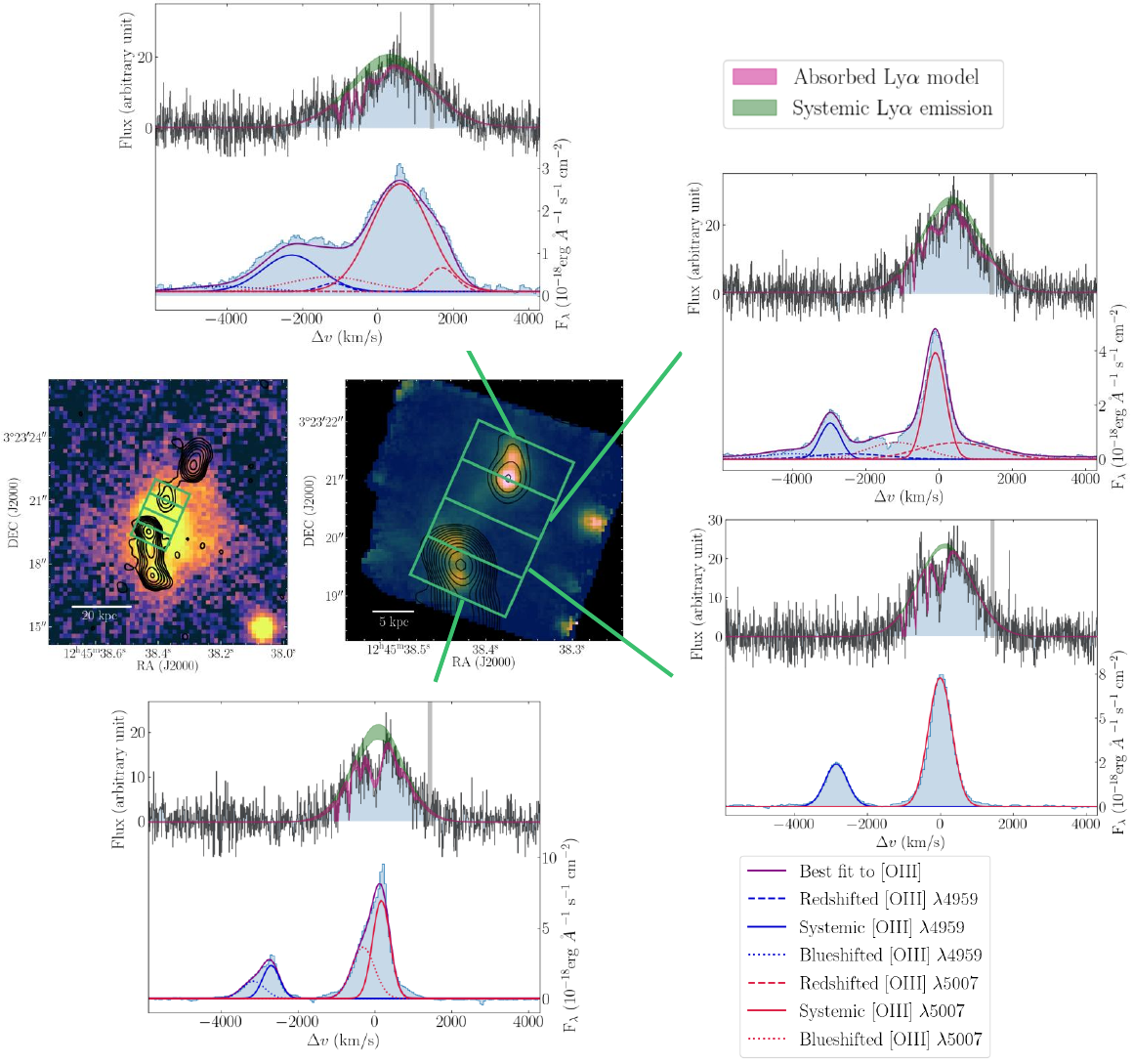}
    \end{adjustbox}
    \begin{tikzpicture}[overlay, remember picture]
        \node at (-6.7,14.4) {1)};
        \node at (1.2,12.2) {2)};
        \node at (1.2,7.4) {3)};
        \node at (-6.6,4.8) {4)};
    \end{tikzpicture}
    
    \caption[{Comparison between UVES Ly$\alpha$ and JWST [\ion{O}{iii}] of 4C+03.24}]{Comparison between the UVES Ly$\alpha$ emission profiles and the JWST/NIRSpec [\ion{O}{iii}] emission profiles of 4C+03.24. The radio contours (black) indicate the positions of the radio jets and the central AGN. The different spatial regions from which we extract the spectra are shown with green rectangles on the MUSE Ly$\alpha$ (left) and NIRSpec [\ion{O}{iii}] (right) surface brightness maps. We number the regions and the corresponding spectral panels from 1 to 4, as shown in the plot. In each panel, the UVES Ly$\alpha$ spectrum is shown at the top and the NIRSpec [\ion{O}{iii}] spectrum at the bottom.  }
    \label{fig:4C0324_jwst_uves_muse}
\end{figure*}
\begin{figure*}[!ht]
    \centering
    \begin{adjustbox}{width=0.85\linewidth,center}
        \includegraphics{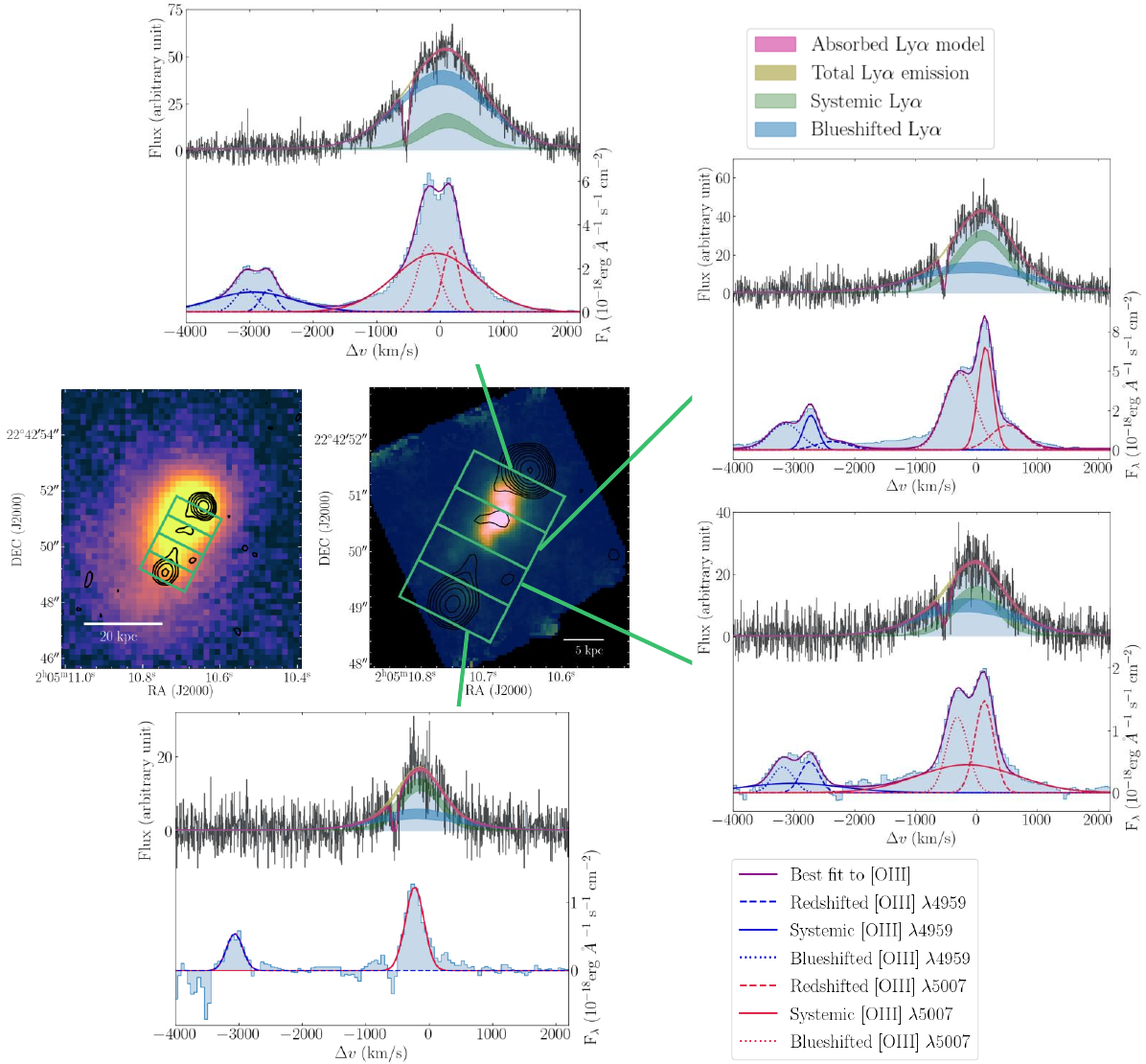}
    \end{adjustbox}
    \begin{tikzpicture}[overlay, remember picture]
        \node at (-6.5,14.1) {1)};
        \node at (1.2,12.1) {2)};
        \node at (1.2,7.3) {3)};
        \node at (-6.5,4.7) {4)};
    \end{tikzpicture}
    \caption [{Comparison between UVES Ly$\alpha$ and JWST [\ion{O}{iii}] of TN\,J0205}]{Comparison between the UVES Ly$\alpha$ emission profiles and the JWST/NIRSpec IFU [\ion{O}{iii}] emission profiles of TN\,J0205. The radio contours (black) indicate the positions of the radio jets and the central AGN. The different spatial regions from which we extract the spectra are shown with green rectangles on the MUSE Ly$\alpha$ (left) and NIRSpec [\ion{O}{iii}] (right) surface brightness maps. We number the regions and the corresponding spectral panels from 1 to 4, as shown in the plot. In each panel, the UVES Ly$\alpha$ spectrum is shown at the top and the NIRSpec [\ion{O}{iii}] spectrum at the bottom.  }
    \label{fig:TNJ0205_jwst_uves_muse}
\end{figure*}
The UVES and NIRSpec spectra extracted for this comparison and the Ly$\alpha$ and [\ion{O}{iii}] surface brightness maps showing the corresponding regions on the sky are presented in Figures \ref{fig:4C0324_jwst_uves_muse} and \ref{fig:TNJ0205_jwst_uves_muse}. The [\ion{O}{iii}] emission lines are fitted with least-squares fitting with up to three Gaussian components and the continuum with a zero-order polynomial. For the UVES spectra we perform MCMC sampling to explore the posterior distribution.
The components closest to $0\,\text{km}\,{\text{s}}^{-1}$, calculated from the systemic \ion{He}{ii} or [O\,\textsc{iii}] redshifts, are called the systemic component and the other components are called blueshifted and redshifted. 

When comparing the intrinsic Ly$\alpha$ emission with the [\ion{O}{iii}] emission, it is apparent that Ly$\alpha$ shows less spectral variation along the jet axis compared to [\ion{O}{iii}], where many different narrow, broad, blueshifted and redshifted components appear in different spatial regions, that are indicative of outflows and inflows. 
The Ly$\alpha$ emission on the other hand is more uniform and homogeneous over the extent of the radio galaxy. While the NIRSpec data have higher spatial resolution than seeing-limited UVES observations, the main reason for the observed discrepancy can likely be attributed to the different physical nature of the lines. Ly$\alpha$ is optically thick and resonant, leading to diffusion of photons and smoother observed structures \citep[e.g.][]{OsterbrockFerland2006}, whereas [\ion{O}{iii}] traces lower surface brightness clumps.

\section{Discussion}\label{sec:discussion}
In this section we discuss the possible origins of the absorption troughs identified in the Ly$\alpha$ spectra of HzRGs, compare to observations of other quasar types, examine a possible connection to the ionized ISM traced by [\ion{O}{iii}] and discuss simulations of Ly$\alpha$ radiative transfer.

\subsection{Origin of the absorbers} \label{sec:Origin_absorbers}
The Ly$\alpha$ emission and absorption probe the gas exclusively along our line of sight. A key question is therefore whether the absorbers are physically associated with the HzRG systems or instead trace large-scale intergalactic structures, such as the Ly$\alpha$ forest \citep[e.g.][]{Rauch1998}.

The spatial fitting along the length of the slit shows that many absorbers (about 66\,\%), especially the deep and broad ones (with typical column densities in our case of $\gtrsim 10^{14}\,\text{cm}^{-2}$) are detectable over large projected spatial extents along the jet, of about 30\,kpc for 4C+03.24, 14\,kpc for 4C+04.11, 22\,kpc for MRC\,0316 and 23\,kpc for TN\,J0205 (Fig. \ref{fig:spatial_column_densities}). Over these spatial extents, the absorbers exhibit remarkably coherent column densities (see Figure \ref{fig:spatial_column_densities}), indicating that the absorbing gas is distributed over large, contiguous structures. The absorbers cover almost the entire Ly$\alpha$ emission region and are found to be predominantly blueshifted with respect to the center of the Ly$\alpha$ emission. If we assume that the velocity offsets between the absorbers and the systemic redshift of the host galaxies trace physical separations along the line of sight, we infer comoving transverse distances of approximately $4 - 24\,\mathrm{cMpc}$ for the majority of absorbers. For comparison, we estimate virial radii for the HzRGs of between $\sim 450$ and $640\,\mathrm{ckpc}$ using the relation from \citealt{Dekel2013}\footnote{$R_{\text{vir}} \approx 100\,\text{kpc} (M_{\text{vir}}/10^{12}\,\text{M}_{\odot})^{1/3}\left(\frac{1+z}{3}\right)^{-1}$.}, assuming a conservative stellar mass to halo mass ratio of 0.02 \citealt{Behroozi2013}) and adopting the stellar masses from \citet{DeBreuck2010} and \citet{Wang2021}.

Interpreted purely as cosmological distances, the absorbers would therefore lie far beyond the virial radii of the host halos and would not be physically associated with the HzRGs, placing the systems in the intergalactic medium (IGM). Our inferred column densities and Doppler parameters are consistent with typical values measured for Ly$\alpha$ forest absorbers \citep[e.g.][]{Hu1995, Kim2002, Kim2013}. Moreover, studies of closely separated quasar sightlines at redshifts $z\lesssim3$ have shown that intergalactic hydrogen absorbers can exhibit transverse coherence over tens to hundreds of kiloparsecs \citep{Bechtold1994, Dinshaw1997, Rorai2017}. The extended absorbers observed in our data could thus trace comparable large-scale intergalactic structures. In particular, our highest-redshift target 4C+04.11 (at $z\sim4.5$) exhibits the largest number of absorption features. This is in line with the known increase in the abundance of low column density \ion{H}{i} absorbers ($\sim 10^{14}-10^{15}\,\text{cm}^{-2}$) toward higher redshift, driven by the rising neutral fraction of the IGM \citep[e.g.][]{Kim2002, Kim2013, Kim2021, Rahmati2013}. Thus, a substantial contribution from Ly$\alpha$ forest absorption is expected for this system.

Alternatively, if the observed velocity offsets of the absorbers relative to the systemic redshift are interpreted as reflecting line-of-sight gas motions rather than cosmological separations, this would indicate that the neutral clouds are outflowing structures in the CGM of the HzRGs. This interpretation has been proposed and studied in previous works. \citet{Binette2000} first interpreted the Ly$\alpha$ absorbers in HzRGs as expanding shells of neutral hydrogen intersecting the emission-line region, motivated by their blueshifted velocities, large spatial scales, and metallicity. Subsequent studies have shown that such extended hydrogen absorbers are a common feature of Ly$\alpha$ halos surrounding HzRGs, further developing and supporting this scenario \citep[e.g.][]{Wilman2004, Swinbank2015, Silva2018b, Kolwa2019, Wang2021}. We note here that we do not expect to observe any of the outflowing or inflowing gas behind the AGN and the bright Ly$\alpha$ emission nebulae due to attenuation effects. The kinematics and column densities of most absorbers in our sample remain remarkably consistent over large spatial extents. A similar behavior is for example seen in the Makani galaxy, where the large-scale outflow shows coherent kinematics over tens of kiloparsecs \citep{Rupke2019, Rupke2023}. Coherent, shell-like outflows driven by AGN feedback are also predicted in hydrodynamic simulations \citep[e.g.][]{Krause2002, Nelson2019, Ehlert2021, Ayromlou2024}.

Previous studies of absorption in the Ly$\alpha$ emission of HzRGs have identified kinematic links between the absorbing gas and the radio jets. Using MUSE, \citet{Wang2021} found that one absorber in 4C+04.11 - also included in our UVES sample (absorber 3 in our analysis) - exhibits a velocity pattern aligned with the jet axis, consistent with jet–gas interactions. Similarly, \citet{Swinbank2015} reported spatial variations in the velocity structure of the absorption that follow that of the underlying Ly$\alpha$ emission, while the column densities remain largely uniform.

In addition to these kinematic signatures, several works have revealed that extended absorbers associated with HzRGs show significant metal enrichment (same absorbers identified in \ion{C}{iv}, \ion{S}{ii} and \ion{N}{v}) (e.g. \citealt{Kolwa2019, Wang2021}), implying chemical abundances well above those expected for pristine or weakly enriched intergalactic gas. Such metallicity levels point to an origin within the host galaxy, with the enriched material subsequently transported to large radii. We stress here that the HzRG 4C+04.11, which shows metal-enriched absorption (for absorber 3 in our UVES analysis) as reported by \citet{Wang2021}, is also included in our sample of UVES observations. These results demonstrate that absorbers with properties comparable to those in our sample can be directly influenced by the host galaxy, reinforcing the possibility that some of the systems we observe are likewise associated with the HzRGs.

Furthermore, cosmological hydrodynamical simulations suggest that the environments of massive halos at these epochs contain significant excess \ion{H}{i} absorption relative to the ambient IGM \citep{Meiksin2015, Meiksin2017, Conaboy2025}. This excess arises from a combination of dense gas which can be ejected in the form of feedback-driven shells and large-scale filaments feeding the halo, which is also supported by observations \citep{Prochaska2013, Matthee2024}.

We note that several weaker and narrower absorbers, particularly those with column densities $< 10^{14}\,\text{cm}^{-2}$, are not detected across the full spatial extent of the Ly$\alpha$ nebulae. This can be explained by a combination of lower Ly$\alpha$ surface brightness and reduced SNR in the spatially resolved spectra which limits our ability to detect small or narrow features in particular in the outer wings of the Ly$\alpha$ emission line. Alternatively, their limited spatial coverage may indicate that these systems are intrinsically not spatially extended or represent fragments of larger shell-like structures associated with the host system.

In summary, several lines of evidence indicate that at least some of the hydrogen absorbers in our sample could be associated with the HzRGs themselves, tracing outflowing shells of gas. At the same time, we cannot exclude that some absorbers arise in the intergalactic medium at Mpc-scale distances from the host galaxies, particularly for the lower-density absorbers with large velocity offsets with respect to the Ly$\alpha$ emission. Deep, high-resolution spectroscopic observations of associated metal lines will be crucial for distinguishing between these scenarios.

\subsection{Comparison to other quasar studies} \label{sec:otherquasars}
In addition to HzRGs (which are effectively obscured type-2 quasars), quasars at high redshift ubiquitously show extended Ly$\alpha$ emission halos that have been studied extensively in the past using IFU observations. Some examples are the Ly$\alpha$ nebulae around type-I quasars at redshifts z$\sim$3 \citep[e.g.][]{Borisova2016, ArrigoniBattaia2019} or around ultra-luminous type-1 quasars at redshifts z$\sim$2 \citep{Cai2019}, as well as around quasars at z$\sim6$ \citep{Farina2019}. 
Ly$\alpha$ nebulae around quasars typically exhibit fewer and less prominent absorption features compared to HzRGs, and their spectra can generally be reasonably fitted with a Gaussian profile without the need to introduce absorption features \citep[e.g.][]{ArrigoniBattaia2018, Vayner2023}.
This difference can be naturally explained by the strong, unobscured ionizing continuum in type-I quasars, which creates large proximity zones (that can extend over thousands of km\,s$^{-1}$, corresponding to distances of several physical Mpc;  e.g. \citealt{Satyavolu2023}). These ionized regions suppress the amount of neutral  gas capable of producing absorption lines and therefore prevent the appearance of strong absorption close to the Ly$\alpha$ peak. In contrast, HzRGs lack observable quasar proximity zones, because their ionizing radiation is blocked along the line of sight by obscuring material. As a result, neutral gas, potentially from Ly$\alpha$ forest absorption systems in the IGM, can persist close to the systemic redshift.

In addition to the proximity zone effects, HzRGs and the considered type-I quasars also differ in their typical Mpc-scale environments and feedback regimes. HzRGs are known to occupy overdense, gas-rich protocluster environments \citep[e.g.][]{Wylezalek2013, Noirot2018} and shocks driven by their powerful radio jets as well as strong AGN winds are expected to produce large-scale outflowing gas shells \citep[e.g.][]{Nesvadba2008, Wagner2012, Mukherjee2018, Meenakshi2022, HeckmanBest2023, Roy2025}. 
The type-I quasars discussed above are predominantly radio-quiet objects and comparable jet-driven shells are not expected to be prevalent in these systems. If some of the absorbers seen in the Ly$\alpha$ profiles of HzRGs are associated with jet-driven gas structures, their lower incidence toward radio-quiet quasars may also reflect differences in the dominant feedback mode \citep[e.g.][]{Wang2023}. It could therefore be of interest to systematically target the Ly$\alpha$ emission of radio-loud quasars at high spectral resolution in future studies.

\subsection{Connection between Ly$\alpha$ emission and ionized gas as revealed by JWST} \label{sec:jwstdiscussion}
The complicated velocity structure of the [\ion{O}{iii}] emission, with narrow, broad, blueshifted, and redshifted components (see Figures \ref{fig:4C0324_narrowband_OIII} and \ref{fig:TNJ0205_narrowband_OIII}), provide evidence for AGN-driven outflows and potentially inflows in the central region. \citet{Wang2024} analyzed the NIRSpec IFU data of the ionized ISM of another HzRG in the JWST program (PI:Wuji Wang) and found evidence for a radiatively-driven outflow that is only weakly coupled to the quasar. The jet may therefore dominate the outflow \citep{Nesvadba2017a}.

For both 4C+03.24 and TNJ\,0205, there is a spatial correlation between the broadening of [\ion{O}{iii}] and Ly$\alpha$, with broader [\ion{O}{iii}] emission appearing in regions with broad Ly$\alpha$. This suggests a shared underlying kinematic structure between [\ion{O}{iii}] and  Ly$\alpha$, potentially driven by outflows or turbulent gas motions. In the case of Ly$\alpha$, however, additional line broadening arises due to resonant scattering, as photons in the line wings can escape more easily than those near the line center \citep{Dijkstra2014}. 

There is known jet-gas interaction south of the AGN of 4C+03.24, which can be seen from the bent structure of the radio jet contours (Figure \ref{fig:MUSE_SB}) and which has also been discussed in \citet{vanOjik1996}. \citet{Wang2025} discovered several companions surrounding both 4C+03.24 and TNJ\,0205 in the NIRSpec observations of [\ion{O}{iii}] and ALMA observations of [\ion{C}{ii}]. 4C+03.24 in particular shows [\ion{C}{ii}] companions, tracing cold gas systems, as well as optical continuum emission tracing the stellar component, to the south of the AGN, which might have been stripped during previous merger activity. The merger could have triggered the radio jet and its debris, the dense gas systems in the south, may have consequently deflected the jet, which would explain the bent radio structure \citep{Wang2025}.

Since the absorbers seen in the Ly$\alpha$ emission could potentially correspond to outflowing shells of gas that were ejected from the host galaxy at earlier times, they might be connected to outflows detected in the ISM. A comparison between the velocity shifts of the H\,\textsc{i} absorbers and the velocity shifts of the [\ion{O}{iii}] emission components does not show a clear correlation. The lack of correlation in velocity shifts suggests that the gas may not be kinematically linked. Furthermore, the significant spatial variation in the [\ion{O}{iii}] emission contrasts with the consistent identification of H\,\textsc{i} absorbers across the entire spatial extent of the UVES spectra. This difference can be attributed to the fact that the H\,\textsc{i} absorbers and the [\ion{O}{iii}] emission trace distinct gas phases, which are subject to different physical conditions and kinematic behaviors. 

When comparing the morphologies of Ly$\alpha$ observed with VLT/MUSE and [\ion{O}{iii}], we find that the Ly$\alpha$ emission is more extended and exhibits a rounder distribution, while [\ion{O}{iii}] is confined to more compact, localized regions (see also \citealt{Peng2025}). This likely reflects distinct physical processes, with Ly$\alpha$ emission shaped by resonant scattering and diffusion over large scales, while [\ion{O}{iii}] emission traces more localized ionization mechanisms (see also Section~\ref{sec:jwst}).

\subsection{Absorption fitting and Ly$\alpha$ radiative transfer} \label{sec:simulations}
We note here that our modeling of the Ly$\alpha$ profiles is based on the assumption that the Ly$\alpha$ photons are mostly absorbed or scattered out of our line of sight in gas shells that are not co-spatial with the emission line region (as also done in past works, see e.g. \citealt{Binette2000, Jarvis2003, Wilman2004, Swinbank2015, Silva2018b, Kolwa2019, Wang2021}). As Ly$\alpha$ is a resonant line, the Ly$\alpha$ photons will scatter many times within the ISM and CGM of the galaxies before reaching the observer. This means that in principle, scattered photons also contribute to the emergent line profiles, and a full radiative transfer treatment is needed to properly predict the line shapes. Pioneering works by \citet{Verhamme2006} and \citet{Dijkstra2006} (and other works following them like \citealt{Gronke2015}, \citealt{GronkeDijkstra2016} or \citealt{Park2022}) revealed that Monte Carlo radiative Ly$\alpha$ transfer modeling of outflowing or infalling gas shells can produce asymmetric, double-peaked line profiles which agree with observations at high redshift. 3d radiative transfer simulations additionally allow to account for the spatial variations in Ly$\alpha$ \citep{SeonKim2020, Chang2023}.

However, the very complicated line profiles and deep absorption troughs split into multiple components that we observe in our sample of HzRGs cannot be reproduced by current radiative transfer simulations. Our approach of fitting Voigt absorption troughs superimposed on a Gaussian emission profile (that might be shaped by radiative transfer effects) is therefore assumed to provide a sufficiently accurate representation of the observed line shapes for the purposes of our analysis. Nevertheless, incorporating radiative transfer effects into future studies of HzRG Ly$\alpha$ halos or comparisons with radiative transfer simulations could provide valuable insights into the nature of the line-emitting gas.

\section{Conclusions}\label{sec:conclusions}
In this paper, we investigate the neutral hydrogen gas detected in absorption against the Ly$\alpha$ nebulae of four high-redshift radio galaxies (z$\approx$3.1-4.5). We complement our UVES absorption results with comparisons to MUSE and JWST/NIRSpec observations. 
The main conclusions can be summarized as follows.
   \begin{enumerate}
      \item We detect between 6 and 14 absorbers in the spectra of the four high-redshift radio galaxies in our sample. We find low column densities between $N_{\text{H\,\textsc{i}}} = 10^{12}-10^{17}\,\text{cm}^{-2}$ and a wide range of Doppler parameters between $\sim 14$ and $300\,\text{km}\,\text{s}^{-1}$. 
      \item Most of the absorbers (about 66\,\%) are visible over large spatial extents (up to 30 kpc along the radio jets), while showing overall little variation in column densities. 
      Taken together, our results indicate that a fraction of the absorbing gas is likely physically associated with the host halos. However, absorption from the IGM might contaminate our sample of absorbers. 
      \item  Comparison between the UVES spectra and pseudolongslit spectra extracted from MUSE data show that high spectral resolution is required for a complete identification of absorption features in the halos of HzRGs. For the HzRG MRC\,0316, the deep absorber observed with MUSE splits into multiple components when observed at higher spectral resolution using UVES. The UVES data thus allow better constraints of the kinematics and morphology of the absorbing gas.
      \item JWST/NIRSpec data of two of the four HzRGs in our sample reveal complicated morphologies and kinematics of the ionized, interstellar gas traced by [\ion{O}{iii}] within about 20 kpc surrounding the AGN. 
      We do not find clear evidence for a relation between outflows in the ISM and the absorbers seen in the Ly$\alpha$ profiles. However, the kinematics of the [\ion{O}{iii}] and Ly$\alpha$ emission indicate a possible connection between the ionized and cool gas phases.
    \end{enumerate}
HzRGs are strong cosmological probes and present ideal targets to study
the evolution of massive galaxies and the role of AGN feedback. The analysis of Ly$\alpha$ absorption features provides valuable insights into the distribution of neutral hydrogen gas in the environments of these systems. Various datasets, including MUSE, UVES, JWST, and ALMA have been acquired for our sample of high-redshift radio galaxies. Future work will focus on integrating these datasets to study the CGM in multiple phases and explore the connections between different gas tracers, ultimately contributing to our understanding of the (re)distribution of gas in massive galaxies.

\section*{Data availability}
The JWST data used in this work are publicly available from the Mikulski Archive for Space Telescopes (MAST) under programme JWST-GO-01970 and can be accessed via \url{https://dx.doi.org/10.17909/ataa-hn19}. The corner plots associated with the MCMC analysis are additionally made available as supplementary online material on Zenodo at \url{https://doi.org/10.5281/zenodo.19737893}.

\begin{acknowledgements}
    We thank the anonymous referee for valuable comments, that helped improve this manuscript.  
    J.R. thanks Seok-Jun Chang and Aniket Bhagwat for useful discussions. 
    This work is based [in part] on observations made with the NASA/ESA/CSA James Webb Space Telescope. The data were obtained from the Mikulski Archive for Space Telescopes at the Space Telescope Science Institute, which is operated by the Association of Universities for Research in Astronomy, Inc., under NASA contract NAS 5-03127 for JWST. W.W. acknowledges the grant support through JWST programs. Support for programs JWST-GO-03045 and JWST-GO-03950 was provided by NASA through a grant from the Space Telescope Science Institute, which is operated by the Association of Universities for Research in Astronomy, Inc., under NASA contract NAS 5-03127. 
    This work uses a number of open source software other than the aforementioned ones such as Jupyter notebook \citep{Kluyver2016}; matplotlib \citep{Hunter2007}; SciPy \citep{Virtanen2020}; NumPy \citep{Harris2020}; Astropy \citep{Astropy2018}.
\end{acknowledgements}

\bibliography{references}

@ARTICLE{Kolwa2023,
       author = {{Kolwa}, S. and {De Breuck}, C. and {Vernet}, J. and {Wylezalek}, D. and {Wang}, W. and {Popping}, G. and {Man}, A.~W.~S. and {Harrison}, C.~M. and {Andreani}, P.},
        title = "{Faint [C I](1-0) emission in z   3.5 radio galaxies}",
      journal = {\mnras},
         year = 2023,
        month = nov,
       volume = {525},
       number = {4},
        pages = {5831-5845},
          doi = {10.1093/mnras/stad2647},
archivePrefix = {arXiv},
       eprint = {2309.00459},
 primaryClass = {astro-ph.GA},
       adsurl = {https://ui.adsabs.harvard.edu/abs/2023MNRAS.525.5831K}
}

@ARTICLE{Wang2023,
       author = {{Wang}, Wuji and {Wylezalek}, Dominika and {Vernet}, Jo{\"e}l and {De Breuck}, Carlos and {Gullberg}, Bitten and {Swinbank}, Mark and {Villar Mart{\'\i}n}, Montserrat and {Lehnert}, Matthew D. and {Drouart}, Guillaume and {Arrigoni Battaia}, Fabrizio and {Humphrey}, Andrew and {Noirot}, Ga{\"e}l and {Kolwa}, Sthabile and {Seymour}, Nick and {Lagos}, Patricio},
        title = "{3D tomography of the giant Ly{\ensuremath{\alpha}} nebulae of z {\ensuremath{\approx}} 3-5 radio-loud AGN}",
      journal = {\aap},
         year = 2023,
        month = dec,
       volume = {680},
          eid = {A70},
        pages = {A70},
          doi = {10.1051/0004-6361/202346415},
archivePrefix = {arXiv},
       eprint = {2309.15144},
 primaryClass = {astro-ph.GA},
       adsurl = {https://ui.adsabs.harvard.edu/abs/2023A&A...680A..70W}
}

@article{LA-cosmic,
author = {Dokkum, Pieter},
year = {2001},
month = {08},
pages = {},
title = {Cosmic-Ray Rejection by Laplacian Edge Detection},
volume = {113},
journal = {\pasp},
doi = {10.1086/323894}
}

@article{astro-scrappy,
author = {McCully, Curtis},
year = {2014},
pages = {},
title = {Astro-SCRAPPY},
journal = {\pasp},
doi = {10.1086/323894}
}

@misc{Esorex2015,
       author = {{ESO CPL Development Team}},
        title = "{EsoRex: ESO Recipe Execution Tool}",
 howpublished = {Astrophysics Source Code Library, record ascl:1504.003},
         year = 2015,
        month = apr,
          eid = {ascl:1504.003},
       adsurl = {https://ui.adsabs.harvard.edu/abs/2015ascl.soft04003E}
}

@ARTICLE{Kolwa2019,
       author = {{Kolwa}, S. and {Vernet}, J. and {De Breuck}, C. and {Villar-Mart{\'\i}n}, M. and {Humphrey}, A. and {Arrigoni-Battaia}, F. and {Gullberg}, B. and {Falkendal}, T. and {Drouart}, G. and {Lehnert}, M.~D. and {Wylezalek}, D. and {Man}, A.},
        title = "{MUSE unravels the ionisation and origin of metal-enriched absorbers in the gas halo of a z = 2.92 radio galaxy}",
      journal = {\aap},
         year = 2019,
        month = may,
       volume = {625},
          eid = {A102},
        pages = {A102},
          doi = {10.1051/0004-6361/201935437},
archivePrefix = {arXiv},
       eprint = {1904.05114},
 primaryClass = {astro-ph.GA},
       adsurl = {https://ui.adsabs.harvard.edu/abs/2019A&A...625A.102K}
}

@ARTICLE{Wang2021,
       author = {{Wang}, Wuji and {Wylezalek}, Dominika and {De Breuck}, Carlos and {Vernet}, Jo{\"e}l and {Humphrey}, Andrew and {Villar Mart{\'\i}n}, Montserrat and {Lehnert}, Matthew D. and {Kolwa}, Sthabile},
        title = "{Mapping the ``invisible'' circumgalactic medium around a z {\ensuremath{\sim}} 4.5 radio galaxy with MUSE}",
      journal = {\aap},
         year = 2021,
        month = oct,
       volume = {654},
          eid = {A88},
        pages = {A88},
          doi = {10.1051/0004-6361/202141558},
archivePrefix = {arXiv},
       eprint = {2107.09066},
 primaryClass = {astro-ph.GA},
       adsurl = {https://ui.adsabs.harvard.edu/abs/2021A&A...654A..88W}
}

@ARTICLE{Jarvis2003,
       author = {{Jarvis}, M.~J. and {Wilman}, R.~J. and {R{\"o}ttgering}, H.~J.~A. and {Binette}, L.},
        title = "{Probing the absorbing haloes around two high-redshift radio galaxies with VLT-UVES$^{*}$}",
      journal = {\mnras},
         year = 2003,
        month = jan,
       volume = {338},
       number = {1},
        pages = {263-272},
          doi = {10.1046/j.1365-8711.2003.06053.x},
archivePrefix = {arXiv},
       eprint = {astro-ph/0209159},
 primaryClass = {astro-ph},
       adsurl = {https://ui.adsabs.harvard.edu/abs/2003MNRAS.338..263J}
}

@ARTICLE{Swinbank2015,
       author = {{Swinbank}, A.~M. and {Vernet}, J.~D.~R. and {Smail}, Ian and {De Breuck}, C. and {Bacon}, R. and {Contini}, T. and {Richard}, J. and {R{\"o}ttgering}, H.~J.~A. and {Urrutia}, T. and {Venemans}, B.},
        title = "{Mapping the dynamics of a giant Ly {\ensuremath{\alpha}} halo at z = 4.1 with MUSE: the energetics of a large-scale AGN-driven outflow around a massive, high-redshift galaxy}",
      journal = {\mnras},
         year = 2015,
        month = may,
       volume = {449},
       number = {2},
        pages = {1298-1308},
          doi = {10.1093/mnras/stv366},
archivePrefix = {arXiv},
       eprint = {1502.05998},
 primaryClass = {astro-ph.GA},
       adsurl = {https://ui.adsabs.harvard.edu/abs/2015MNRAS.449.1298S}
}

@ARTICLE{Silva2018b,
       author = {{Silva}, M. and {Humphrey}, A. and {Lagos}, P. and {Villar-Mart{\'\i}n}, M. and {Morais}, S.~G. and {di Serego Alighieri}, S. and {Cimatti}, A. and {Fosbury}, R. and {Overzier}, R.~A. and {Vernet}, J. and {Binette}, L.},
        title = "{The MUSE 3D view of feedback in a high-metallicity radio galaxy at z = 2.9}",
      journal = {\mnras},
         year = 2018,
        month = mar,
       volume = {474},
       number = {3},
        pages = {3649-3672},
          doi = {10.1093/mnras/stx3019},
archivePrefix = {arXiv},
       eprint = {1711.10601},
 primaryClass = {astro-ph.GA},
       adsurl = {https://ui.adsabs.harvard.edu/abs/2018MNRAS.474.3649S}
}

@ARTICLE{Tepper-Garcia2006,
       author = {{Tepper-Garc{\'\i}a}, Thorsten},
        title = "{Voigt profile fitting to quasar absorption lines: an analytic approximation to the Voigt-Hjerting function}",
      journal = {\mnras},
         year = 2006,
        month = jul,
       volume = {369},
       number = {4},
        pages = {2025-2035},
          doi = {10.1111/j.1365-2966.2006.10450.x},
archivePrefix = {arXiv},
       eprint = {astro-ph/0602124},
 primaryClass = {astro-ph},
       adsurl = {https://ui.adsabs.harvard.edu/abs/2006MNRAS.369.2025T}
}

@misc{Kramida2023,
  title = {{NIST} Atomic Spectra Database ({version} 5.11)},
  author = {Kramida, A. and Ralchenko, Yu. and Reader, J. and {NIST ASD Team}},
  year = {2023},
  howpublished = {Online},
  note = { Available: \url{https://physics.nist.gov/asd} [Accessed: Mon May 06 2024]. National Institute of Standards and Technology, Gaithersburg, MD},
  url = {https://physics.nist.gov/asd},
  urldate = {2024-05-06},
  doi = {10.18434/T4W30F}
}

@misc{Newville_lmfit,
       author = {{Newville}, Matt and {Nelson}, Andrew and {Ingargiola}, Antonino and {Stensitzki}, Till and {Allan}, Dan and {Micha{\l}} and {Glenn} and {Ram}, Yoav and {MerlinSmiles} and {Li}, Li and {Pasquevich}, Gustavo and {Deil}, Christoph and {Fobes}, David M and {Stuermer} and {Spillane}, Tim and {Ampolloreno} and {Stonebig} and {Brodtkorb}, Per A. and {Clarken}, Robbie and {Anagnostopoulos}, Kostis and {Almarza}, Anthony and {Gamari}, Ben},
        title = "{lmfit-py 0.9.5}",
         year = 2016,
        month = jul,
          eid = {10.5281/zenodo.58759},
          doi = {10.5281/zenodo.58759},
      version = {0.9.5},
    publisher = {Zenodo},
       adsurl = {https://ui.adsabs.harvard.edu/abs/2016zndo.....58759N}
}

@ARTICLE{Liddle2007,
       author = {{Liddle}, Andrew R.},
        title = "{Information criteria for astrophysical model selection}",
      journal = {\mnras},
         year = 2007,
        month = may,
       volume = {377},
       number = {1},
        pages = {L74-L78},
          doi = {10.1111/j.1745-3933.2007.00306.x},
archivePrefix = {arXiv},
       eprint = {astro-ph/0701113},
 primaryClass = {astro-ph},
       adsurl = {https://ui.adsabs.harvard.edu/abs/2007MNRAS.377L..74L}
}

@ARTICLE{Foreman-Mackey2013,
       author = {{Foreman-Mackey}, Daniel and {Hogg}, David W. and {Lang}, Dustin and {Goodman}, Jonathan},
        title = "{emcee: The MCMC Hammer}",
      journal = {\pasp},
         year = 2013,
        month = mar,
       volume = {125},
       number = {925},
        pages = {306},
          doi = {10.1086/670067},
archivePrefix = {arXiv},
       eprint = {1202.3665},
 primaryClass = {astro-ph.IM},
       adsurl = {https://ui.adsabs.harvard.edu/abs/2013PASP..125..306F}}

@ARTICLE{Goodman-Weare2010,
       author = {{Goodman}, Jonathan and {Weare}, Jonathan},
        title = "{Ensemble samplers with affine invariance}",
      journal = {Commun. Appl. Math. Comput. Sci.},
         year = 2010,
        month = jan,
       volume = {5},
       number = {1},
        pages = {65-80},
          doi = {10.2140/camcos.2010.5.65},
       adsurl = {https://ui.adsabs.harvard.edu/abs/2010CAMCS...5...65G}
}

@ARTICLE{Wilman2004,
       author = {{Wilman}, R.~J. and {Jarvis}, M.~J. and {R{\"o}ttgering}, H.~J.~A. and {Binette}, L.},
        title = "{HI in the protocluster environment at z > 2: absorbing haloes and the Ly{\ensuremath{\alpha}} forest}",
      journal = {\mnras},
         year = 2004,
        month = jul,
       volume = {351},
       number = {3},
        pages = {1109-1119},
          doi = {10.1111/j.1365-2966.2004.07857.x},
archivePrefix = {arXiv},
       eprint = {astro-ph/0403519},
 primaryClass = {astro-ph},
       adsurl = {https://ui.adsabs.harvard.edu/abs/2004MNRAS.351.1109W}
}

@ARTICLE{vanOjik1997,
       author = {{van Ojik}, R. and {Roettgering}, H.~J.~A. and {Miley}, G.~K. and {Hunstead}, R.~W.},
        title = "{The gaseous environments of radio galaxies in the early Universe: kinematics of the Lyman {\ensuremath{\alpha}} emission and spatially resolved H I absorption.}",
      journal = {\aap},
         year = 1997,
        month = jan,
       volume = {317},
        pages = {358-384},
          doi = {10.48550/arXiv.astro-ph/9608092},
archivePrefix = {arXiv},
       eprint = {astro-ph/9608092},
 primaryClass = {astro-ph},
       adsurl = {https://ui.adsabs.harvard.edu/abs/1997A&A...317..358V}
}

@ARTICLE{vanOjik1996,
       author = {{van Ojik}, R. and {Roettgering}, H.~J.~A. and {Carilli}, C.~L. and {Miley}, G.~K. and {Bremer}, M.~N. and {Macchetto}, F.},
        title = "{A powerful radio galaxy at z=3.6 in a giant rotating Lyman {\ensuremath{\alpha}} halo.}",
      journal = {\aap},
         year = 1996,
        month = sep,
       volume = {313},
        pages = {25-44},
          doi = {10.48550/arXiv.astro-ph/9608099},
archivePrefix = {arXiv},
       eprint = {astro-ph/9608099},
 primaryClass = {astro-ph},
       adsurl = {https://ui.adsabs.harvard.edu/abs/1996A&A...313...25V}
}

@ARTICLE{Humphrey2008b,
       author = {{Humphrey}, A. and {Villar-Mart{\'\i}n}, M. and {S{\'a}nchez}, S.~F. and {di Serego Alighieri}, S. and {De Breuck}, C. and {Binette}, L. and {Tadhunter}, C. and {Vernet}, J. and {Fosbury}, R. and {Stasielak}, J.},
        title = "{Spatially extended absorption around the z = 2.63 radio galaxy MRC 2025-218: outflow or infall?}",
      journal = {\mnras},
         year = 2008,
        month = nov,
       volume = {390},
       number = {4},
        pages = {1505-1516},
          doi = {10.1111/j.1365-2966.2008.13826.x},
archivePrefix = {arXiv},
       eprint = {0809.1267},
 primaryClass = {astro-ph},
       adsurl = {https://ui.adsabs.harvard.edu/abs/2008MNRAS.390.1505H}
}

@ARTICLE{Rottgering1995,
       author = {{R{\"o}ttgering}, H.~J.~A. and {Hunstead}, R.~W. and {Miley}, G.~K. and {van Ojik}, R. and {Wieringa}, M.~H.},
        title = "{Spatially resolved Ly-alpha absorption in the z=2.9 radio galaxy 0943-242.}",
      journal = {\mnras},
         year = 1995,
        month = nov,
       volume = {277},
        pages = {389-396},
          doi = {10.1093/mnras/277.2.389},
       adsurl = {https://ui.adsabs.harvard.edu/abs/1995MNRAS.277..389R}
}

@INPROCEEDINGS{Rottgering1999,
       author = {{R{\"o}ttgering}, Huub and {de Bruyn}, Ger and {Pentericci}, Laura},
        title = "{HI and high redshift radio galaxies}",
    booktitle = {The Most Distant Radio Galaxies},
         year = 1999,
       editor = {{R{\"o}ttgering}, H.~J.~A. and {Best}, P.~N. and {Lehnert}, M.~D.},
        month = jan,
        pages = {113},
       adsurl = {https://ui.adsabs.harvard.edu/abs/1999mdrg.conf..113R}
}

@ARTICLE{Cody2003,
       author = {{Cody}, A.~M. and {Braun}, R.},
        title = "{Observations of the high-redshift galaxy B2 0902+343 at 92 cm}",
      journal = {\aap},
         year = 2003,
        month = mar,
       volume = {400},
        pages = {871-875},
          doi = {10.1051/0004-6361:20030082},
archivePrefix = {arXiv},
       eprint = {astro-ph/0301418},
 primaryClass = {astro-ph},
       adsurl = {https://ui.adsabs.harvard.edu/abs/2003A&A...400..871C}
}

@ARTICLE{Chandra2004,
       author = {{Chandra}, P. and {Swarup}, G. and {Kulkarni}, V.~K. and {Kantharia}, N.~G.},
        title = "{Associated HI absorption in the z=3.4 radio galaxy B2 0902+343 observed with the GMRT.}",
      journal = {Journal of Astrophysics and Astronomy},
         year = 2004,
        month = mar,
       volume = {25},
        pages = {57-65},
          doi = {10.1007/BF02702288},
archivePrefix = {arXiv},
       eprint = {astro-ph/0407380},
 primaryClass = {astro-ph},
       adsurl = {https://ui.adsabs.harvard.edu/abs/2004JApA...25...57C}
}

@ARTICLE{Silva2018a,
       author = {{Silva}, M. and {Humphrey}, A. and {Lagos}, P. and {Guimar{\~a}es}, R. and {Scott}, T. and {Papaderos}, P. and {Morais}, S.~G.},
        title = "{Detection of large-scale Ly {\ensuremath{\alpha}} absorbers at large angles to the radio axis of high-redshift radio galaxies using SOAR}",
      journal = {\mnras},
         year = 2018,
        month = nov,
       volume = {481},
       number = {1},
        pages = {1401-1415},
          doi = {10.1093/mnras/sty2351},
archivePrefix = {arXiv},
       eprint = {1809.03340},
 primaryClass = {astro-ph.GA},
       adsurl = {https://ui.adsabs.harvard.edu/abs/2018MNRAS.481.1401S}
}

@ARTICLE{Krause2002,
       author = {{Krause}, M.},
        title = "{Absorbers and globular cluster formation in powerful high-redshift radio galaxies}",
      journal = {\aap},
         year = 2002,
        month = may,
       volume = {386},
        pages = {L1-L4},
          doi = {10.1051/0004-6361:20020135},
archivePrefix = {arXiv},
       eprint = {astro-ph/0203372},
 primaryClass = {astro-ph},
       adsurl = {https://ui.adsabs.harvard.edu/abs/2002A&A...386L...1K}
}

@ARTICLE{Krause2005,
       author = {{Krause}, M.},
        title = "{Galactic wind shells and high redshift radio galaxies. On the nature of associated absorbers}",
      journal = {\aap},
         year = 2005,
        month = jun,
       volume = {436},
       number = {3},
        pages = {845-851},
          doi = {10.1051/0004-6361:20042450},
archivePrefix = {arXiv},
       eprint = {astro-ph/0503322},
 primaryClass = {astro-ph},
       adsurl = {https://ui.adsabs.harvard.edu/abs/2005A&A...436..845K}
}

@ARTICLE{Binette2000,
       author = {{Binette}, L. and {Kurk}, J.~D. and {Villar-Mart{\'\i}n}, M. and {R{\"o}ttgering}, H.~J.~A.},
        title = "{A vestige low metallicity gas shell surrounding the radio galaxy 0943-242 at z=2.92}",
      journal = {\aap},
         year = 2000,
        month = apr,
       volume = {356},
        pages = {23-32},
          doi = {10.48550/arXiv.astro-ph/0002210},
archivePrefix = {arXiv},
       eprint = {astro-ph/0002210},
 primaryClass = {astro-ph},
       adsurl = {https://ui.adsabs.harvard.edu/abs/2000A&A...356...23B}
}

@ARTICLE{Gullberg2016,
       author = {{Gullberg}, Bitten and {De Breuck}, Carlos and {Lehnert}, Matthew D. and {Vernet}, Jo{\"e}l and {Bacon}, Roland and {Drouart}, Guillaume and {Emonts}, Bjorn and {Galametz}, Audrey and {Ivison}, Rob and {Nesvadba}, Nicole P.~H. and {Richard}, Johan and {Seymour}, Nick and {Stern}, Daniel and {Wylezalek}, Dominika},
        title = "{The mysterious morphology of MRC0943-242 as revealed by ALMA and MUSE}",
      journal = {\aap},
         year = 2016,
        month = feb,
       volume = {586},
          eid = {A124},
        pages = {A124},
          doi = {10.1051/0004-6361/201526858},
archivePrefix = {arXiv},
       eprint = {1510.03442},
 primaryClass = {astro-ph.GA},
       adsurl = {https://ui.adsabs.harvard.edu/abs/2016A&A...586A.124G}
}

@INPROCEEDINGS{RöttgeringPentericci1999,
       author = {{R{\"o}ttgering}, Huub and {Pentericci}, Laura},
        title = "{Ly{\ensuremath{\alpha}} emitting gas in distant radio galaxies: an evolutionary probe?}",
    booktitle = {The Most Distant Radio Galaxies},
         year = 1999,
       editor = {{R{\"o}ttgering}, H.~J.~A. and {Best}, P.~N. and {Lehnert}, M.~D.},
        month = jan,
        pages = {85},
       adsurl = {https://ui.adsabs.harvard.edu/abs/1999mdrg.conf...85R}
}

@ARTICLE{Vernet2017,
       author = {{Vernet}, J. and {Lehnert}, M.~D. and {De Breuck}, C. and {Villar-Mart{\'\i}n}, M. and {Wylezalek}, D. and {Falkendal}, T. and {Drouart}, G. and {Kolwa}, S. and {Humphrey}, A. and {Venemans}, B.~P. and {Boulanger}, F.},
        title = "{Are we seeing accretion flows in a 250 kpc Ly{\ensuremath{\alpha}} halo at z = 3?}",
      journal = {\aap},
         year = 2017,
        month = jun,
       volume = {602},
          eid = {L6},
        pages = {L6},
          doi = {10.1051/0004-6361/201730865},
archivePrefix = {arXiv},
       eprint = {1705.07125},
 primaryClass = {astro-ph.GA},
       adsurl = {https://ui.adsabs.harvard.edu/abs/2017A&A...602L...6V}
}

@ARTICLE{Nesvadba2007,
       author = {{Nesvadba}, N.~P.~H. and {Lehnert}, M.~D. and {De Breuck}, C. and {Gilbert}, A. and {van Breugel}, W.},
        title = "{Compact radio sources and jet-driven AGN feedback in the early universe: constraints from integral-field spectroscopy}",
      journal = {\aap},
         year = 2007,
        month = nov,
       volume = {475},
       number = {1},
        pages = {145-153},
          doi = {10.1051/0004-6361:20078175},
archivePrefix = {arXiv},
       eprint = {0708.4150},
 primaryClass = {astro-ph},
       adsurl = {https://ui.adsabs.harvard.edu/abs/2007A&A...475..145N}
}

@ARTICLE{Nesvadba2008,
       author = {{Nesvadba}, N.~P.~H. and {Lehnert}, M.~D. and {De Breuck}, C. and {Gilbert}, A.~M. and {van Breugel}, W.},
        title = "{Evidence for powerful AGN winds at high redshift: dynamics of galactic outflows in radio galaxies during the ``Quasar Era''}",
      journal = {\aap},
         year = 2008,
        month = nov,
       volume = {491},
       number = {2},
        pages = {407-424},
          doi = {10.1051/0004-6361:200810346},
archivePrefix = {arXiv},
       eprint = {0809.5171},
 primaryClass = {astro-ph},
       adsurl = {https://ui.adsabs.harvard.edu/abs/2008A&A...491..407N}
}

@ARTICLE{Nesvadba2017a,
       author = {{Nesvadba}, N.~P.~H. and {De Breuck}, C. and {Lehnert}, M.~D. and {Best}, P.~N. and {Collet}, C.},
        title = "{The SINFONI survey of powerful radio galaxies at z 2: Jet-driven AGN feedback during the Quasar Era}",
      journal = {\aap},
         year = 2017,
        month = mar,
       volume = {599},
          eid = {A123},
        pages = {A123},
          doi = {10.1051/0004-6361/201528040},
archivePrefix = {arXiv},
       eprint = {1610.02057},
 primaryClass = {astro-ph.GA},
       adsurl = {https://ui.adsabs.harvard.edu/abs/2017A&A...599A.123N}
}

@ARTICLE{Parijskij2014,
       author = {{Parijskij}, Yu. N. and {Thomasson}, P. and {Kopylov}, A.~I. and {Zhelenkova}, O.~P. and {Muxlow}, T.~W.~B. and {Beswick}, R. and {Soboleva}, N.~S. and {Temirova}, A.~V. and {Verkhodanov}, O.~V.},
        title = "{Observations of the z = 4.514 radio galaxy RC J0311+0507}",
      journal = {\mnras},
         year = 2014,
        month = apr,
       volume = {439},
       number = {3},
        pages = {2314-2322},
          doi = {10.1093/mnras/stu047},
       adsurl = {https://ui.adsabs.harvard.edu/abs/2014MNRAS.439.2314P}
}

@ARTICLE{HeckmannBest2014,
       author = {{Heckman}, Timothy M. and {Best}, Philip N.},
        title = "{The Coevolution of Galaxies and Supermassive Black Holes: Insights from Surveys of the Contemporary Universe}",
      journal = {\araa},
         year = 2014,
        month = aug,
       volume = {52},
        pages = {589-660},
          doi = {10.1146/annurev-astro-081913-035722},
archivePrefix = {arXiv},
       eprint = {1403.4620},
 primaryClass = {astro-ph.GA},
       adsurl = {https://ui.adsabs.harvard.edu/abs/2014ARA&A..52..589H}
}

@ARTICLE{Tumlinson2017,
       author = {{Tumlinson}, Jason and {Peeples}, Molly S. and {Werk}, Jessica K.},
        title = "{The Circumgalactic Medium}",
      journal = {\araa},
         year = 2017,
        month = aug,
       volume = {55},
       number = {1},
        pages = {389-432},
          doi = {10.1146/annurev-astro-091916-055240},
archivePrefix = {arXiv},
       eprint = {1709.09180},
 primaryClass = {astro-ph.GA},
       adsurl = {https://ui.adsabs.harvard.edu/abs/2017ARA&A..55..389T}
}

@INPROCEEDINGS{Seymour2007,
       author = {{Seymour}, N. and {Stern}, D. and {De Breuck}, C.},
        title = "{Spitzer Observations of High Redshift Radio Galaxies}",
    booktitle = {Deepest Astronomical Surveys},
         year = 2007,
        editor = {{Afonso}, J. and {Ferguson}, H.~C. and {Mobasher}, B. and {Norris}, R.},
       series = {Astronomical Society of the Pacific Conference Series},
       volume = {380},
        month = dec,
        pages = {393},
          doi = {10.48550/arXiv.astro-ph/0604226},
archivePrefix = {arXiv},
       eprint = {astro-ph/0604226},
 primaryClass = {astro-ph},
       adsurl = {https://ui.adsabs.harvard.edu/abs/2007ASPC..380..393S}
}

@ARTICLE{Rocca-Volmerange2004,
       author = {{Rocca-Volmerange}, B. and {Le Borgne}, D. and {De Breuck}, C. and {Fioc}, M. and {Moy}, E.},
        title = "{The radio galaxy K-z relation: The {}10$^{12}$ M$_{{\ensuremath{\odot}}}$ mass limit. Masses of galaxies from the L$_{K}$ luminosity, up to z > 4}",
      journal = {\aap},
         year = 2004,
        month = mar,
       volume = {415},
        pages = {931-940},
          doi = {10.1051/0004-6361:20031717},
archivePrefix = {arXiv},
       eprint = {astro-ph/0311490},
 primaryClass = {astro-ph},
       adsurl = {https://ui.adsabs.harvard.edu/abs/2004A&A...415..931R}
}

@INPROCEEDINGS{Wylezalek2013,
       author = {{Wylezalek}, D. and {De Breuck}, C. and {Vernet}, J. and {Galametz}, A. and {Stern}, D. and {Brodwin}, M. and {Eisenhardt}, P. and {Gonzalez}, A. and {Hatch}, N. and {Jarvis}, M. and {Rettura}, A. and {Seymour}, N. and {Stanford}, A. and {Stevens}, J.},
        title = "{Galaxy Clusters around Radio-Loud AGN at 1.3<z<3.2}",
    booktitle = {Tracing Cosmic Evolution with Clusters of Galaxies},
         year = 2013,
        month = jul,
          eid = {87},
        pages = {87},
       adsurl = {https://ui.adsabs.harvard.edu/abs/2013tcec.confE..87W}
}

@ARTICLE{Noirot2018,
       author = {{Noirot}, Ga{\"e}l and {Stern}, Daniel and {Mei}, Simona and {Wylezalek}, Dominika and {Cooke}, Elizabeth A. and {De Breuck}, Carlos and {Galametz}, Audrey and {Hatch}, Nina A. and {Vernet}, Jo{\"e}l and {Brodwin}, Mark and {Eisenhardt}, Peter and {Gonzalez}, Anthony H. and {Jarvis}, Matt and {Rettura}, Alessandro and {Seymour}, Nick and {Stanford}, S.~A.},
        title = "{HST Grism Confirmation of 16 Structures at 1.4 < z < 2.8 from the Clusters Around Radio-Loud AGN (CARLA) Survey}",
      journal = {\apj},
         year = 2018,
        month = may,
       volume = {859},
       number = {1},
          eid = {38},
        pages = {38},
          doi = {10.3847/1538-4357/aabadb},
archivePrefix = {arXiv},
       eprint = {1804.01500},
 primaryClass = {astro-ph.GA},
       adsurl = {https://ui.adsabs.harvard.edu/abs/2018ApJ...859...38N}
}

@ARTICLE{Venemans2002,
       author = {{Venemans}, B.~P. and {Kurk}, J.~D. and {Miley}, G.~K. and {R{\"o}ttgering}, H.~J.~A. and {van Breugel}, W. and {Carilli}, C.~L. and {De Breuck}, C. and {Ford}, H. and {Heckman}, T. and {McCarthy}, P. and {Pentericci}, L.},
        title = "{The Most Distant Structure of Galaxies Known: A Protocluster at z=4.1}",
      journal = {\apjl},
         year = 2002,
        month = apr,
       volume = {569},
       number = {1},
        pages = {L11-L14},
          doi = {10.1086/340563},
archivePrefix = {arXiv},
       eprint = {astro-ph/0203249},
 primaryClass = {astro-ph},
       adsurl = {https://ui.adsabs.harvard.edu/abs/2002ApJ...569L..11V}
}

@ARTICLE{MileyDeBreuck2008,
       author = {{Miley}, George and {De Breuck}, Carlos},
        title = "{Distant radio galaxies and their environments}",
      journal = {\aapr},
         year = 2008,
        month = feb,
       volume = {15},
       number = {2},
        pages = {67-144},
          doi = {10.1007/s00159-007-0008-z},
archivePrefix = {arXiv},
       eprint = {0802.2770},
 primaryClass = {astro-ph},
       adsurl = {https://ui.adsabs.harvard.edu/abs/2008A&ARv..15...67M}
}

@ARTICLE{McCarthy1993_review,
       author = {{McCarthy}, Patrick J.},
        title = "{High redshift radio galaxies.}",
      journal = {\araa},
         year = 1993,
        month = jan,
       volume = {31},
        pages = {639-688},
          doi = {10.1146/annurev.aa.31.090193.003231},
       adsurl = {https://ui.adsabs.harvard.edu/abs/1993ARA&A..31..639M}
}

@ARTICLE{Rauch1998,
       author = {{Rauch}, Michael},
        title = "{The Lyman Alpha Forest in the Spectra of QSOs}",
      journal = {\araa},
         year = 1998,
        month = jan,
       volume = {36},
        pages = {267-316},
          doi = {10.1146/annurev.astro.36.1.267},
archivePrefix = {arXiv},
       eprint = {astro-ph/9806286},
 primaryClass = {astro-ph},
       adsurl = {https://ui.adsabs.harvard.edu/abs/1998ARA&A..36..267R}
}

@ARTICLE{Dekel2013,
       author = {{Dekel}, A. and {Zolotov}, A. and {Tweed}, D. and {Cacciato}, M. and {Ceverino}, D. and {Primack}, J.~R.},
        title = "{Toy models for galaxy formation versus simulations}",
      journal = {\mnras},
         year = 2013,
        month = oct,
       volume = {435},
       number = {2},
        pages = {999-1019},
          doi = {10.1093/mnras/stt1338},
archivePrefix = {arXiv},
       eprint = {1303.3009},
 primaryClass = {astro-ph.CO},
       adsurl = {https://ui.adsabs.harvard.edu/abs/2013MNRAS.435..999D}
}

@ARTICLE{Behroozi2013,
       author = {{Behroozi}, Peter S. and {Wechsler}, Risa H. and {Conroy}, Charlie},
        title = "{The Average Star Formation Histories of Galaxies in Dark Matter Halos from z = 0-8}",
      journal = {\apj},
         year = 2013,
        month = jun,
       volume = {770},
       number = {1},
          eid = {57},
        pages = {57},
          doi = {10.1088/0004-637X/770/1/57},
archivePrefix = {arXiv},
       eprint = {1207.6105},
 primaryClass = {astro-ph.CO},
       adsurl = {https://ui.adsabs.harvard.edu/abs/2013ApJ...770...57B}
}

@ARTICLE{DeBreuck2010,
       author = {{De Breuck}, Carlos and {Seymour}, Nick and {Stern}, Daniel and {Willner}, S.~P. and {Eisenhardt}, P.~R.~M. and {Fazio}, G.~G. and {Galametz}, Audrey and {Lacy}, Mark and {Rettura}, Alessandro and {Rocca-Volmerange}, Brigitte and {Vernet}, Jo{\"e}l},
        title = "{The Spitzer High-redshift Radio Galaxy Survey}",
      journal = {\apj},
         year = 2010,
        month = dec,
       volume = {725},
       number = {1},
        pages = {36-62},
          doi = {10.1088/0004-637X/725/1/36},
archivePrefix = {arXiv},
       eprint = {1010.1385},
 primaryClass = {astro-ph.CO},
       adsurl = {https://ui.adsabs.harvard.edu/abs/2010ApJ...725...36D}
}

@ARTICLE{Cai2019,
       author = {{Cai}, Zheng and {Cantalupo}, Sebastiano and {Prochaska}, J. Xavier and {Arrigoni Battaia}, Fabrizio and {Burchett}, Joe and {Li}, Qiong and {Chisholm}, John and {Bundy}, Kevin and {Hennawi}, Joseph F.},
        title = "{Evolution of the Cool Gas in the Circumgalactic Medium of Massive Halos: A Keck Cosmic Web Imager Survey of Ly{\ensuremath{\alpha}} Emission around QSOs at z {\ensuremath{\approx}} 2}",
      journal = {\apjs},
         year = 2019,
        month = dec,
       volume = {245},
       number = {2},
          eid = {23},
        pages = {23},
          doi = {10.3847/1538-4365/ab4796},
archivePrefix = {arXiv},
       eprint = {1909.11098},
 primaryClass = {astro-ph.GA},
       adsurl = {https://ui.adsabs.harvard.edu/abs/2019ApJS..245...23C}
}

@ARTICLE{ArrigoniBattaia2019,
       author = {{Arrigoni Battaia}, Fabrizio and {Hennawi}, Joseph F. and {Prochaska}, J. Xavier and {O{\~n}orbe}, Jose and {Farina}, Emanuele P. and {Cantalupo}, Sebastiano and {Lusso}, Elisabeta},
        title = "{QSO MUSEUM I: a sample of 61 extended Ly {\ensuremath{\alpha}}-emission nebulae surrounding z {\ensuremath{\sim}} 3 quasars}",
      journal = {\mnras},
         year = 2019,
        month = jan,
       volume = {482},
       number = {3},
        pages = {3162-3205},
          doi = {10.1093/mnras/sty2827},
archivePrefix = {arXiv},
       eprint = {1808.10857},
 primaryClass = {astro-ph.GA},
       adsurl = {https://ui.adsabs.harvard.edu/abs/2019MNRAS.482.3162A}
}

@ARTICLE{Borisova2016,
       author = {{Borisova}, Elena and {Cantalupo}, Sebastiano and {Lilly}, Simon J. and {Marino}, Raffaella A. and {Gallego}, Sofia G. and {Bacon}, Roland and {Blaizot}, Jeremy and {Bouch{\'e}}, Nicolas and {Brinchmann}, Jarle and {Carollo}, C. Marcella and {Caruana}, Joseph and {Finley}, Hayley and {Herenz}, Edmund C. and {Richard}, Johan and {Schaye}, Joop and {Straka}, Lorrie A. and {Turner}, Monica L. and {Urrutia}, Tanya and {Verhamme}, Anne and {Wisotzki}, Lutz},
        title = "{Ubiquitous Giant Ly{\ensuremath{\alpha}} Nebulae around the Brightest Quasars at z {\ensuremath{\sim}} 3.5 Revealed with MUSE}",
      journal = {\apj},
         year = 2016,
        month = nov,
       volume = {831},
       number = {1},
          eid = {39},
        pages = {39},
          doi = {10.3847/0004-637X/831/1/39},
archivePrefix = {arXiv},
       eprint = {1605.01422},
 primaryClass = {astro-ph.GA},
       adsurl = {https://ui.adsabs.harvard.edu/abs/2016ApJ...831...39B}
}

@ARTICLE{Farina2019,
       author = {{Farina}, Emanuele Paolo and {Arrigoni-Battaia}, Fabrizio and {Costa}, Tiago and {Walter}, Fabian and {Hennawi}, Joseph F. and {Drake}, Alyssa B. and {Decarli}, Roberto and {Gutcke}, Thales A. and {Mazzucchelli}, Chiara and {Neeleman}, Marcel and {Georgiev}, Iskren and {Eilers}, Anna-Christina and {Davies}, Frederick B. and {Ba{\~n}ados}, Eduardo and {Fan}, Xiaohui and {Onoue}, Masafusa and {Schindler}, Jan-Torge and {Venemans}, Bram P. and {Wang}, Feige and {Yang}, Jinyi and {Rabien}, Sebastian and {Busoni}, Lorenzo},
        title = "{The REQUIEM Survey. I. A Search for Extended Ly{\ensuremath{\alpha}} Nebular Emission Around 31 z > 5.7 Quasars}",
      journal = {\apj},
         year = 2019,
        month = dec,
       volume = {887},
       number = {2},
          eid = {196},
        pages = {196},
          doi = {10.3847/1538-4357/ab5847},
archivePrefix = {arXiv},
       eprint = {1911.08498},
 primaryClass = {astro-ph.GA},
       adsurl = {https://ui.adsabs.harvard.edu/abs/2019ApJ...887..196F}
}

@ARTICLE{Vayner2023,
       author = {{Vayner}, Andrey and {Zakamska}, Nadia L. and {Ishikawa}, Yuzo and {Sankar}, Swetha and {Wylezalek}, Dominika and {Rupke}, David S.~N. and {Veilleux}, Sylvain and {Bertemes}, Caroline and {Barrera-Ballesteros}, Jorge K. and {Chen}, Hsiao-Wen and {Diachenko}, Nadiia and {Goulding}, Andy D. and {Greene}, Jenny E. and {Hainline}, Kevin N. and {Hamann}, Fred and {Heckman}, Timothy and {Johnson}, Sean D. and {Grace Lim}, Hui Xian and {Liu}, Weizhe and {Lutz}, Dieter and {L{\"u}tzgendorf}, Nora and {Mainieri}, Vincenzo and {McCrory}, Ryan and {Murphree}, Grey and {Nesvadba}, Nicole P.~H. and {Ogle}, Patrick and {Sturm}, Eckhard and {Whitesell}, Lillian},
        title = "{First Results from the JWST Early Release Science Program Q3D: Ionization Cone, Clumpy Star Formation, and Shocks in a z = 3 Extremely Red Quasar Host}",
      journal = {\apj},
         year = 2023,
        month = oct,
       volume = {955},
       number = {2},
          eid = {92},
        pages = {92},
          doi = {10.3847/1538-4357/ace784},
archivePrefix = {arXiv},
       eprint = {2303.06970},
 primaryClass = {astro-ph.GA},
       adsurl = {https://ui.adsabs.harvard.edu/abs/2023ApJ...955...92V}
}

@ARTICLE{Wylezalek2017,
       author = {{Wylezalek}, Dominika and {Schnorr M{\"u}ller}, Allan and {Zakamska}, Nadia L. and {Storchi-Bergmann}, Thaisa and {Greene}, Jenny E. and {M{\"u}ller-S{\'a}nchez}, Francisco and {Kelly}, Michael and {Liu}, Guilin and {Law}, David R. and {Barrera-Ballesteros}, Jorge K. and {Riffel}, Rogemar A. and {Thomas}, Daniel},
        title = "{Zooming into local active galactic nuclei: the power of combining SDSS-IV MaNGA with higher resolution integral field unit observations}",
      journal = {\mnras},
         year = 2017,
        month = may,
       volume = {467},
       number = {3},
        pages = {2612-2624},
          doi = {10.1093/mnras/stx246},
archivePrefix = {arXiv},
       eprint = {1610.01602},
 primaryClass = {astro-ph.GA},
       adsurl = {https://ui.adsabs.harvard.edu/abs/2017MNRAS.467.2612W}
}

@ARTICLE{Harrison2014,
       author = {{Harrison}, C.~M. and {Alexander}, D.~M. and {Mullaney}, J.~R. and {Swinbank}, A.~M.},
        title = "{Kiloparsec-scale outflows are prevalent among luminous AGN: outflows and feedback in the context of the overall AGN population}",
      journal = {\mnras},
         year = 2014,
        month = jul,
       volume = {441},
       number = {4},
        pages = {3306-3347},
          doi = {10.1093/mnras/stu515},
archivePrefix = {arXiv},
       eprint = {1403.3086},
 primaryClass = {astro-ph.GA},
       adsurl = {https://ui.adsabs.harvard.edu/abs/2014MNRAS.441.3306H}
}

@ARTICLE{Morton2000,
       author = {{Morton}, Donald C.},
        title = "{Atomic Data for Resonance Absorption Lines. II. Wavelengths Longward of the Lyman Limit for Heavy Elements}",
      journal = {\apjs},
         year = 2000,
        month = oct,
       volume = {130},
       number = {2},
        pages = {403-436},
          doi = {10.1086/317349},
       adsurl = {https://ui.adsabs.harvard.edu/abs/2000ApJS..130..403M}
}

@ARTICLE{Wang2024,
       author = {{Wang}, Wuji and {Wylezalek}, Dominika and {De Breuck}, Carlos and {Vernet}, Jo{\"e}l and {Rupke}, David S.~N. and {Zakamska}, Nadia L. and {Vayner}, Andrey and {Lehnert}, Matthew D. and {Nesvadba}, Nicole P.~H. and {Stern}, Daniel},
        title = "{JWST discovers an AGN ionization cone but only weak radiatively driven feedback in a powerful z {\ensuremath{\approx}} 3.5 radio-loud AGN}",
      journal = {\aap},
         year = 2024,
        month = mar,
       volume = {683},
          eid = {A169},
        pages = {A169},
          doi = {10.1051/0004-6361/202348531},
archivePrefix = {arXiv},
       eprint = {2401.02479},
 primaryClass = {astro-ph.GA},
       adsurl = {https://ui.adsabs.harvard.edu/abs/2024A&A...683A.169W}
}

@ARTICLE{Böker2022,
       author = {{B{\"o}ker}, T. and {Arribas}, S. and {L{\"u}tzgendorf}, N. and {Alves de Oliveira}, C. and {Beck}, T.~L. and {Birkmann}, S. and {Bunker}, A.~J. and {Charlot}, S. and {de Marchi}, G. and {Ferruit}, P. and {Giardino}, G. and {Jakobsen}, P. and {Kumari}, N. and {L{\'o}pez-Caniego}, M. and {Maiolino}, R. and {Manjavacas}, E. and {Marston}, A. and {Moseley}, S.~H. and {Muzerolle}, J. and {Ogle}, P. and {Pirzkal}, N. and {Rauscher}, B. and {Rawle}, T. and {Rix}, H. -W. and {Sabbi}, E. and {Sargent}, B. and {Sirianni}, M. and {te Plate}, M. and {Valenti}, J. and {Willott}, C.~J. and {Zeidler}, P.},
        title = "{The Near-Infrared Spectrograph (NIRSpec) on the James Webb Space Telescope. III. Integral-field spectroscopy}",
      journal = {\aap},
         year = 2022,
        month = may,
       volume = {661},
          eid = {A82},
        pages = {A82},
          doi = {10.1051/0004-6361/202142589},
archivePrefix = {arXiv},
       eprint = {2202.03308},
 primaryClass = {astro-ph.IM},
       adsurl = {https://ui.adsabs.harvard.edu/abs/2022A&A...661A..82B}
}

@ARTICLE{Chang2023,
       author = {{Chang}, Seok-Jun and {Yang}, Yujin and {Seon}, Kwang-Il and {Zabludoff}, Ann and {Lee}, Hee-Won},
        title = "{Radiative Transfer in Ly{\ensuremath{\alpha}} Nebulae. I. Modeling a Continuous or Clumpy Spherical Halo with a Central Source}",
      journal = {\apj},
         year = 2023,
        month = mar,
       volume = {945},
       number = {2},
          eid = {100},
        pages = {100},
          doi = {10.3847/1538-4357/acac98},
archivePrefix = {arXiv},
       eprint = {2212.09630},
 primaryClass = {astro-ph.GA},
       adsurl = {https://ui.adsabs.harvard.edu/abs/2023ApJ...945..100C}
}

@ARTICLE{SeonKim2020,
       author = {{Seon}, Kwang-il and {Kim}, Chang-Goo},
        title = "{Ly{\ensuremath{\alpha}} Radiative Transfer: Monte Carlo Simulation of the Wouthuysen-Field Effect}",
      journal = {\apjs},
         year = 2020,
        month = sep,
       volume = {250},
       number = {1},
          eid = {9},
        pages = {9},
          doi = {10.3847/1538-4365/aba2d6},
archivePrefix = {arXiv},
       eprint = {2005.00238},
 primaryClass = {astro-ph.GA},
       adsurl = {https://ui.adsabs.harvard.edu/abs/2020ApJS..250....9S}
}

@ARTICLE{Verhamme2006,
       author = {{Verhamme}, A. and {Schaerer}, D. and {Maselli}, A.},
        title = "{3D Ly{\ensuremath{\alpha}} radiation transfer. I. Understanding Ly{\ensuremath{\alpha}} line profile morphologies}",
      journal = {\aap},
         year = 2006,
        month = dec,
       volume = {460},
       number = {2},
        pages = {397-413},
          doi = {10.1051/0004-6361:20065554},
archivePrefix = {arXiv},
       eprint = {astro-ph/0608075},
 primaryClass = {astro-ph},
       adsurl = {https://ui.adsabs.harvard.edu/abs/2006A&A...460..397V}
}

@ARTICLE{GronkeDijkstra2016,
       author = {{Gronke}, M. and {Dijkstra}, M.},
        title = "{Lyman-{\ensuremath{\alpha}} Spectra from Multiphase Outflows, and their Connection to Shell Models}",
      journal = {\apj},
         year = 2016,
        month = jul,
       volume = {826},
       number = {1},
          eid = {14},
        pages = {14},
          doi = {10.3847/0004-637X/826/1/14},
archivePrefix = {arXiv},
       eprint = {1604.06805},
 primaryClass = {astro-ph.GA},
       adsurl = {https://ui.adsabs.harvard.edu/abs/2016ApJ...826...14G}
}

@ARTICLE{Gronke2015,
       author = {{Gronke}, M. and {Bull}, P. and {Dijkstra}, M.},
        title = "{A Systematic Study of Lyman-{\ensuremath{\alpha}} Transfer through Outflowing Shells: Model Parameter Estimation}",
      journal = {\apj},
         year = 2015,
        month = oct,
       volume = {812},
       number = {2},
          eid = {123},
        pages = {123},
          doi = {10.1088/0004-637X/812/2/123},
archivePrefix = {arXiv},
       eprint = {1506.03836},
 primaryClass = {astro-ph.GA},
       adsurl = {https://ui.adsabs.harvard.edu/abs/2015ApJ...812..123G}
}

@ARTICLE{Park2022,
       author = {{Park}, Hyunbae and {Kim}, Hyo Jeong and {Ahn}, Kyungjin and {Song}, Hyunmi and {Jung}, Intae and {Ocvirk}, Pierre and {Shapiro}, Paul R. and {Dawoodbhoy}, Taha and {Sorce}, Jenny G. and {Iliev}, Ilian T.},
        title = "{Scattering of Ly{\ensuremath{\alpha}} Photons through the Reionizing Intergalactic Medium: I. Spectral Energy Distribution}",
      journal = {\apj},
         year = 2022,
        month = jun,
       volume = {931},
       number = {2},
          eid = {126},
        pages = {126},
          doi = {10.3847/1538-4357/ac69e4},
archivePrefix = {arXiv},
       eprint = {2202.06277},
 primaryClass = {astro-ph.CO},
       adsurl = {https://ui.adsabs.harvard.edu/abs/2022ApJ...931..126P}
}

@ARTICLE{Dijkstra2014,
       author = {{Dijkstra}, Mark},
        title = "{Ly{\ensuremath{\alpha}} Emitting Galaxies as a Probe of Reionisation}",
      journal = {\pasa},
         year = 2014,
        month = oct,
       volume = {31},
          eid = {e040},
        pages = {e040},
          doi = {10.1017/pasa.2014.33},
archivePrefix = {arXiv},
       eprint = {1406.7292},
 primaryClass = {astro-ph.CO},
       adsurl = {https://ui.adsabs.harvard.edu/abs/2014PASA...31...40D}
}

@ARTICLE{Dijkstra2006,
       author = {{Dijkstra}, Mark and {Haiman}, Zolt{\'a}n and {Spaans}, Marco},
        title = "{Ly{\ensuremath{\alpha}} Radiation from Collapsing Protogalaxies. I. Characteristics of the Emergent Spectrum}",
      journal = {\apj},
         year = 2006,
        month = sep,
       volume = {649},
       number = {1},
        pages = {14-36},
          doi = {10.1086/506243},
archivePrefix = {arXiv},
       eprint = {astro-ph/0510407},
 primaryClass = {astro-ph},
       adsurl = {https://ui.adsabs.harvard.edu/abs/2006ApJ...649...14D}
}

@misc{Bushouse2023,
       author = {{Bushouse}, Howard and {Eisenhamer}, Jonathan and {Dencheva}, Nadia and {Davies}, James and {Greenfield}, Perry and {Morrison}, Jane and {Hodge}, Phil and {Simon}, Bernie and {Grumm}, David and {Droettboom}, Michael and {Slavich}, Edward and {Sosey}, Megan and {Pauly}, Tyler and {Miller}, Todd and {Jedrzejewski}, Robert and {Hack}, Warren and {Davis}, David and {Crawford}, Steven and {Law}, David and {Gordon}, Karl and {Regan}, Michael and {Cara}, Mihai and {MacDonald}, Ken and {Bradley}, Larry and {Shanahan}, Clare and {Jamieson}, William and {Teodoro}, Mairan and {Williams}, Thomas and {Pena-Guerrero}, Maria},
        title = "{JWST Calibration Pipeline}",
         year = 2023,
        month = oct,
          eid = {10.5281/zenodo.10022973},
          doi = {10.5281/zenodo.10022973},
      version = {1.12.5},
    publisher = {Zenodo},
       adsurl = {https://ui.adsabs.harvard.edu/abs/2023zndo..10022973B}
}

@ARTICLE{ArrigoniBattaia2018,
       author = {{Arrigoni Battaia}, Fabrizio and {Prochaska}, J. Xavier and {Hennawi}, Joseph F. and {Obreja}, Aura and {Buck}, Tobias and {Cantalupo}, Sebastiano and {Dutton}, Aaron A. and {Macci{\`o}}, Andrea V.},
        title = "{Inspiraling halo accretion mapped in Ly {\ensuremath{\alpha}} emission around a z {\ensuremath{\sim}} 3 quasar}",
      journal = {\mnras},
         year = 2018,
        month = jan,
       volume = {473},
       number = {3},
        pages = {3907-3940},
          doi = {10.1093/mnras/stx2465},
archivePrefix = {arXiv},
       eprint = {1709.08228},
 primaryClass = {astro-ph.GA},
       adsurl = {https://ui.adsabs.harvard.edu/abs/2018MNRAS.473.3907A}
}

@ARTICLE{Wang2025,
       author = {{Wang}, Wuji and {De Breuck}, Carlos and {Wylezalek}, Dominika and {Vernet}, Jo{\"e}l and {Lehnert}, Matthew D. and {Stern}, Daniel and {Rupke}, David S.~N. and {Nesvadba}, Nicole P.~H. and {Vayner}, Andrey and {Zakamska}, Nadia L. and {Lin}, Lingrui and {Kukreti}, Pranav and {Dall'Agnol de Oliveira}, Bruno and {Groth}, Julian T.},
        title = "{JWST + ALMA ubiquitously discover companion systems within {\ensuremath{\lesssim}}18 kpc around four z ≍ 3.5 luminous radio-loud AGN}",
      journal = {\aap},
         year = 2025,
        month = apr,
       volume = {696},
          eid = {A88},
        pages = {A88},
          doi = {10.1051/0004-6361/202553668},
archivePrefix = {arXiv},
       eprint = {2502.20442},
 primaryClass = {astro-ph.GA},
       adsurl = {https://ui.adsabs.harvard.edu/abs/2025A&A...696A..88W}
}

@ARTICLE{Kim2002,
       author = {{Kim}, T. -S. and {Carswell}, R.~F. and {Cristiani}, S. and {D'Odorico}, S. and {Giallongo}, E.},
        title = "{The physical properties of the Ly{\ensuremath{\alpha}} forest at z > 1.5}",
      journal = {\mnras},
         year = 2002,
        month = sep,
       volume = {335},
       number = {3},
        pages = {555-573},
          doi = {10.1046/j.1365-8711.2002.05599.x},
archivePrefix = {arXiv},
       eprint = {astro-ph/0205237},
 primaryClass = {astro-ph},
       adsurl = {https://ui.adsabs.harvard.edu/abs/2002MNRAS.335..555K}
}

@ARTICLE{Rahmati2013,
       author = {{Rahmati}, Alireza and {Pawlik}, Andreas H. and {Rai{\v{c}}evi{\'c}}, Milan and {Schaye}, Joop},
        title = "{On the evolution of the H I column density distribution in cosmological simulations}",
      journal = {\mnras},
         year = 2013,
        month = apr,
       volume = {430},
       number = {3},
        pages = {2427-2445},
          doi = {10.1093/mnras/stt066},
archivePrefix = {arXiv},
       eprint = {1210.7808},
 primaryClass = {astro-ph.CO},
       adsurl = {https://ui.adsabs.harvard.edu/abs/2013MNRAS.430.2427R}
}

@article{Hu1995,
   title={The Distribution of Column Densities and B Values in the Lyman-Alpha Forest},
   volume={110},
   ISSN={0004-6256},
   url={http://dx.doi.org/10.1086/117625},
   DOI={10.1086/117625},
   journal={The Astronomical Journal},
   publisher={American Astronomical Society},
   author={Hu, Esther M. and Kim, Tae-Sun and Cowie, Lennox L. and Songaila, Antoinette and Rauch, Michael},
   year={1995},
   month=oct, pages={1526} }

@ARTICLE{Villar-Martin2007,
       author = {{Villar-Mart{\'\i}n}, M. and {S{\'a}nchez}, S.~F. and {Humphrey}, A. and {Dijkstra}, M. and {di Serego Alighieri}, S. and {De Breuck}, C. and {Gonz{\'a}lez Delgado}, R.},
        title = "{VIMOS-VLT spectroscopy of the giant Ly{\ensuremath{\alpha}} nebulae associated with three z \raisebox{-0.5ex}\textasciitilde 2.5 radio galaxies}",
      journal = {\mnras},
         year = 2007,
        month = jun,
       volume = {378},
       number = {2},
        pages = {416-428},
          doi = {10.1111/j.1365-2966.2007.11811.x},
archivePrefix = {arXiv},
       eprint = {0704.1116},
 primaryClass = {astro-ph},
       adsurl = {https://ui.adsabs.harvard.edu/abs/2007MNRAS.378..416V}
}

@ARTICLE{Shukla2022,
       author = {{Shukla}, Gitika and {Srianand}, Raghunathan and {Gupta}, Neeraj and {Petitjean}, Patrick and {Baker}, Andrew J. and {Krogager}, Jens-Kristian and {Noterdaeme}, Pasquier},
        title = "{Spatially resolved Lyman-{\ensuremath{\alpha}} emission around radio bright quasars}",
      journal = {\mnras},
         year = 2022,
        month = feb,
       volume = {510},
       number = {1},
        pages = {786-806},
          doi = {10.1093/mnras/stab3467},
archivePrefix = {arXiv},
       eprint = {2109.00576},
 primaryClass = {astro-ph.GA},
       adsurl = {https://ui.adsabs.harvard.edu/abs/2022MNRAS.510..786S}
}

@ARTICLE{Emonts2023,
       author = {{Emonts}, Bjorn H.~C. and {Lehnert}, Matthew D. and {Lebowitz}, Sophie and {Miley}, George K. and {Villar-Mart{\'\i}n}, Montserrat and {Norris}, Ray and {De Breuck}, Carlos and {Carilli}, Chris and {Feain}, Ilana},
        title = "{CO Survey of High-z Radio Galaxies, Revisited with the Atacama Large Millimeter/submillimeter Array: Jet-Cloud Alignments and Synchrotron Brightening by Molecular Gas in the Circumgalactic Environment}",
      journal = {\apj},
         year = 2023,
        month = aug,
       volume = {952},
       number = {2},
          eid = {148},
        pages = {148},
          doi = {10.3847/1538-4357/acde53},
archivePrefix = {arXiv},
       eprint = {2306.12636},
 primaryClass = {astro-ph.GA},
       adsurl = {https://ui.adsabs.harvard.edu/abs/2023ApJ...952..148E}
}

@ARTICLE{Li2021,
       author = {{Li}, Jianrui and {Emonts}, Bjorn H.~C. and {Cai}, Zheng and {Prochaska}, J. Xavier and {Yoon}, Ilsang and {Lehnert}, Matthew D. and {Zhang}, Shiwu and {Wu}, Yunjing and {Li}, Jianan and {Li}, Mingyu and {Lacy}, Mark and {Villar-Mart{\'\i}n}, Montserrat},
        title = "{Massive Molecular Outflow and 100 kpc Extended Cold Halo Gas in the Enormous Ly{\ensuremath{\alpha}} Nebula of QSO 1228+3128}",
      journal = {\apjl},
         year = 2021,
        month = dec,
       volume = {922},
       number = {2},
          eid = {L29},
        pages = {L29},
          doi = {10.3847/2041-8213/ac390d},
archivePrefix = {arXiv},
       eprint = {2111.06409},
 primaryClass = {astro-ph.GA},
       adsurl = {https://ui.adsabs.harvard.edu/abs/2021ApJ...922L..29L}
}

@ARTICLE{Falkendal2021,
       author = {{Falkendal}, Theresa and {Lehnert}, Matthew D. and {Vernet}, Jo{\"e}l and {De Breuck}, Carlos and {Wang}, Wuji},
        title = "{ALMA and MUSE observations reveal a quiescent multi-phase circumgalactic medium around the z ≃ 3.6 radio galaxy 4C 19.71}",
      journal = {\aap},
         year = 2021,
        month = jan,
       volume = {645},
          eid = {A120},
        pages = {A120},
          doi = {10.1051/0004-6361/201935237},
archivePrefix = {arXiv},
       eprint = {2007.10061},
 primaryClass = {astro-ph.GA},
       adsurl = {https://ui.adsabs.harvard.edu/abs/2021A&A...645A.120F}
}

@ARTICLE{DeBreuck2022,
       author = {{De Breuck}, C. and {Lundgren}, A. and {Emonts}, B. and {Kolwa}, S. and {Dannerbauer}, H. and {Lehnert}, M.},
        title = "{Feeding the spider with carbon. [CII] emission from the circumgalactic medium and active galactic nucleus}",
      journal = {\aap},
         year = 2022,
        month = feb,
       volume = {658},
          eid = {L2},
        pages = {L2},
          doi = {10.1051/0004-6361/202141853},
archivePrefix = {arXiv},
       eprint = {2201.05064},
 primaryClass = {astro-ph.GA},
       adsurl = {https://ui.adsabs.harvard.edu/abs/2022A&A...658L...2D}
}

@ARTICLE{Emonts2015,
       author = {{Emonts}, B.~H.~C. and {Mao}, M.~Y. and {Stroe}, A. and {Pentericci}, L. and {Villar-Mart{\'\i}n}, M. and {Norris}, R.~P. and {Miley}, G. and {De Breuck}, C. and {van Moorsel}, G.~A. and {Lehnert}, M.~D. and {Carilli}, C.~L. and {R{\"o}ttgering}, H.~J.~A. and {Seymour}, N. and {Sadler}, E.~M. and {Ekers}, R.~D. and {Drouart}, G. and {Feain}, I. and {Colina}, L. and {Stevens}, J. and {Holt}, J.},
        title = "{A CO-rich merger shaping a powerful and hyperluminous infrared radio galaxy at z = 2: the Dragonfly Galaxy}",
      journal = {\mnras},
         year = 2015,
        month = jul,
       volume = {451},
       number = {1},
        pages = {1025-1035},
          doi = {10.1093/mnras/stv930},
archivePrefix = {arXiv},
       eprint = {1505.00949},
 primaryClass = {astro-ph.GA},
       adsurl = {https://ui.adsabs.harvard.edu/abs/2015MNRAS.451.1025E}
}

@ARTICLE{Emonts2016,
       author = {{Emonts}, B.~H.~C. and {Lehnert}, M.~D. and {Villar-Mart{\'\i}n}, M. and {Norris}, R.~P. and {Ekers}, R.~D. and {van Moorsel}, G.~A. and {Dannerbauer}, H. and {Pentericci}, L. and {Miley}, G.~K. and {Allison}, J.~R. and {Sadler}, E.~M. and {Guillard}, P. and {Carilli}, C.~L. and {Mao}, M.~Y. and {R{\"o}ttgering}, H.~J.~A. and {De Breuck}, C. and {Seymour}, N. and {Gullberg}, B. and {Ceverino}, D. and {Jagannathan}, P. and {Vernet}, J. and {Indermuehle}, B.~T.},
        title = "{Molecular gas in the halo fuels the growth of a massive cluster galaxy at high redshift}",
      journal = {Science},
         year = 2016,
        month = dec,
       volume = {354},
       number = {6316},
        pages = {1128-1130},
          doi = {10.1126/science.aag0512},
archivePrefix = {arXiv},
       eprint = {1612.00387},
 primaryClass = {astro-ph.GA},
       adsurl = {https://ui.adsabs.harvard.edu/abs/2016Sci...354.1128E}
}

@ARTICLE{Emonts2014,
       author = {{Emonts}, B.~H.~C. and {Norris}, R.~P. and {Feain}, I. and {Mao}, M.~Y. and {Ekers}, R.~D. and {Miley}, G. and {Seymour}, N. and {R{\"o}ttgering}, H.~J.~A. and {Villar-Mart{\'\i}n}, M. and {Sadler}, E.~M. and {Carilli}, C.~L. and {Mahony}, E.~K. and {de Breuck}, C. and {Stroe}, A. and {Pentericci}, L. and {van Moorsel}, G.~A. and {Drouart}, G. and {Ivison}, R.~J. and {Greve}, T.~R. and {Humphrey}, A. and {Wylezalek}, D. and {Tadhunter}, C.~N.},
        title = "{CO(1-0) survey of high-z radio galaxies: alignment of molecular halo gas with distant radio sources}",
      journal = {\mnras},
         year = 2014,
        month = mar,
       volume = {438},
       number = {4},
        pages = {2898-2915},
          doi = {10.1093/mnras/stt2398},
archivePrefix = {arXiv},
       eprint = {1312.4785},
 primaryClass = {astro-ph.GA},
       adsurl = {https://ui.adsabs.harvard.edu/abs/2014MNRAS.438.2898E}
}

@ARTICLE{Ivison2012,
       author = {{Ivison}, R.~J. and {Smail}, Ian and {Amblard}, A. and {Arumugam}, V. and {De Breuck}, C. and {Emonts}, B.~H.~C. and {Feain}, I. and {Greve}, T.~R. and {Haas}, M. and {Ibar}, E. and {Jarvis}, M.~J. and {Kov{\'a}cs}, A. and {Lehnert}, M.~D. and {Nesvadba}, N.~P.~H. and {R{\"o}ttgering}, H.~J.~A. and {Seymour}, N. and {Wylezalek}, D.},
        title = "{Gas-rich mergers and feedback are ubiquitous amongst starbursting radio galaxies, as revealed by the VLA, IRAM PdBI and Herschel}",
      journal = {\mnras},
         year = 2012,
        month = sep,
       volume = {425},
       number = {2},
        pages = {1320-1331},
          doi = {10.1111/j.1365-2966.2012.21544.x},
archivePrefix = {arXiv},
       eprint = {1206.4046},
 primaryClass = {astro-ph.CO},
       adsurl = {https://ui.adsabs.harvard.edu/abs/2012MNRAS.425.1320I}
}

@ARTICLE{Klamer2005,
       author = {{Klamer}, I.~J. and {Ekers}, R.~D. and {Sadler}, E.~M. and {Weiss}, A. and {Hunstead}, R.~W. and {De Breuck}, C.},
        title = "{CO (1-0) and CO (5-4) Observations of the Most Distant Known Radio Galaxy at z=5.2}",
      journal = {\apjl},
         year = 2005,
        month = mar,
       volume = {621},
       number = {1},
        pages = {L1-L4},
          doi = {10.1086/429147},
archivePrefix = {arXiv},
       eprint = {astro-ph/0501447},
 primaryClass = {astro-ph},
       adsurl = {https://ui.adsabs.harvard.edu/abs/2005ApJ...621L...1K}
}

@ARTICLE{Noirot2016,
       author = {{Noirot}, Ga{\"e}l and {Vernet}, Jo{\"e}l and {De Breuck}, Carlos and {Wylezalek}, Dominika and {Galametz}, Audrey and {Stern}, Daniel and {Mei}, Simona and {Brodwin}, Mark and {Cooke}, Elizabeth A. and {Gonzalez}, Anthony H. and {Hatch}, Nina A. and {Rettura}, Alessandro and {Stanford}, Spencer Adam},
        title = "{HST Grism Confirmation of Two z {\ensuremath{\sim}} 2 Structures from the Clusters around Radio-loud AGN (CARLA) Survey}",
      journal = {\apj},
         year = 2016,
        month = oct,
       volume = {830},
       number = {2},
          eid = {90},
        pages = {90},
          doi = {10.3847/0004-637X/830/2/90},
archivePrefix = {arXiv},
       eprint = {1609.04162},
 primaryClass = {astro-ph.GA},
       adsurl = {https://ui.adsabs.harvard.edu/abs/2016ApJ...830...90N}
}

@ARTICLE{DeBreuck2000,
       author = {{De Breuck}, C. and {R{\"o}ttgering}, H. and {Miley}, G. and {van Breugel}, W. and {Best}, P.},
        title = "{A statistical study of emission lines from high redshift radio galaxies}",
      journal = {\aap},
         year = 2000,
        month = oct,
       volume = {362},
        pages = {519-543},
          doi = {10.48550/arXiv.astro-ph/0008264},
archivePrefix = {arXiv},
       eprint = {astro-ph/0008264},
 primaryClass = {astro-ph},
       adsurl = {https://ui.adsabs.harvard.edu/abs/2000A&A...362..519D}
}

@ARTICLE{Villar-Martin2007_SF,
       author = {{Villar-Mart{\'\i}n}, M. and {Humphrey}, A. and {De Breuck}, C. and {Fosbury}, R. and {Binette}, L. and {Vernet}, J.},
        title = "{Ly{\ensuremath{\alpha}} excess in high-redshift radio galaxies: a signature of star formation}",
      journal = {\mnras},
         year = 2007,
        month = mar,
       volume = {375},
       number = {4},
        pages = {1299-1310},
          doi = {10.1111/j.1365-2966.2006.11371.x},
archivePrefix = {arXiv},
       eprint = {astro-ph/0612116},
 primaryClass = {astro-ph},
       adsurl = {https://ui.adsabs.harvard.edu/abs/2007MNRAS.375.1299V}
}

@ARTICLE{Humphrey2008,
       author = {{Humphrey}, A. and {Villar-Mart{\'\i}n}, M. and {Vernet}, J. and {Fosbury}, R. and {di Serego Alighieri}, S. and {Binette}, L.},
        title = "{Deep spectroscopy of the FUV-optical emission lines from a sample of radio galaxies at z \raisebox{-0.5ex}\textasciitilde 2.5: metallicity and ionization}",
      journal = {\mnras},
         year = 2008,
        month = jan,
       volume = {383},
       number = {1},
        pages = {11-40},
          doi = {10.1111/j.1365-2966.2007.12506.x},
archivePrefix = {arXiv},
       eprint = {0710.5324},
 primaryClass = {astro-ph},
       adsurl = {https://ui.adsabs.harvard.edu/abs/2008MNRAS.383...11H}
}

@ARTICLE{Carilli1997,
       author = {{Carilli}, C.~L. and {R{\"o}ttgering}, H.~J.~A. and {van Ojik}, R. and {Miley}, G.~K. and {van Breugel}, W.~J.~M.},
        title = "{Radio Continuum Imaging of High-Redshift Radio Galaxies}",
      journal = {\apjs},
         year = 1997,
        month = jan,
       volume = {109},
       number = {1},
        pages = {1-44},
          doi = {10.1086/312973},
archivePrefix = {arXiv},
       eprint = {astro-ph/9610157},
 primaryClass = {astro-ph},
       adsurl = {https://ui.adsabs.harvard.edu/abs/1997ApJS..109....1C}
}

@ARTICLE{HarrisonRamosAlmeida2024,
       author = {{Harrison}, Chris M. and {Ramos Almeida}, Cristina},
        title = "{Observational Tests of Active Galactic Nuclei Feedback: An Overview of Approaches and Interpretation}",
      journal = {Galaxies},
         year = 2024,
        month = apr,
       volume = {12},
       number = {2},
          eid = {17},
        pages = {17},
          doi = {10.3390/galaxies12020017},
archivePrefix = {arXiv},
       eprint = {2404.08050},
 primaryClass = {astro-ph.GA},
       adsurl = {https://ui.adsabs.harvard.edu/abs/2024Galax..12...17H}
}

@INCOLLECTION{Kluyver2016,
       author = {{Kluyver}, Thomas and {Ragan-Kelley}, Benjain and {P{\'e}rez}, Fernando and {Granger}, Brian and {Bussonnier}, Matthias and {Frederic}, Jonathan and {Kelley}, Kyle and {Hamrick}, Jessica and {Grout}, Jason and {Corlay}, Sylvain and {Ivanov}, Paul and {Avila}, Dami{\'a}n and {Abdalla}, Safia and {Willing}, Carol and {Jupyter Development Team}},
        title = "{Jupyter Notebooks{\textemdash}a publishing format for reproducible computational workflows}",
    booktitle = {IOS Press},
         year = 2016,
        pages = {87-90},
          doi = {10.3233/978-1-61499-649-1-87},
       adsurl = {https://ui.adsabs.harvard.edu/abs/2016ppap.book...87K}
}

@Article{Hunter2007,
  Author    = {Hunter, J. D.},
  Title     = {Matplotlib: A 2D graphics environment},
  Journal   = {Computing in Science \& Engineering},
  Volume    = {9},
  Number    = {3},
  Pages     = {90--95},
  publisher = {IEEE COMPUTER SOC},
  doi       = {10.1109/MCSE.2007.55},
  year      = 2007
}

@ARTICLE{Virtanen2020,
  author  = {Virtanen, Pauli and Gommers, Ralf and Oliphant, Travis E. and
            Haberland, Matt and Reddy, Tyler and Cournapeau, David and
            Burovski, Evgeni and Peterson, Pearu and Weckesser, Warren and
            Bright, Jonathan and {van der Walt}, St{\'e}fan J. and
            Brett, Matthew and Wilson, Joshua and Millman, K. Jarrod and
            Mayorov, Nikolay and Nelson, Andrew R. J. and Jones, Eric and
            Kern, Robert and Larson, Eric and Carey, C J and
            Polat, {\.I}lhan and Feng, Yu and Moore, Eric W. and
            {VanderPlas}, Jake and Laxalde, Denis and Perktold, Josef and
            Cimrman, Robert and Henriksen, Ian and Quintero, E. A. and
            Harris, Charles R. and Archibald, Anne M. and
            Ribeiro, Ant{\^o}nio H. and Pedregosa, Fabian and
            {van Mulbregt}, Paul and {SciPy 1.0 Contributors}},
  title   = {{{SciPy} 1.0: Fundamental Algorithms for Scientific
            Computing in Python}},
  journal = {Nature Methods},
  year    = {2020},
  volume  = {17},
  pages   = {261--272},
  adsurl  = {https://rdcu.be/b08Wh},
  doi     = {10.1038/s41592-019-0686-2},
}

@Article{Harris2020,
 title         = {Array programming with {NumPy}},
 author        = {Charles R. Harris and K. Jarrod Millman and St{\'{e}}fan J.
                 van der Walt and Ralf Gommers and Pauli Virtanen and David
                 Cournapeau and Eric Wieser and Julian Taylor and Sebastian
                 Berg and Nathaniel J. Smith and Robert Kern and Matti Picus
                 and Stephan Hoyer and Marten H. van Kerkwijk and Matthew
                 Brett and Allan Haldane and Jaime Fern{\'{a}}ndez del
                 R{\'{i}}o and Mark Wiebe and Pearu Peterson and Pierre
                 G{\'{e}}rard-Marchant and Kevin Sheppard and Tyler Reddy and
                 Warren Weckesser and Hameer Abbasi and Christoph Gohlke and
                 Travis E. Oliphant},
 year          = {2020},
 month         = sep,
 journal       = {Nature},
 volume        = {585},
 number        = {7825},
 pages         = {357--362},
 doi           = {10.1038/s41586-020-2649-2},
 publisher     = {Springer Science and Business Media {LLC}},
 url           = {https://doi.org/10.1038/s41586-020-2649-2}
}

@article{Astropy2018,
  title={The Astropy Project: Building an open-science project and status of the v2. 0 core package},
  author={Price-Whelan, Adrian M and Sip{\H{o}}cz, BM and G{\"u}nther, HM and Lim, PL and Crawford, SM and Conseil, S and Shupe, DL and Craig, MW and Dencheva, N and Ginsburg, A and others},
  journal={The Astronomical Journal},
  volume={156},
  number={3},
  pages={123},
  year={2018},
  publisher={IOP Publishing}
}

@ARTICLE{Satyavolu2023,
       author = {{Satyavolu}, Sindhu and {Eilers}, Anna-Christina and {Kulkarni}, Girish and {Ryan-Weber}, Emma and {Davies}, Rebecca L. and {Becker}, George D. and {Bosman}, Sarah E.~I. and {Greig}, Bradley and {Mazzucchelli}, Chiara and {Ba{\~n}ados}, Eduardo and {Bischetti}, Manuela and {D'Odorico}, Valentina and {Fan}, Xiaohui and {Farina}, Emanuele Paolo and {Haehnelt}, Martin G. and {Keating}, Laura C. and {Lai}, Samuel and {Walter}, Fabian},
        title = "{New quasar proximity zone size measurements at z   6 using the enlarged XQR-30 sample}",
      journal = {\mnras},
         year = 2023,
        month = jul,
       volume = {522},
       number = {4},
        pages = {4918-4933},
          doi = {10.1093/mnras/stad1326},
archivePrefix = {arXiv},
       eprint = {2305.00998},
 primaryClass = {astro-ph.GA},
       adsurl = {https://ui.adsabs.harvard.edu/abs/2023MNRAS.522.4918S}
}

@ARTICLE{Bechtold1994,
       author = {{Bechtold}, Jill and {Crotts}, Arlin P.~S. and {Duncan}, Robert C. and {Fang}, Yihu},
        title = "{Spectroscopy of the Double Quasars Q1343+266A, B: A New Determination of the Size of Lyman-Alpha Forest Absorbers}",
      journal = {\apjl},
         year = 1994,
        month = dec,
       volume = {437},
        pages = {L83},
          doi = {10.1086/187688},
archivePrefix = {arXiv},
       eprint = {astro-ph/9409007},
 primaryClass = {astro-ph},
       adsurl = {https://ui.adsabs.harvard.edu/abs/1994ApJ...437L..83B}
}

@ARTICLE{Dinshaw1997,
       author = {{Dinshaw}, Nadine and {Weymann}, Ray J. and {Impey}, Chris D. and {Foltz}, Craig B. and {Morris}, Simon L. and {Ake}, Tom},
        title = "{Additional Observations and Analysis of the Lyman-{\ensuremath{\alpha}} Absorption Lines toward the QSO Pair Q0107-025A,B}",
      journal = {\apj},
         year = 1997,
        month = dec,
       volume = {491},
       number = {1},
        pages = {45-68},
          doi = {10.1086/304926},
       adsurl = {https://ui.adsabs.harvard.edu/abs/1997ApJ...491...45D}
}

@ARTICLE{Matthee2024,
       author = {{Matthee}, Jorryt and {Golling}, Christopher and {Mackenzie}, Ruari and {Pezzulli}, Gabriele and {Lilly}, Simon and {Schaye}, Joop and {Bacon}, Roland and {Kusakabe}, Haruka and {Urrutia}, Tanya and {Boogaard}, Leindert and {Brinchmann}, Jarle and {Maseda}, Michael V. and {Garel}, Thibault and {Bouch{\'e}}, Nicolas F. and {Wisotzki}, Lutz},
        title = "{Large-scale excess H I absorption around z {\ensuremath{\approx}} 4 galaxies detected in a background galaxy spectrum in the MUSE eXtremely deep field}",
      journal = {\mnras},
         year = 2024,
        month = apr,
       volume = {529},
       number = {3},
        pages = {2794-2806},
          doi = {10.1093/mnras/stae673},
archivePrefix = {arXiv},
       eprint = {2305.15346},
 primaryClass = {astro-ph.GA},
       adsurl = {https://ui.adsabs.harvard.edu/abs/2024MNRAS.529.2794M}
}

@ARTICLE{Rupke2019,
       author = {{Rupke}, David S.~N. and {Coil}, Alison and {Geach}, James E. and {Tremonti}, Christy and {Diamond-Stanic}, Aleksandar M. and {George}, Erin R. and {Hickox}, Ryan C. and {Kepley}, Amanda A. and {Leung}, Gene and {Moustakas}, John and {Rudnick}, Gregory and {Sell}, Paul H.},
        title = "{A 100-kiloparsec wind feeding the circumgalactic medium of a massive compact galaxy}",
      journal = {\nat},
         year = 2019,
        month = oct,
       volume = {574},
       number = {7780},
        pages = {643-646},
          doi = {10.1038/s41586-019-1686-1},
archivePrefix = {arXiv},
       eprint = {1910.13507},
 primaryClass = {astro-ph.GA},
       adsurl = {https://ui.adsabs.harvard.edu/abs/2019Natur.574..643R}
}

@ARTICLE{Rupke2023,
       author = {{Rupke}, David S.~N. and {Coil}, Alison L. and {Perrotta}, Serena and {Davis}, Julie D. and {Diamond-Stanic}, Aleksandar M. and {Geach}, James E. and {Hickox}, Ryan C. and {Moustakas}, John and {Petter}, Grayson C. and {Rudnick}, Gregory H. and {Sell}, Paul H. and {Tremonti}, Christy A. and {Whalen}, Kelly E.},
        title = "{The Ionization and Dynamics of the Makani Galactic Wind}",
      journal = {\apj},
         year = 2023,
        month = apr,
       volume = {947},
       number = {1},
          eid = {33},
        pages = {33},
          doi = {10.3847/1538-4357/acbfae},
archivePrefix = {arXiv},
       eprint = {2303.00194},
 primaryClass = {astro-ph.GA},
       adsurl = {https://ui.adsabs.harvard.edu/abs/2023ApJ...947...33R}
}

@ARTICLE{Conaboy2025,
       author = {{Conaboy}, Luke and {Bolton}, James S. and {Keating}, Laura C. and {Haehnelt}, Martin G. and {Kulkarni}, Girish and {Puchwein}, Ewald},
        title = "{The connection between high-redshift galaxies and Lyman {\ensuremath{\alpha}} transmission in the Sherwood{\textendash}Relics simulations of patchy reionization}",
      journal = {\mnras},
         year = 2025,
        month = may,
       volume = {539},
       number = {3},
        pages = {2790-2805},
          doi = {10.1093/mnras/staf648},
archivePrefix = {arXiv},
       eprint = {2502.02983},
 primaryClass = {astro-ph.CO},
       adsurl = {https://ui.adsabs.harvard.edu/abs/2025MNRAS.539.2790C}
}

@ARTICLE{Meiksin2015,
       author = {{Meiksin}, Avery and {Bolton}, James S. and {Tittley}, Eric R.},
        title = "{Gas around galaxy haloes - II. Hydrogen absorption signatures from the environments of galaxies at redshifts 2 < z < 3}",
      journal = {\mnras},
         year = 2015,
        month = oct,
       volume = {453},
       number = {1},
        pages = {899-913},
          doi = {10.1093/mnras/stv1682},
archivePrefix = {arXiv},
       eprint = {1505.02101},
 primaryClass = {astro-ph.GA},
       adsurl = {https://ui.adsabs.harvard.edu/abs/2015MNRAS.453..899M}
}

@ARTICLE{Meiksin2017,
       author = {{Meiksin}, Avery and {Bolton}, James S. and {Puchwein}, Ewald},
        title = "{Gas around galaxy haloes - III: hydrogen absorption signatures around galaxies and QSOs in the Sherwood simulation suite}",
      journal = {\mnras},
         year = 2017,
        month = jun,
       volume = {468},
       number = {2},
        pages = {1893-1901},
          doi = {10.1093/mnras/stx191},
archivePrefix = {arXiv},
       eprint = {1701.06948},
 primaryClass = {astro-ph.GA},
       adsurl = {https://ui.adsabs.harvard.edu/abs/2017MNRAS.468.1893M}
}

@ARTICLE{Prochaska2013,
       author = {{Prochaska}, J. Xavier and {Hennawi}, Joseph F. and {Lee}, Khee-Gan and {Cantalupo}, Sebastiano and {Bovy}, Jo and {Djorgovski}, S.~G. and {Ellison}, Sara L. and {Lau}, Marie Wingyee and {Martin}, Crystal L. and {Myers}, Adam and {Rubin}, Kate H.~R. and {Simcoe}, Robert A.},
        title = "{Quasars Probing Quasars. VI. Excess H I Absorption within One Proper Mpc of z \raisebox{-0.5ex}\textasciitilde 2 Quasars}",
      journal = {\apj},
         year = 2013,
        month = oct,
       volume = {776},
       number = {2},
          eid = {136},
        pages = {136},
          doi = {10.1088/0004-637X/776/2/136},
archivePrefix = {arXiv},
       eprint = {1308.6222},
 primaryClass = {astro-ph.CO},
       adsurl = {https://ui.adsabs.harvard.edu/abs/2013ApJ...776..136P}
}

@ARTICLE{Mukherjee2018,
       author = {{Mukherjee}, Dipanjan and {Bicknell}, Geoffrey V. and {Wagner}, Alexander Y. and {Sutherland}, Ralph S. and {Silk}, Joseph},
        title = "{Relativistic jet feedback - III. Feedback on gas discs}",
      journal = {\mnras},
         year = 2018,
        month = oct,
       volume = {479},
       number = {4},
        pages = {5544-5566},
          doi = {10.1093/mnras/sty1776},
archivePrefix = {arXiv},
       eprint = {1803.08305},
 primaryClass = {astro-ph.HE},
       adsurl = {https://ui.adsabs.harvard.edu/abs/2018MNRAS.479.5544M}
}

@ARTICLE{Roy2025,
       author = {{Roy}, Namrata and {Heckman}, Timothy and {Henry}, Alaina},
        title = "{Mapping Jet-Gas Coupling and energetic ionized outflows in High-Redshift Radio Galaxies with JWST/NIRSpec}",
      journal = {arXiv e-prints},
         year = 2025,
        month = aug,
          eid = {arXiv:2508.06707},
        pages = {arXiv:2508.06707},
          doi = {10.48550/arXiv.2508.06707},
archivePrefix = {arXiv},
       eprint = {2508.06707},
 primaryClass = {astro-ph.GA},
       adsurl = {https://ui.adsabs.harvard.edu/abs/2025arXiv250806707R}
}

@ARTICLE{HeckmanBest2023,
       author = {{Heckman}, Timothy M. and {Best}, Philip N.},
        title = "{A Global Inventory of Feedback}",
      journal = {Galaxies},
         year = 2023,
        month = jan,
       volume = {11},
       number = {1},
          eid = {21},
        pages = {21},
          doi = {10.3390/galaxies11010021},
archivePrefix = {arXiv},
       eprint = {2301.11960},
 primaryClass = {astro-ph.GA},
       adsurl = {https://ui.adsabs.harvard.edu/abs/2023Galax..11...21H}
}

@ARTICLE{Wagner2012,
       author = {{Wagner}, A.~Y. and {Bicknell}, G.~V. and {Umemura}, M.},
        title = "{Driving Outflows with Relativistic Jets and the Dependence of Active Galactic Nucleus Feedback Efficiency on Interstellar Medium Inhomogeneity}",
      journal = {\apj},
         year = 2012,
        month = oct,
       volume = {757},
       number = {2},
          eid = {136},
        pages = {136},
          doi = {10.1088/0004-637X/757/2/136},
archivePrefix = {arXiv},
       eprint = {1205.0542},
 primaryClass = {astro-ph.CO},
       adsurl = {https://ui.adsabs.harvard.edu/abs/2012ApJ...757..136W}
}

@ARTICLE{Meenakshi2022,
       author = {{Meenakshi}, Moun and {Mukherjee}, Dipanjan and {Wagner}, Alexander Y. and {Nesvadba}, Nicole P.~H. and {Bicknell}, Geoffrey V. and {Morganti}, Raffaella and {Janssen}, Reinier M.~J. and {Sutherland}, Ralph S. and {Mandal}, Ankush},
        title = "{Modelling observable signatures of jet-ISM interaction: thermal emission and gas kinematics}",
      journal = {\mnras},
         year = 2022,
        month = oct,
       volume = {516},
       number = {1},
        pages = {766-786},
          doi = {10.1093/mnras/stac2251},
archivePrefix = {arXiv},
       eprint = {2203.10251},
 primaryClass = {astro-ph.GA},
       adsurl = {https://ui.adsabs.harvard.edu/abs/2022MNRAS.516..766M}
}

@ARTICLE{Kim2013,
       author = {{Kim}, T.-S. and {Partl}, A.~M. and {Carswell}, R.~F. and {M{\"u}ller}, V.},
        title = "{The evolution of H I and C IV quasar absorption line systems at 1.9 < z < 3.2}",
      journal = {\aap},
         year = 2013,
        month = apr,
       volume = {552},
          eid = {A77},
        pages = {A77},
          doi = {10.1051/0004-6361/201220042},
archivePrefix = {arXiv},
       eprint = {1302.6622},
 primaryClass = {astro-ph.CO},
       adsurl = {https://ui.adsabs.harvard.edu/abs/2013A&A...552A..77K}
}

@ARTICLE{Nelson2019,
       author = {{Nelson}, Dylan and {Pillepich}, Annalisa and {Springel}, Volker and {Pakmor}, R{\"u}diger and {Weinberger}, Rainer and {Genel}, Shy and {Torrey}, Paul and {Vogelsberger}, Mark and {Marinacci}, Federico and {Hernquist}, Lars},
        title = "{First results from the TNG50 simulation: galactic outflows driven by supernovae and black hole feedback}",
      journal = {\mnras},
         year = 2019,
        month = dec,
       volume = {490},
       number = {3},
        pages = {3234-3261},
          doi = {10.1093/mnras/stz2306},
archivePrefix = {arXiv},
       eprint = {1902.05554},
 primaryClass = {astro-ph.GA},
       adsurl = {https://ui.adsabs.harvard.edu/abs/2019MNRAS.490.3234N}
}

@ARTICLE{Ayromlou2024,
       author = {{Ayromlou}, Mohammadreza and {Nelson}, Dylan and {Pillepich}, Annalisa and {Rohr}, Eric and {Truong}, Nhut and {Li}, Yuan and {Simionescu}, Aurora and {Lehle}, Katrin and {Lee}, Wonki},
        title = "{An atlas of gas motions in the TNG-Cluster simulation: From cluster cores to the outskirts}",
      journal = {\aap},
         year = 2024,
        month = oct,
       volume = {690},
          eid = {A20},
        pages = {A20},
          doi = {10.1051/0004-6361/202348612},
archivePrefix = {arXiv},
       eprint = {2311.06339},
 primaryClass = {astro-ph.GA},
       adsurl = {https://ui.adsabs.harvard.edu/abs/2024A&A...690A..20A}
}

@ARTICLE{Ehlert2021,
       author = {{Ehlert}, K. and {Weinberger}, R. and {Pfrommer}, C. and {Springel}, V.},
        title = "{Connecting turbulent velocities and magnetic fields in galaxy cluster simulations with active galactic nuclei jets}",
      journal = {\mnras},
         year = 2021,
        month = may,
       volume = {503},
       number = {1},
        pages = {1327-1344},
          doi = {10.1093/mnras/stab551},
archivePrefix = {arXiv},
       eprint = {2011.13964},
 primaryClass = {astro-ph.GA},
       adsurl = {https://ui.adsabs.harvard.edu/abs/2021MNRAS.503.1327E}
}

@ARTICLE{Rorai2017,
       author = {{Rorai}, Alberto and {Hennawi}, Joseph F. and {O{\~n}orbe}, Jose and {White}, Martin and {Prochaska}, J. Xavier and {Kulkarni}, Girish and {Walther}, Michael and {Luki{\'c}}, Zarija and {Lee}, Khee-Gan},
        title = "{Measurement of the small-scale structure of the intergalactic medium using close quasar pairs}",
      journal = {Science},
         year = 2017,
        month = apr,
       volume = {356},
       number = {6336},
        pages = {418-422},
          doi = {10.1126/science.aaf9346},
archivePrefix = {arXiv},
       eprint = {1704.08366},
 primaryClass = {astro-ph.CO},
       adsurl = {https://ui.adsabs.harvard.edu/abs/2017Sci...356..418R}
}

@ARTICLE{Kim2021,
       author = {{Kim}, T.-S. and {Wakker}, B.~P. and {Nasir}, F. and {Carswell}, R.~F. and {Savage}, B.~D. and {Bolton}, J.~S. and {Fox}, A.~J. and {Viel}, M. and {Haehnelt}, M.~G. and {Charlton}, J.~C. and {Rosenwasser}, B.~E.},
        title = "{The evolution of the low-density H I> intergalactic medium from z = 3.6 to 0: data, transmitted flux, and H I> column density,}",
      journal = {\mnras},
         year = 2021,
        month = mar,
       volume = {501},
       number = {4},
        pages = {5811-5833},
          doi = {10.1093/mnras/staa3844},
archivePrefix = {arXiv},
       eprint = {2012.05861},
 primaryClass = {astro-ph.CO},
       adsurl = {https://ui.adsabs.harvard.edu/abs/2021MNRAS.501.5811K}
}

@ARTICLE{Peng2025,
       author = {{Peng}, B. and {Arrigoni Battaia}, F. and {Vishwas}, A. and {Li}, M. and {Iani}, E. and {Sun}, F. and {Li}, Q. and {Ferkinhoff}, C. and {Stacey}, G. and {Cai}, Z. and {Ivison}, R.},
        title = "{Direct high-resolution observation of feedback and chemical enrichment in the circumgalactic medium at redshift z {\ensuremath{\sim}} 2.8}",
      journal = {\aap},
         year = 2025,
        month = feb,
       volume = {694},
          eid = {L1},
        pages = {L1},
          doi = {10.1051/0004-6361/202452610},
archivePrefix = {arXiv},
       eprint = {2410.10993},
 primaryClass = {astro-ph.GA},
       adsurl = {https://ui.adsabs.harvard.edu/abs/2025A&A...694L...1P}
}

@BOOK{OsterbrockFerland2006,
       author = {{Osterbrock}, Donald E. and {Ferland}, Gary J.},
        title = "{Astrophysics of gaseous nebulae and active galactic nuclei}",
         year = 2006,
       adsurl = {https://ui.adsabs.harvard.edu/abs/2006agna.book.....O}
}

@ARTICLE{Fabian2012,
       author = {{Fabian}, A.~C.},
        title = "{Observational Evidence of Active Galactic Nuclei Feedback}",
      journal = {\araa},
         year = 2012,
        month = sep,
       volume = {50},
        pages = {455-489},
          doi = {10.1146/annurev-astro-081811-125521},
archivePrefix = {arXiv},
       eprint = {1204.4114},
 primaryClass = {astro-ph.CO},
       adsurl = {https://ui.adsabs.harvard.edu/abs/2012ARA&A..50..455F}
}

@ARTICLE{Harrison2017,
       author = {{Harrison}, C.~M.},
        title = "{Impact of supermassive black hole growth on star formation}",
      journal = {Nature Astronomy},
         year = 2017,
        month = jul,
       volume = {1},
          eid = {0165},
        pages = {0165},
          doi = {10.1038/s41550-017-0165},
archivePrefix = {arXiv},
       eprint = {1703.06889},
 primaryClass = {astro-ph.GA},
       adsurl = {https://ui.adsabs.harvard.edu/abs/2017NatAs...1E.165H}
}

@ARTICLE{Nesvadba2017b,
       author = {{Nesvadba}, N.~P.~H. and {Drouart}, G. and {De Breuck}, C. and {Best}, P. and {Seymour}, N. and {Vernet}, J.},
        title = "{Gas kinematics in powerful radio galaxies at z   2: Energy supply from star formation, AGN, and radio jets}",
      journal = {\aap},
         year = 2017,
        month = apr,
       volume = {600},
          eid = {A121},
        pages = {A121},
          doi = {10.1051/0004-6361/201629357},
archivePrefix = {arXiv},
       eprint = {1610.01627},
 primaryClass = {astro-ph.GA},
       adsurl = {https://ui.adsabs.harvard.edu/abs/2017A&A...600A.121N}
}

@ARTICLE{KingPounds2015,
       author = {{King}, Andrew and {Pounds}, Ken},
        title = "{Powerful Outflows and Feedback from Active Galactic Nuclei}",
      journal = {\araa},
         year = 2015,
        month = aug,
       volume = {53},
        pages = {115-154},
          doi = {10.1146/annurev-astro-082214-122316},
archivePrefix = {arXiv},
       eprint = {1503.05206},
 primaryClass = {astro-ph.GA},
       adsurl = {https://ui.adsabs.harvard.edu/abs/2015ARA&A..53..115K}
}

@ARTICLE{Curran2013,
       author = {{Curran}, S.~J. and {Whiting}, M.~T. and {Sadler}, E.~M. and {Bignell}, C.},
        title = "{A survey for the missing hydrogen in high-redshift radio sources}",
      journal = {\mnras},
         year = 2013,
        month = jan,
       volume = {428},
       number = {3},
        pages = {2053-2063},
          doi = {10.1093/mnras/sts171},
archivePrefix = {arXiv},
       eprint = {1210.1886},
 primaryClass = {astro-ph.CO},
       adsurl = {https://ui.adsabs.harvard.edu/abs/2013MNRAS.428.2053C}
}

@ARTICLE{Morganti2018,
       author = {{Morganti}, Raffaella and {Oosterloo}, Tom},
        title = "{The interstellar and circumnuclear medium of active nuclei traced by H i 21 cm absorption}",
      journal = {\aapr},
         year = 2018,
        month = jul,
       volume = {26},
       number = {1},
          eid = {4},
        pages = {4},
          doi = {10.1007/s00159-018-0109-x},
archivePrefix = {arXiv},
       eprint = {1807.01475},
 primaryClass = {astro-ph.GA},
       adsurl = {https://ui.adsabs.harvard.edu/abs/2018A&ARv..26....4M}
}
\bibliographystyle{aa}

\onecolumn
\begin{appendix}

\section{Additional material of the MCMC sampling} \label{chap:appendix_MCMC}
\begin{figure*}[!ht]
    \centering
    \includegraphics[width= 0.9\hsize]{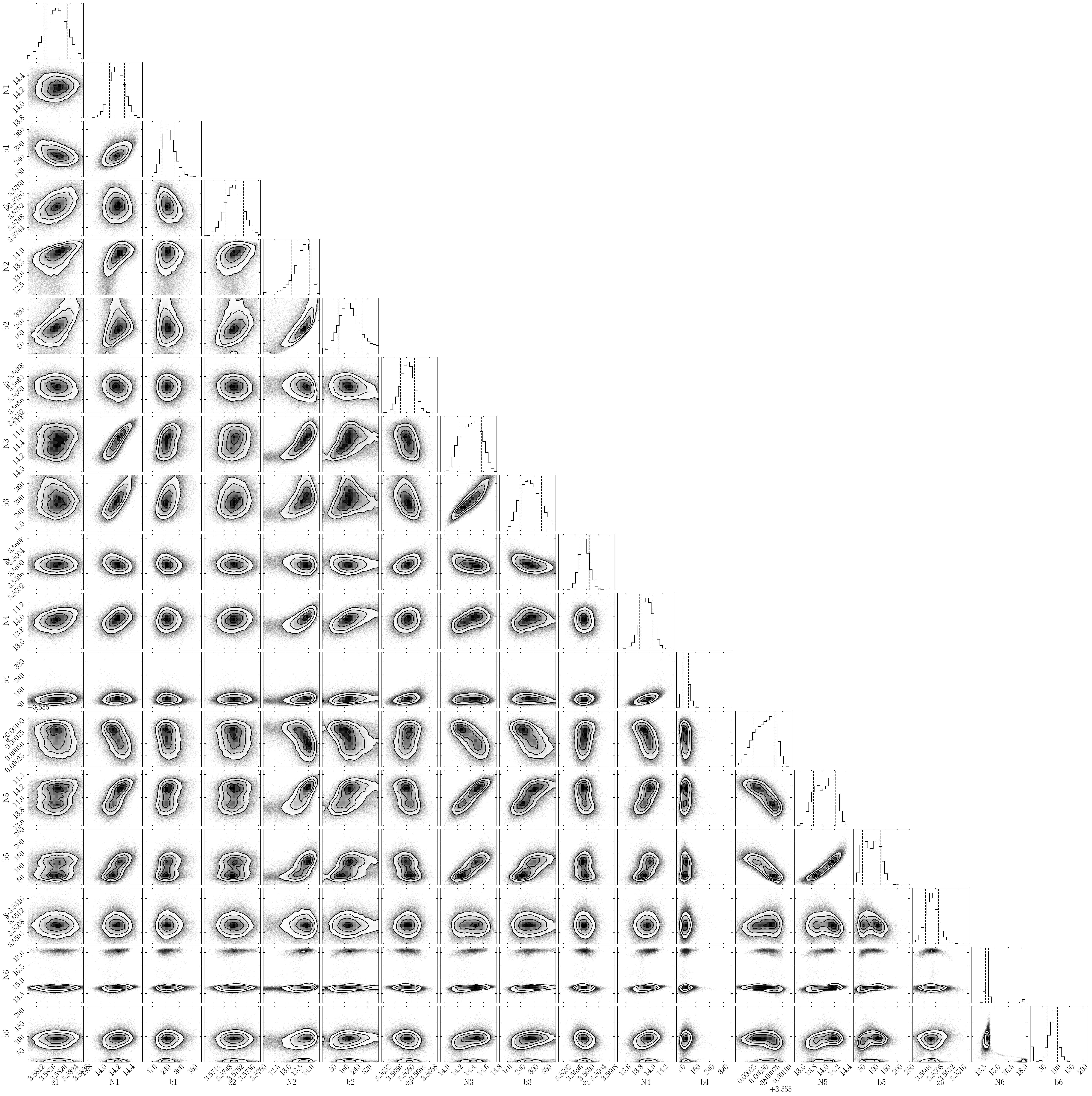}
    \caption{Corner plot derived from MCMC sampling of Ly$\alpha$ of 4C+03.24. The H\,\textsc{i} absorber redshifts $z_i$, logarithmic column densities $N_i$ and Doppler parameters $b_i$ (in $\text{km}\,\text{s}^{-1}$) are shown. The black dotted lines in the histograms indicate the uncertainty ranges, namely the 16th and 84th percentiles. The mean acceptance fraction of the sampling is 0.301. }
    \label{fig:corner_4C0324}
\end{figure*}

\begin{figure*}[!ht]
    \centering
    \includegraphics[width= 0.9\textwidth]{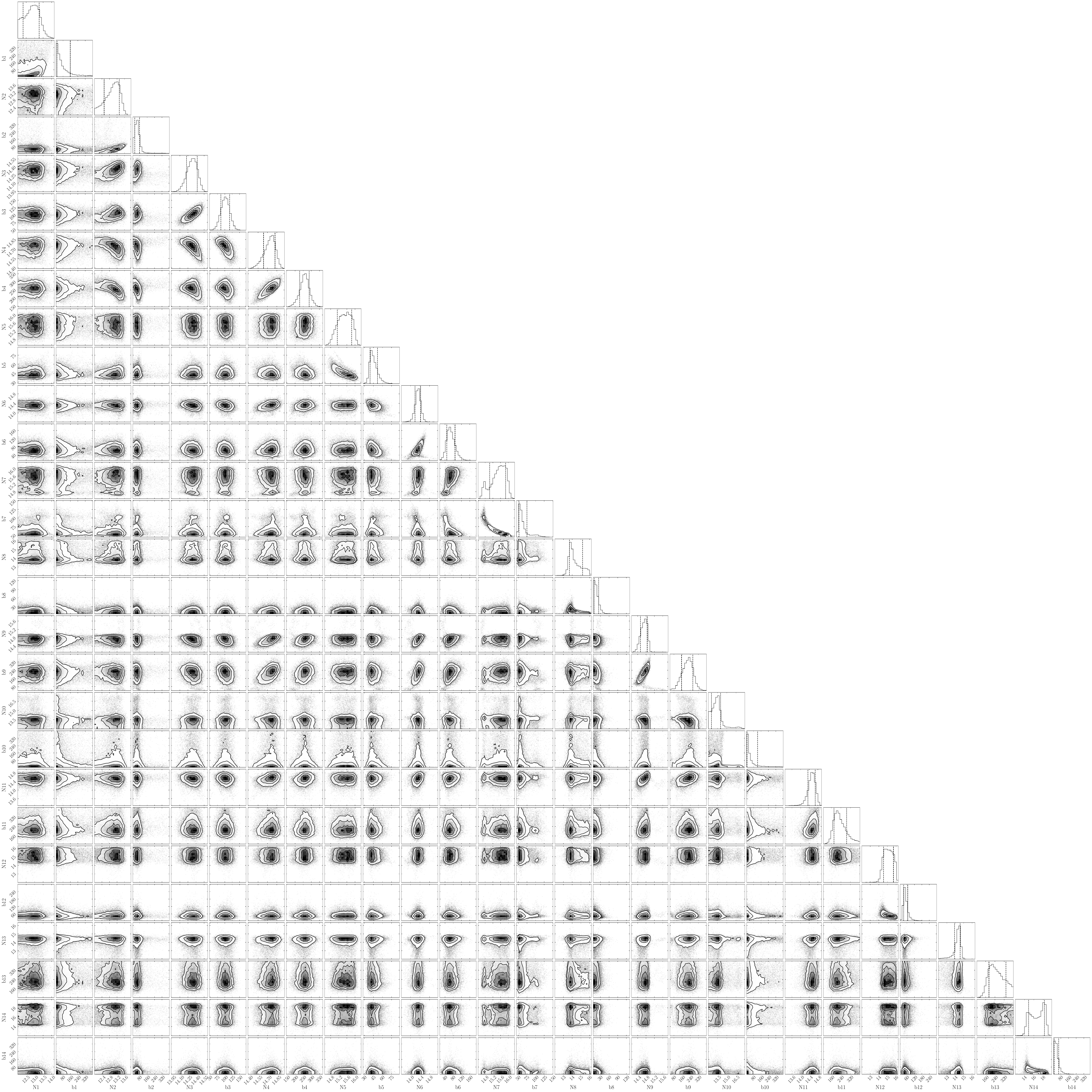}
    \caption{Corner plot derived from MCMC sampling of Ly$\alpha$ of 4C+04.11. We only show the H\,\textsc{i} logarithmic column densities $N_i$ and Doppler parameters $b_i$ (in $\text{km}\,\text{s}^{-1}$) because of the large number of absorbers identified for this target. The black dotted lines in the histograms indicate the uncertainty ranges, namely the 16th and 84th percentiles. The blue lines mark the median, i.e. the reported fit values. The mean acceptance fraction of the sampling is 0.140. }
    \label{fig:corner_4C0411}
\end{figure*}

\begin{figure*}[!ht]
    \centering
    \includegraphics[width= 0.9\textwidth]{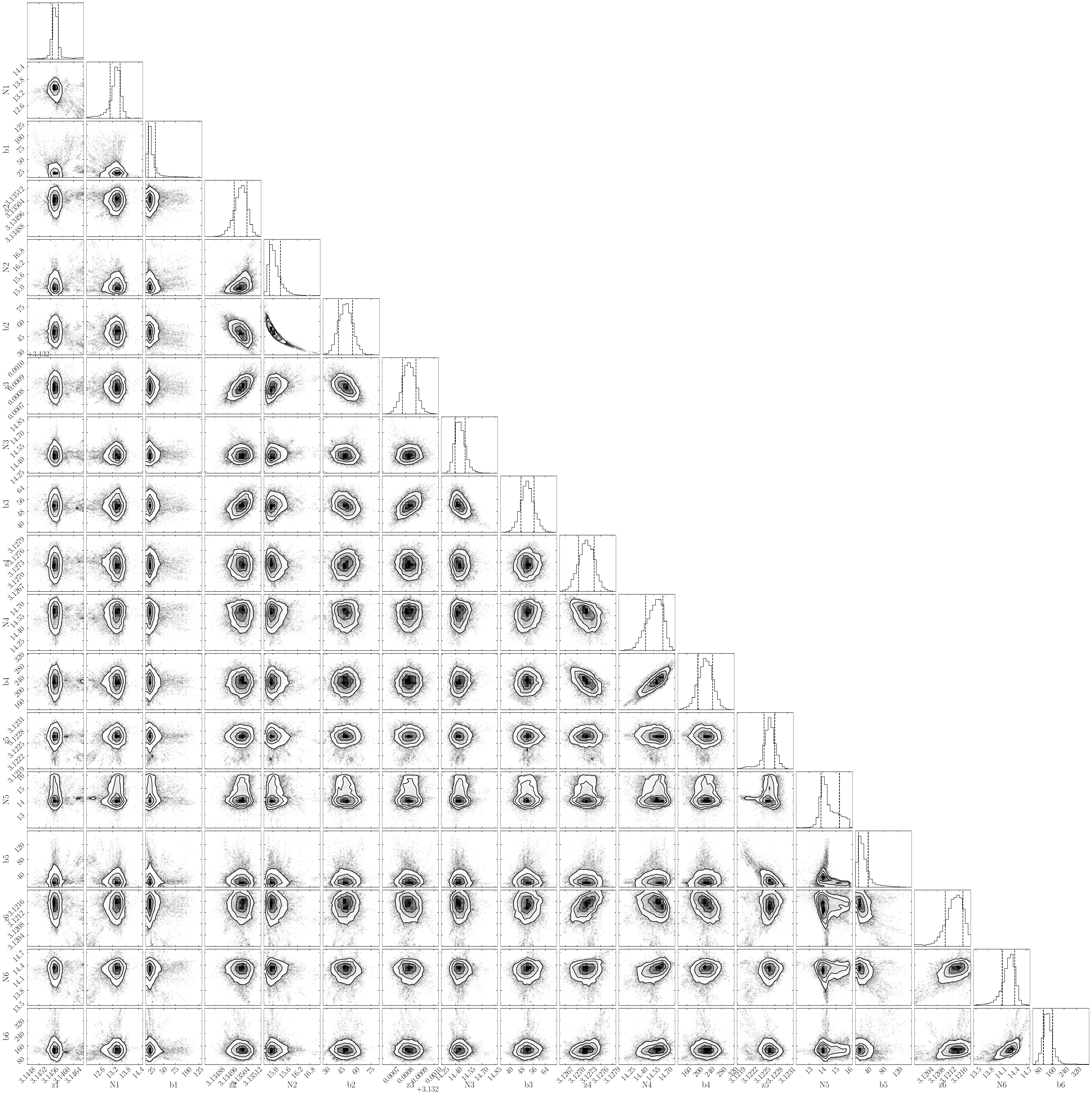}
    \caption{Corner plot derived from MCMC sampling of Ly$\alpha$ of MRC\,0316. The H\,\textsc{i} absorber redshifts $z_i$, logarithmic column densities $N_i$ and Doppler parameters $b_i$ (in $\text{km}\,\text{s}^{-1}$) are shown. The black dotted lines in the histograms indicate the uncertainty ranges, namely the 16th and 84th percentiles. The blue lines mark the median, i.e. the reported fit values. The mean acceptance fraction of the sampling is 0.270.}
    \label{fig:corner_MRC0316}
\end{figure*}

\begin{figure*}[!ht]
    \centering
    \includegraphics[width= 0.9\textwidth]{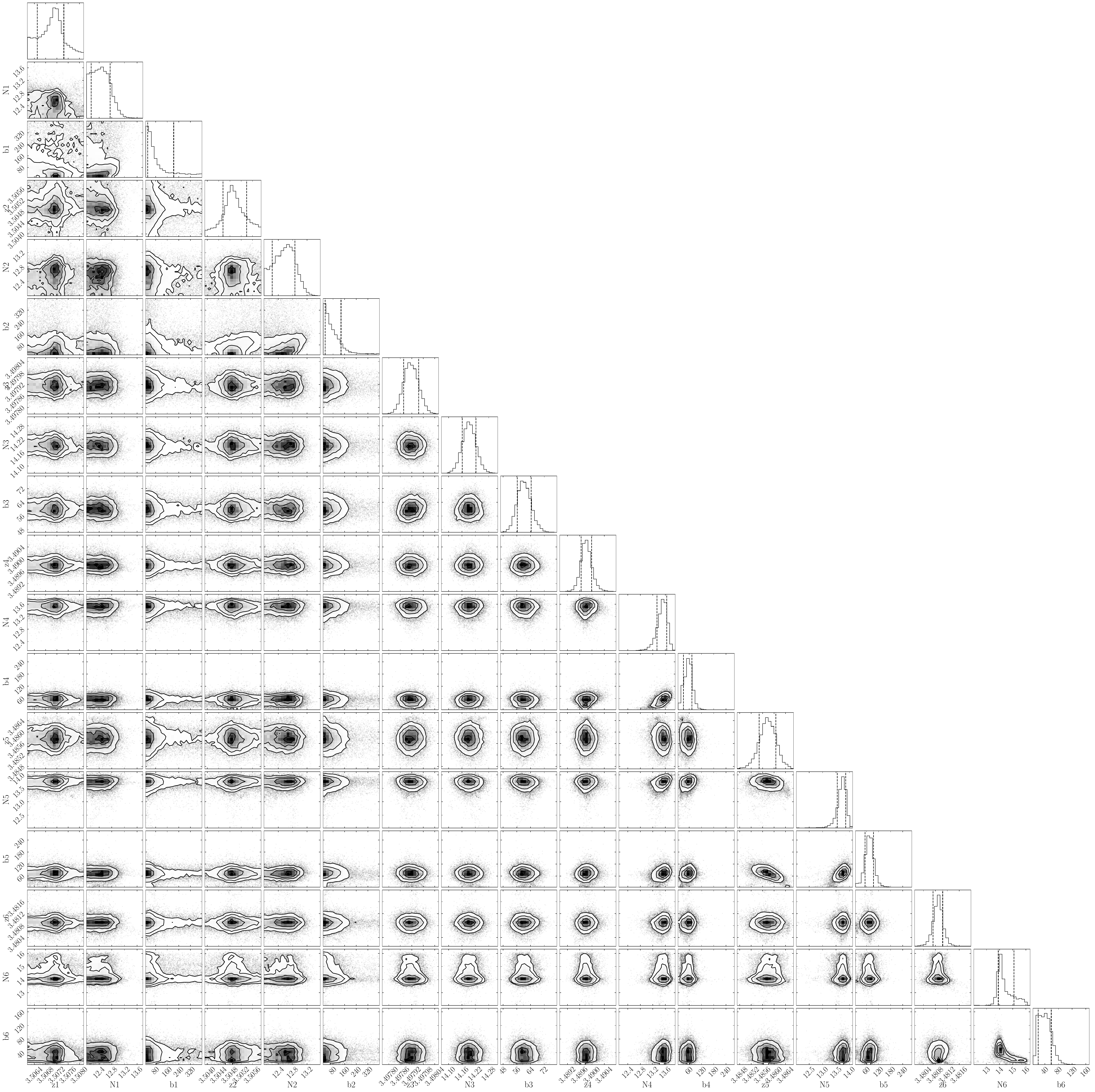}
    \caption{Corner plot derived from MCMC sampling of Ly$\alpha$ of TN\,J0205. The H\,\textsc{i} absorber redshifts $z_i$, logarithmic column densities $N_i$ and Doppler parameters $b_i$ (in $\text{km}\,\text{s}^{-1}$) are shown. The black dotted lines in the histograms indicate the uncertainty ranges, namely the 16th and 84th percentiles. The blue lines mark the median, i.e. the reported fit values. The mean acceptance fraction of the sampling is 0.305.}
    \label{fig:corner_TNJ0205}
\end{figure*}

In this Appendix, we present auxiliary material related to the MCMC sampling used to estimate the statistical uncertainties on the fitted Ly$\alpha$ line parameters from the UVES master spectra. In table \ref{tab:MCMC_constraints}, the priors for the MCMC sampling of the UVES master spectra are listed. Figures \ref{fig:corner_4C0324}, \ref{fig:corner_4C0411}, \ref{fig:corner_MRC0316} and \ref{fig:corner_TNJ0205} show the corner plots obtained from the MCMC sampling. These plots are additionally made available as supplementary online material on Zenodo. The corner plots display the probability distribution of parameters and their pairwise correlations. The off-diagonal subplots are contour plots of parameter pairs, indicating their joint distributions. Degeneracies between fitted parameters can appear as elongated or curved shapes in the off-diagonal subplots, typically oriented diagonally. We evaluate here the correlations between the Voigt parameters, namely the H\,\textsc{i} absorber redshifts $z_i$, logarithmic column densities $N_i$ and Doppler parameters $b_i$ (in $\text{km}\,\text{s}^{-1}$). As we identify 14 absorbers for 4C+04.11, we only show the corner plots for the column density and Doppler parameter. The majority of parameters seem well constrained, however several are only weakly constrained and show extended posterior tails. This behavior is likely due to a combination of the noisy data, the large number of free parameters and intrinsic degeneracies in the model. In particular, the degeneracy between column density $N$ and Doppler parameter $b$ implies that combinations of small $N$ and large $b$ can produce Voigt profiles similar to those obtained with large $N$ and small $b$. This is an issue that has been already identified by \citet{Silva2018b} in the fitting of the Ly$\alpha$ profile of the HzRG MRC\,0943–242.

\begin{table*}[!ht]
\caption{Fitting constraints for the MCMC sampling of the UVES master spectra.}
\begin{center}
\label{tab:MCMC_constraints}
\begin{tabular}{ |l|c|c| }
 \hline
 Fitting parameter & lower boundary & upper boundary\\
 \hline
 \hline
 \multicolumn{3}{|c|}{Gaussian parameters } \\
 \hline
  Line center, $\lambda_c$ $[\text{\AA}]$ \tnote & $\lambda_{c,i}-10$ & $\lambda_{c,i}+10$ \\
  Line flux, $F$  & - &  - \\
  Line width, $\sigma$  & $0.001$ & 60 \\
  Intercept & -2 & 10 \\
 \hline
 \multicolumn{3}{|c|}{Voigt parameters} \\
 \hline
 Column density, $\log(N_\text{H}/\text{cm}^{-2})$ & 12 & 20 \\
 Doppler parameter, $b\,[\text{km}\,\text{s}^{-1}]$ & 5 & 400 \\
 Redshift, $z$ & $z_i-0.001$ & $z_i+0.001$ \\
 \hline
\end{tabular}
\tablefoot{$\lambda_{c,i}$ and $z_i$ are the initial-guess values for the emission-line center and absorber redshifts, respectively. For multi-component Gaussian fits, we impose an ordering constraint on the line centers. For 4C+04.11, both Gaussian line centers are constrained to lie within $\pm 3$\,\AA\ of the wavelength obtained from a preliminary least-squares fit of the red wing of the emission line, as the blue wing is heavily affected by absorption.}
\end{center}
\end{table*}

\clearpage

\section{JWST NIRSpec narrowband images for 4C+03.24 and TNJ\,0205} \label{chap:JWSTnarrowband}
\begin{figure*}[!ht]
    \centering
    \includegraphics[width= 1.0\textwidth]{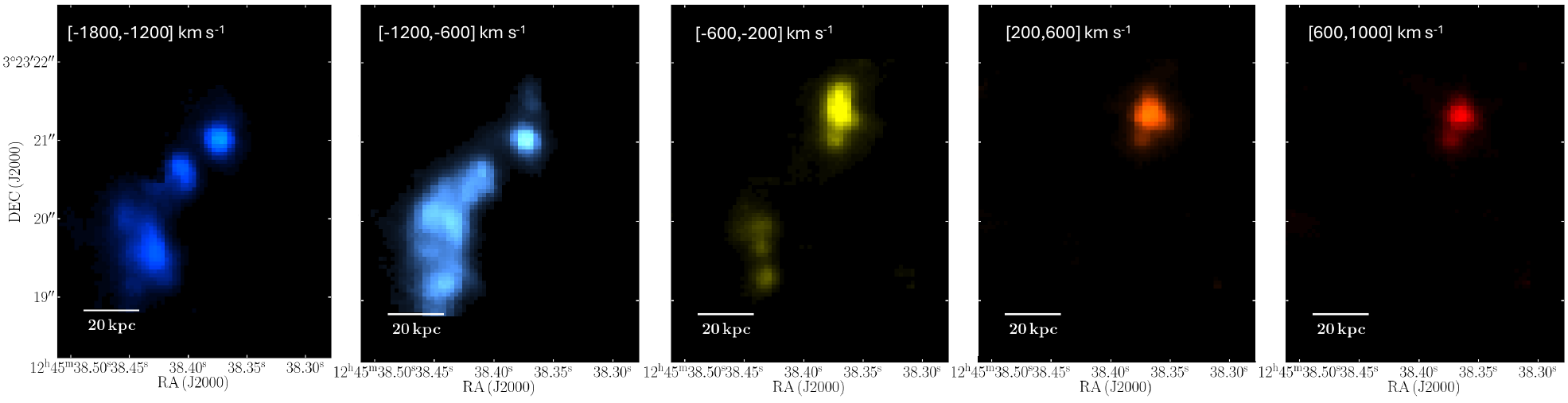}
    \caption[{Narrowband images of the [\ion{O}{iii}]\,$\lambda5007$ emission of 4C+03.24}]{Narrowband images of the [\ion{O}{iii}]\,$\lambda5007$ emission of 4C+03.24, collapsed at different velocity ranges. }
    \label{fig:4C0324_narrowband_OIII}
\end{figure*}

\begin{figure*}[!ht]
    \centering
    \includegraphics[width= 1.0\textwidth]{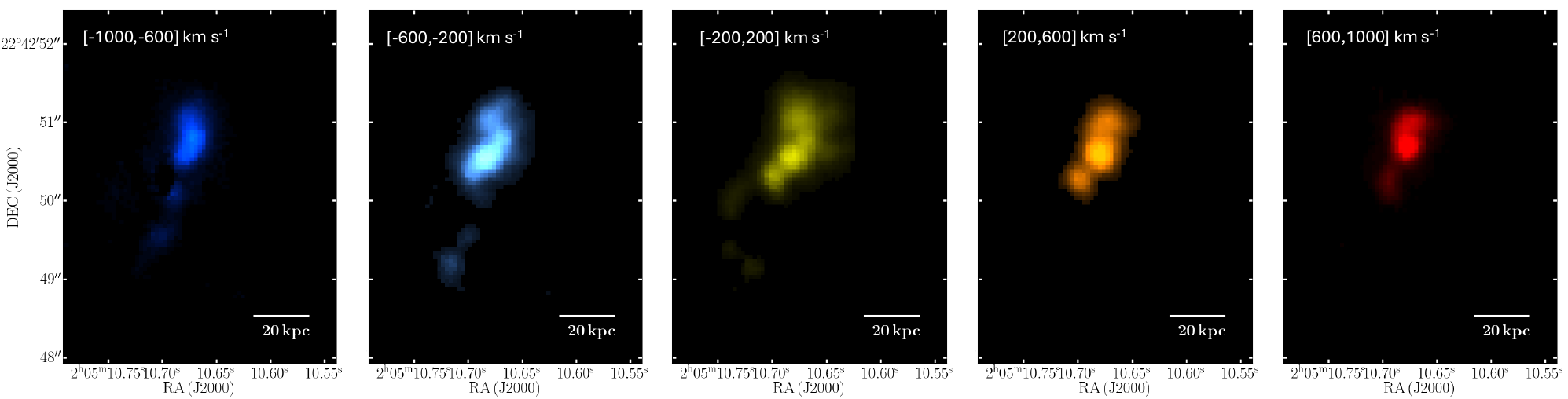}
    \caption[{Narrowband images of the [\ion{O}{iii}]\,$\lambda5007$ emission of TN\,J0205}]{Narrowband images of the [\ion{O}{iii}]\,$\lambda5007$ emission of TN\,J0205, collapsed at different velocity ranges.}
    \label{fig:TNJ0205_narrowband_OIII}
\end{figure*}

In this appendix, we show the NIRSpec narrowband images of the [\ion{O}{iii}]$\lambda5007$ emission for 4C+03.24 (Figure \ref{fig:4C0324_narrowband_OIII}) and TN\,J0205 (Figure \ref{fig:TNJ0205_narrowband_OIII}). They reveal complex kinematics of the ionized interstellar gas. In particular, 4C+03.24 shows blueshifted emission extended towards the south, where jet-gas interactions are taking place (e.g. \citet{vanOjik1996}). \\

We also list the fit results of the comparison between the UVES Ly$\alpha$ profiles and the JWST [\ion{O}{iii}] profiles, as shown in Figures \ref{fig:4C0324_jwst_uves_muse} and \ref{fig:TNJ0205_jwst_uves_muse} in the following. We model the [\ion{O}{iii}] emission lines using up to three Gaussian components and fit the continuum with a zero-order polynomial. The UVES spectra are fitted with the same procedure as used for the UVES master spectra, using MCMC sampling. The velocity shifts $\Delta v$ and the FWHM of the [\ion{O}{iii}]\,$\lambda5007$ and Ly$\alpha$ emission components in the different spatial regions are shown in Tables 
\ref{tab:
4C0324_JWST_UVES} and \ref{tab: TNJ0205_JWST_UVES}.

\begin{table}[htbp]
\caption[{Gaussian fit results of the comparison between UVES Ly$\alpha$ and JWST [\ion{O}{iii}] of 4C+03.24}]{Gaussian fit results for 4C+03.24 in the different regions shown in Figure \ref{fig:4C0324_jwst_uves_muse}.}
\centering
\small
\setlength{\tabcolsep}{5pt}
\renewcommand{\arraystretch}{1.15}
\begin{adjustbox}{width=0.8\textwidth}
\label{tab: 4C0324_JWST_UVES}
\begin{threeparttable}
\begin{tabular}{ |c c||c|c|c||c| }
 \hline
 \multicolumn{2}{|c||}{} &
      \multicolumn{3}{c||}{JWST [\ion{O}{iii}]$\lambda5007$ emission} &
      \multicolumn{1}{c|}{UVES Ly$\alpha$ emission} \\
 \hline
 \hline
  & & Systemic & Blueshifted & Redshifted & Systemic \\
 \hline
 Region 1: & $\Delta v$ [$\text{km}\,\text{s}^{-1}$]& $579 \pm 13$  &$-1389\pm 129$ & $1692 \pm 33$ &\asym{236}{75}{44}\\
   &  FWHM [$\text{km}\,\text{s}^{-1}$]& $1820 \pm 27$ &$2526 \pm 107$ & $853 \pm 33$ & \asym{2351}{116}{107}\\
 \hline
 Region 2: & $\Delta v$ [$\text{km}\,\text{s}^{-1}$]& $-99 \pm 5$  &$-1155\pm 93$ & $105 \pm 82$ & \asym{317}{23}{23}\\
   &  FWHM [$\text{km}\,\text{s}^{-1}$]& $674 \pm 12$ &$2073 \pm 75$ & $2034 \pm 98$ & \asym{1890}{45}{48}\\
 \hline
 Region 3: &$\Delta v$ [$\text{km}\,\text{s}^{-1}$] & $96 \pm 2$ & &  & \asym{132}{14}{17}\\
 &  FWHM [$\text{km}\,\text{s}^{-1}$]& $727 \pm 5$ & &   & \asym{1758}{39}{39}\\
 \hline
 Region 4: & $\Delta v$ [$\text{km}\,\text{s}^{-1}$]& $184 \pm 5$  &$- 286 \pm 13 $& &\asym{67}{31}{31}\\
   &  FWHM [$\text{km}\,\text{s}^{-1}$]& $470  \pm 12$ & $653 \pm 22$& & \asym{1537}{70}{75}\\
 \hline
\end{tabular}
\tablefoot{The velocity shifts and FWHM of the Gaussian components for the JWST [\ion{O}{iii}]$\lambda5007$ emission are shown on the left and for the UVES Ly$\alpha$ emission on the right. The Ly$\alpha$ parameters and uncertainties are derived from the posterior distributions of the MCMC sampling. The velocity shifts are calculated with respect to the systemic redshift of the target as determined from the [\ion{O}{iii}] line \citep{Nesvadba2017a}.}

\end{threeparttable}
\end{adjustbox}
\end{table}

\begin{table}[htbp]
\caption[{Gaussian fit results of the comparison between UVES Ly$\alpha$ and JWST [\ion{O}{iii}] of TN\,J0205}]  {Gaussian fit results for TNJ\,0205 in the different regions shown in Figure \ref{fig:TNJ0205_jwst_uves_muse}.}
\small
\centering
\setlength{\tabcolsep}{5pt}
\renewcommand{\arraystretch}{1.15}
\begin{adjustbox}{width=0.8\textwidth}
\label{tab: TNJ0205_JWST_UVES}
\begin{threeparttable}
\begin{tabular}{ |c c||c|c|c||c|c| }
 \hline
 \multicolumn{2}{|c||}{} &
      \multicolumn{3}{c||}{JWST [\ion{O}{iii}]$\lambda5007$ emission} &
      \multicolumn{2}{c|}{UVES Ly$\alpha$ emission} \\

 \hline
 \hline
  & & Systemic & Blueshifted & Redshifted & Systemic & Blueshifted \\
 \hline
 Region 1: & $\Delta v$ [$\text{km}\,\text{s}^{-1}$]& $-53 \pm 9$  &$-186 \pm 5$ & $169 \pm 4$ &\asym{117}{23}{21} &\asym{33}{14}{19}\\
   &  FWHM [$\text{km}\,\text{s}^{-1}$]& $1448 \pm 25$ &$364 \pm 13$ & $275 \pm 8$ & \asym{1007}{99}{105} & \asym{1773}{85}{68}\\
 \hline
 Region 2: & $\Delta v$ [$\text{km}\,\text{s}^{-1}$]& $151 \pm 4$  &$-263 \pm 9$ & $538\pm 23$ &\asym{108}{12}{12} &\asym{-28}{35}{44}\\
   &  FWHM [$\text{km}\,\text{s}^{-1}$]& $301 \pm 7$ &$645 \pm 16$ & $498 \pm 19$ & \asym{1014}{57}{59} &  \asym{2173}{196}{159}\\
 \hline
 Region 3: &$\Delta v$ [$\text{km}\,\text{s}^{-1}$] & $-201 \pm 44$ & $-313 \pm 8$ & $128 \pm 6$  &\asym{-11}{27}{27} &\asym{-155}{53}{88}\\
 &  FWHM [$\text{km}\,\text{s}^{-1}$]& $1944 \pm 69$ &$410 \pm 13$ & $369 \pm 11$    & \asym{963}{123}{125} & \asym{1885}{239}{171}\\
 \hline
 Region 4: & $\Delta v$ [$\text{km}\,\text{s}^{-1}$]& $-235 \pm 5$  & & &\asym{-144}{19}{19} &\asym{-183}{26}{35}\\
   &  FWHM [$\text{km}\,\text{s}^{-1}$]& $340 \pm 13$ & & &\asym{755}{68}{66}& \asym{1591}{122}{121}\\
 \hline
\end{tabular}
\tablefoot{The velocity shifts and FWHM of the Gaussian components for the JWST [\ion{O}{iii}]$\lambda5007$ emission are shown on the left and for the UVES Ly$\alpha$ emission on the right. The Ly$\alpha$ parameters and uncertainties are derived from the posterior distributions of the MCMC sampling. The velocity shifts are calculated with respect to the systemic redshift of the target as determined from the \ion{He}{ii} line \citep{Kolwa2023}. We note that the naming conventions for the systemic and blueshifted components of the Ly$\alpha$ emission are based on comparison with the UVES master spectra. }

\end{threeparttable}
\end{adjustbox}
\end{table}

\end{appendix}

\label{LastPage}
\end{document}